\documentclass[amsmath, amssymb, floatfix, reprint,onecolumn, notitlepage, 10pt]{revtex4-1}
\usepackage[T1]{fontenc}
\usepackage[utf8]{inputenc}
\usepackage{microtype,bm,bbm,graphicx,booktabs,times}
\usepackage[usenames,dvipsnames]{xcolor}
\usepackage[normalem]{ulem}

\usepackage{booktabs}

\usepackage{setspace}
\selectfont{}

\usepackage{hyperref}
\usepackage{subfiles} 
\hypersetup{colorlinks,allcolors=NavyBlue,linktoc=all}
\usepackage[capitalize]{cleveref}
\crefformat{section}{Sec.~#2#1#3}

\usepackage{multirow}
\usepackage{tabularx}
\usepackage{textgreek}

\newcommand{\dl}[1]{#1}

\begin{document}
\title{Integrated quantum photonics with silicon carbide: challenges and prospects}
\author{Daniil M. Lukin}
\author{Melissa A. Guidry}
\author{Jelena Vu\v{c}kovi\'c}
\email{jela@stanford.edu}
\affiliation{E. L. Ginzton Laboratory, Stanford University, Stanford, CA 94305, USA.}

\date{\today}

\begin{abstract}
\dl{Optically-addressable solid-state spin defects are promising candidates for storing and manipulating quantum information using their long coherence ground state manifold; individual defects can be entangled using photon-photon interactions, offering a path toward large scale quantum photonic networks.
Quantum computing protocols place strict limits on the acceptable photon losses in the system. These low-loss requirements cannot be achieved without photonic engineering, but are attainable if combined with state-of-the-art nanophotonic technologies.
However, most materials that host spin defects are challenging to process: as a result, the performance of quantum photonic devices is orders of magnitude behind that of their classical counterparts.
Silicon carbide (SiC) is well-suited to bridge the classical-quantum photonics gap, since it hosts promising optically-addressable spin defects and can be processed into SiC-on-insulator for scalable, integrated photonics. In this Perspective, we discuss recent progress toward the development of scalable quantum photonic technologies based on solid state spins in silicon carbide, and discuss current challenges and future directions.}

\end{abstract}
\maketitle

\section*{Introduction}

Quantum information processing (QIP) is among the most rapidly developing areas of science and technology. It is perhaps the final frontier in the quest to harness the fundamental properties of 
matter for computation, communication, data processing, and molecular simulation. Any physical system governed by the laws of quantum mechanics can in principle be a candidate for QIP; to date, however, the most advanced QIP demonstrations have been implemented via superconducting qubits \cite{arute2019quantum}, trapped ions and atoms \cite{zhang2017observation, bernien2017probing}, and photons (via linear-optical quantum computing)  \cite{harris2017quantum}. Recently, optically-addressable crystal defects have emerged as a novel platform for QIP \cite{atature2018material, awschalom2018quantum, wang2019integrated, elshaari2020hybrid}, interfacing some of nature's best quantum memories (a protected solid-state spin \cite{zhong2015optically, bradley2019ten, miao2020universal}) with a robust flying qubit (photon) that can transport the quantum information \cite{hensen2015loophole}. Notably, solid-state defects lend themselves to on-chip integration, promising future scalability. Optically-addressable spin defects are thus noteworthy candidates for several QIP proposals, including network-based quantum computing \cite{nemoto2014photonic, nickerson2013topological, nickerson2014freely}, cluster state generation \cite{buterakos2017deterministic, russo2018photonic, schwartz2016deterministic}, and quantum communications \cite{muralidharan2016optimal, borregaard2020one}.

In recent years, the field of defect-based QIP has made extraordinary strides toward realizing such proposals. Breakthroughs include the demonstration of long-distance entanglement of solid state spins \cite{hensen2015loophole}; high fidelity  single-shot readout ($F> 0.9995$) of a color center spin state and memory-enhanced quantum communication \cite{bhaskar2020experimental}; nanophotonic quantum memories based on rare-earth ensembles \cite{zhong2017nanophotonic}
; entanglement distillation between distant electron-nuclear two-qubit nodes \cite{kalb2017entanglement}; and a 10-qubit quantum register based on nuclear spins coupled to a single color center, with single-qubit coherence exceeding one minute \cite{bradley2019ten}. Although scalability is cited as a strength of optically-addressable spin defects, the field has yet to demonstrate a breakthrough toward this end: So far, entanglement has been realized between at most two optically-connected color centers, whether in fiber networks \cite{hensen2015loophole} or on a chip \cite{evans2018photon}. A central issue is the efficient interaction between defects and photons.
The photon emission of a dipole source is difficult to direct into a single optical mode, a prerequisite for photon interference. 
The resulting low collection efficiencies translate into prohibitively low rates of higher-dimensional entanglement generation. However, by integrating a defect with a nanophotonic cavity, one can greatly enhance the photon emission rate into the cavity mode via the Purcell effect, thereby funneling the majority of emitted photonics into the desired optical channel. This powerful technique has, for instance, enabled single-shot readout of single rare-earth ions \cite{kindem2020control, raha2020optical}, which are too dim outside of a cavity to even observe individually. Many of the aforementioned recent breakthroughs in defect-based QIP have been enabled by integration with nanophotonics \cite{bhaskar2020experimental, zhong2017nanophotonic, evans2018photon}. 

Defect-based integrated quantum photonics \cite{wang2019integrated, elshaari2020hybrid} is thus a recent and exciting union of two distinct and rapidly developing fields: the study of quantum spintronics \cite{awschalom2018quantum, atature2018material} and the development of classical integrated photonics \cite{wang2018integrated, heck2014ultra, liu2020monolithic, liu2020high}. In this Perspective, we discuss the challenges and prospects for the field in context of the recent advances in silicon carbide (SiC). Silicon carbide hosts a wide range of optically-active defects \cite{castelletto2020silicon} and is amenable to CMOS-compatible photonics fabrication technologies \cite{Lukin20194H, song2019ultrahigh, Guidry2020Optical}, and is thus a key candidate in implementing fully-integrated quantum photonic circuits. We begin with a brief summary of the fundamental requirements and challenges of defect-based QIP, followed by a summary of the state-of-the-art in SiC defects and in SiC photonics. We then discuss the feasibility of overcoming key challenges to develop scalable integrated quantum photonic circuits.

\begin{figure*}[tp]
\centering
\includegraphics[width=\textwidth]{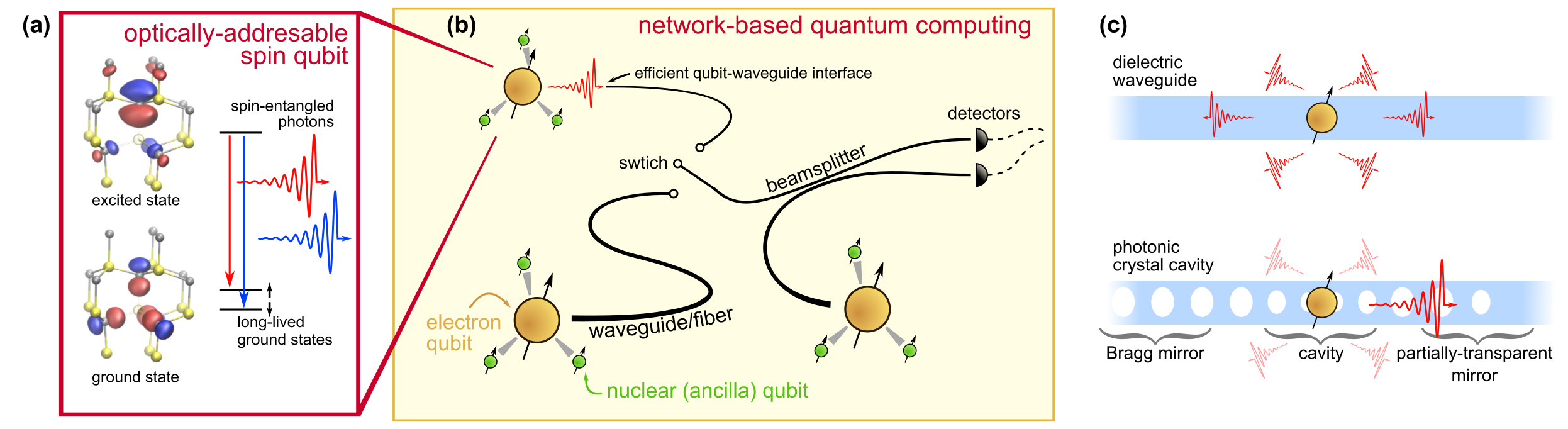}
\caption{Quantum photonics with optically-addressable spins. (a) A suitable defect features a spin-selective optical transition, where a photon degree of freedom (i.e. polarization, frequency, or time-bin) is entangled with a ground-state spin featuring a long coherence time. The defect's optical lifetime is determined by the magnitude of its optical dipole moment, which in turn is dictated by the orbital structure of the excited and ground state. (b) A quantum photonic network consists of multi-qubit registers, each consisting of an optically-addressable electron spin strongly-coupled to nearby nuclear spins. The registered are integrated together in a network via an efficient waveguide or fiber interface. The network is equipped with beamsplitters and switches (which may be one and the same depending on the implementation) for long-distance entanglement and circuit reconfigurability. Low losses at all stages (including efficient photon collection from the defect, low-loss photon propagation in waveguide, and high detector efficiency) are essential for fault-tolerant computation, efficient quantum simulation, and long-distance quantum communications. (c) Due to the weak confinement of optical photons in dielectric structures, light from a quantum emitter does not couple efficiently to a simple dielectric waveguide. Instead, a nanophotonic cavity or a slow-light waveguide mode must be used to enhance the emission into the waveguide mode via the Purcell effect. Orbital graphic in (a) adapted from \cite{nagy2019high}}\label{fig:intro}
\end{figure*}

\section*{Motivation for combining spin defects with nanophotonics}
An optically-addressable defect, illustrated in Fig.~\ref{fig:intro}a, features a ground state manifold with a long coherence time which can emit spin-entangled photons. This manifests as spin-dependent photon emission, where either the polarization or frequency of the photon encodes the electron spin state (alternatively, time-bin entanglement can be used \cite{vasconcelos2020scalable}). The electronic spin can also be coupled to one or more nearby nuclear spins \cite{awschalom2018quantum,son2020developing, bradley2019ten}. Thus the defect can serve as a multi-qubit register, for applications in error-corrected quantum computation (as part of a quantum photonic network, Fig.~\ref{fig:intro}b) \cite{nickerson2014freely} or as a source of photonic cluster states \cite{buterakos2017deterministic, russo2018photonic} for quantum communications and linear-optical quantum computing.

Spin-entangled optical photons are ideal carriers of quantum information for generating remote entanglement: Their high energy renders them insensitive to decoherence at room temperature (enabling routing of quantum information via the same commercial fibers that route classical data) and makes it possible to measure them with high quantum efficiency \cite{esmaeil2020efficient}. However, optical photons are difficult to confine within an integrated circuit.
In contrast with microwave photons whose efficient confinement in metal wires enables photolithographically-defined superconducting qubit circuits, optical photons are only weakly confined in dielectric structures using refractive index contrast.
Distributed Bragg reflectors, such as photonic crystals, can be used to engineer a fully-reflecting boundary to confine and route light \cite{meade1995photonic}, but the high photon energy dictates the small feature size of these confining structures, requiring more sophisticated nanofabrication methods. Furthermore, since three-dimensional photonic crystals \cite{meade1995photonic} (which create a complete $4\pi$ steradian bandgap) are not yet practical, all means to confine light on a chip still rely on weak confinement via total internal reflection along at least one spatial dimension. Although a waveguide based on total internal reflection is a theoretically lossless and experimentally practical structure for routing photons on-chip, it is not straightforward to make a defect efficiently emit photons into the waveguide in the first place. This can be intuitively understood by decomposing the dipole radiation pattern in the plane-wave basis and noting that only a modest fraction of dipole emission goes into those plane wave modes which get totally internally reflected, while the rest are scattered into a free space continuum of modes. In order to increase the fraction of photons emitted into the desired mode (referred to as the $\beta$ factor), one must rely on more advanced techniques, through careful control over the electromagnetic local density of states (LDOS). By embedding an emitter into a nanophotonic cavity that is coupled strongly to a waveguide, or by enhancing the LDOS in the waveguide itself \cite{arcari2014near}, it is possible to enhance the defect's single-mode emission while maximally suppressing all other scattering (Fig.~\ref{fig:intro}c). Using this approach, $\beta$ factors approaching unity (to the point where they are negligible compared to other losses) have been achieved \cite{kindem2020control, bhaskar2020experimental}. Since photon emission is reciprocal to photon absorption, a defect well-coupled to a waveguide is equivalently suitable for reflection-based spin-photon entanglement via dipole-induced transparency \cite{waks2006dipole}.

The control of LDOS of the quantum emitter not only minimizes the undesirable interactions with free-space modes, but also reduces the effects of other emitter nonidealities caused by its interaction with the solid-state environment. There are several mechanisms that degrade a defect's performance as a spin-photon interface: First, environmental fluctuations induce decoherence of the emitted photons, manifesting as the broadening of optical linewidths beyond the Fourier-transform limit (thus reducing the emitted photon indistinguishability). Second, many emitters have non-radiative pathways, via for example phonon-assisted spin-mixing transitions \cite{dong2019spin}. As a result, most emitters have a non-unity \textit{quantum efficiency}, meaning not every excitation yields a photon. Third, coupling of the optical transition to optical phonons creates an additional decay pathway, whereby a photon is emitted in combination with one or more phonons. This emission is broad in spectrum (>0.1~eV) and cannot be used for entanglement purposes. The fraction of ``useful'' --- direct --- emission into the zero-phonon line (ZPL) is referred to as the Debye-Waller Factor (DWF). The DWF depends on the electronic orbital structure of the defect, and thus varies greatly for different defect types. Modification of the LDOS via, for example, integration of the defect into a cavity, enhances the emission rate into the ZPL via the Purcell effect, thus boosting the effective DWF and the quantum efficiency of the defect. Furthermore, since the emission enhancement is accompanied by lifetime reduction, the negative effect of spectral diffusion and homogeneous broadening on photon indistinguishability is also reduced. 

Thus, the key purpose of combining a spin defect with photonic resonators is to increase its interaction with indistinguishable photons to enable efficient entanglement of remote defects. Here, ``remote'' signifies a distance greater than approximately 10~nm \cite{neumann2010quantum}, beyond which direct dipole-dipole interaction between two defects is too weak.



\section*{State-of-the-art in SiC spintronics and photonics}

\subsection*{Optically-addressable spins in SiC}

\begin{figure*}[ht]
\centering
\includegraphics[width=\textwidth]{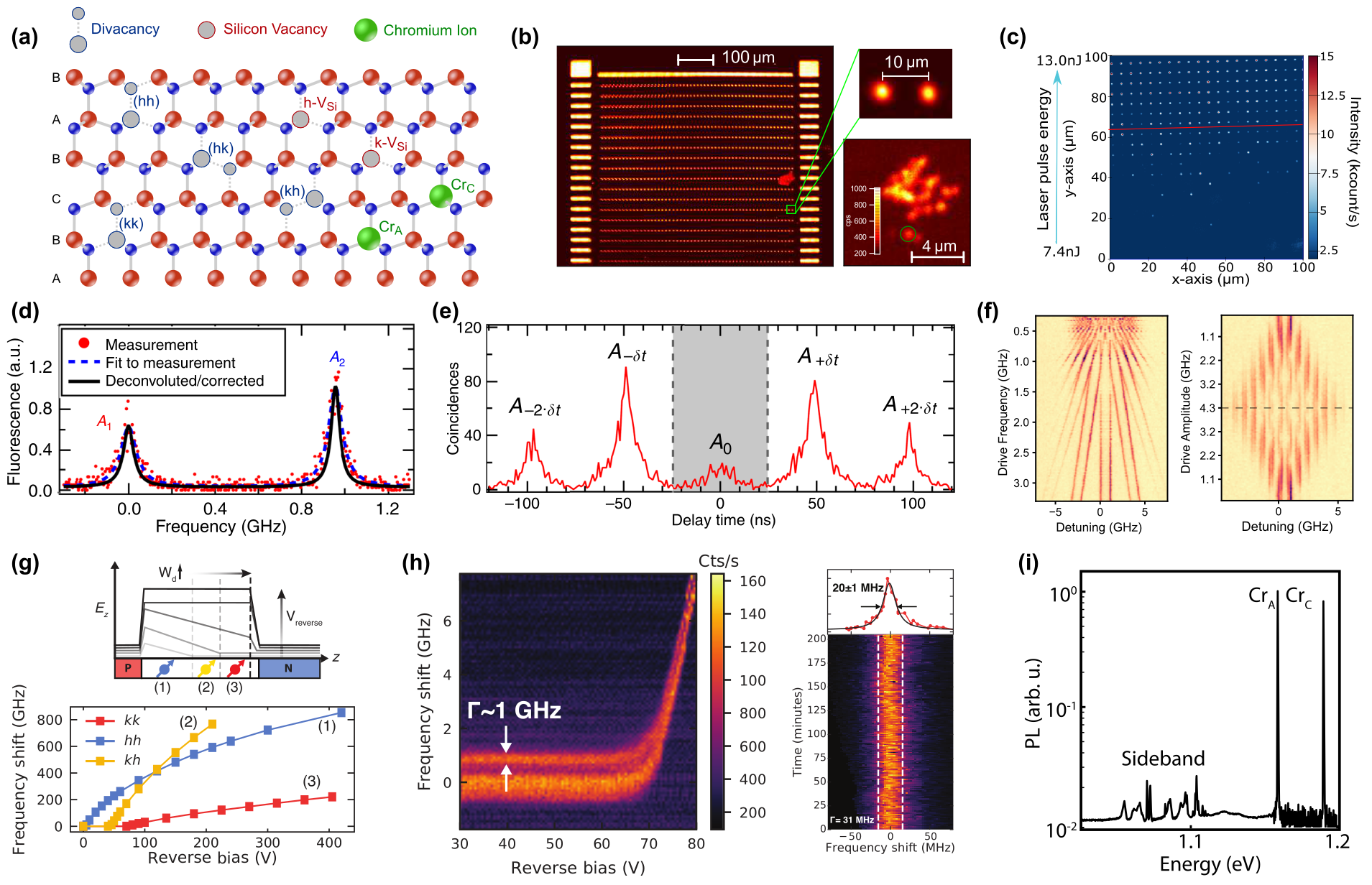}
\caption{Optically-active spins in SiC. (a) The 4H-SiC crystal lattice, showing the inequivalent configurations of the divacancy, silicon vacancy (V\textsubscript{Si}), and chromium ion. (b) Patterning of V\textsubscript{Si} using a focused proton beam enables three-dimensional control over placement of single defects and ensembles. (c) Direct laser writing of single V\textsubscript{Si} for on-demand color center generation in SiC devices under test. (d) Stable, nearly lifetime-limited optical emission from the V\textsubscript{Si} under above-resonant excitation. (e) The demonstration of indistinguishable single-photon emission from the V\textsubscript{Si}. (f) The V\textsubscript{Si} remains spectrally stable even under high amplitude fast Stark shift modulation, enabling the observation of optical Floquet eigenstates. (g) Advanced epitaxy techniques available commercially for SiC made possible the integration of high-quality divacancy color centers into p-i-n junctions, enabling the demonstration of nearly 1~THz of Stark shift. (h) By depleting the charge-trap environment via reverse bias in the p-i-n junction, linewidth narrowing from 1~GHz to 20~MHz was observed, approaching the transform limit. (i) Among the nascent optically addressable defects in SiC, the Chromium ion stands out for, among other characteristics, its high Debye-Waller Factor of 75\%. Reproduced from (b)\cite{kraus2017three},(c)\cite{chen2019laser},(d,e)\cite{morioka2020spin}, (f)\cite{lukin2020spectrally}, (g,h)\cite{anderson2019electrical}, (i)\cite{diler2020coherent}} \label{fig:color_centers}
\end{figure*}

Silicon carbide, in its numerous polytypes, has proven to be a versatile host to optically-addressable, long-lived spin qubits \cite{atature2018material, awschalom2018quantum, castelletto2020silicon}. Here, we briefly review the developments in spin-based quantum technologies in SiC. We focus on the two most well-studied color centers to-date, the silicon vacancy and the divacancy; we also highlight several emergent defects, such as the chromium ion, that may offer new functionalities. A summary of the properties of select defects in SiC is presented in Table~\ref{tab:color_centers}. We note here that the more complex SiC polytypes like 4H and 6H have multiple inequivalent lattice sites within a crystal unit cell (illustrated in Fig.~\ref{fig:color_centers}(a)). Consequently, each defect in Table~\ref{tab:color_centers} is a family of several defect types with similar but not identical properties. In addition to inequivalent lattice sites, rotational symmetries of the crystal give rise to multiple orientations of the same defect, the properties of which are otherwise identical.

\begin{table}[h!] 
  \begin{center}
    \caption{Optically-addressable spin defects in SiC}
    \label{tab:color_centers}
    
    \begin{tabularx}{\textwidth}{X X X X X X X X X l}
    \toprule[1pt]\midrule[0.3pt]
        & & & Debye &   &   &   &  & \\
        & & & Waller & Inverse & Measured & Stark & Spin& \\
        Defect & ZPL (nm)& Polytype & Factor & lifetime & linewidth & shift &    T\textsubscript{2} & \\
             &  &  & (DWF) & (MHz) & (MHz) & (GHz) &(ms) & Refs. \\
        \hline
        
        V$_\text{Si}^-$ & 862-917 & 4H,6H,15R& 0.06-0.09 & 27 & 51& 200 & 20 &\cite{sorman2000silicon, baranov2011silicon, soltamov2015optically, simin2017locking, nagy2019high, banks2019resonant,davidsson2019identification, udvarhelyi2020vibronic, shang2020local, morioka2020spin, ruhl2019stark, lukin2020spectrally}\\[0.75em]
          
        V\textsubscript{Si}V$_\text{C}^0$ & 1078-1132 & 4H, 6H, 3C& 0.07 & 11 & 20 & 850 & 64  & \cite{koehl2011room, falk2013polytype, christle2015isolated, falk2015optical,klimov2015quantum, ivady2016high, christle2017isolated, davidsson2019identification, miao2019electrically, anderson2019electrical, miao2020universal} \\[0.75em]
        
        N\textsubscript{C}V$_\text{Si}^-$  & 1180-1468& 4H, 6H, 3C& - & 75& - & - & 0.001 (T$_2^\text{*}$)& \cite{von2016nv, zargaleh2018nitrogen, zargaleh2016evidence, csore2017characterization, wang2020coherent}\\
        
        Cr$^{4+}$ & 1042, 1070 & 4H & 0.75 & 0.002 & 31 & - & 0.081 &\cite{koehl2017resonant, diler2020coherent}\\[0.75em]
        
        V$^{4+}$ &  1278-1388 & 4H, 6H & $<0.50$ & 0.9-14 & 750 & - & - & \cite{spindlberger2019optical, wolfowicz2019vanadium}\\[0.75em]
    \midrule[0.3pt]\bottomrule[1pt]
    \end{tabularx}
    
  \end{center}
\end{table}

\subsubsection*{The Silicon Vacancy}
The negatively-charged silicon vacancy (V$^-_\text{Si}$, written V\textsubscript{Si} henceforth), a single missing silicon atom with an extra electron at the vacancy site, has been observed in the 4H, 6H, and 15R polytypes of SiC. Its electronic configuration is modeled by five active electrons (three holes) resulting in a unique spin-3/2 system \cite{kraus2014room}, which has enabled novel sensing protocols \cite{simin2015high, simin2016all} and the realization of a spin qudit \cite{soltamov2019excitation}.  In single isolated defects, a spin-coherence time of $0.8$~ms has been measured at 4~K \cite{nagy2019high}. In V\textsubscript{Si} ensembles, spin-coherence time as high as $T_2 = 20$~ms has been observed using dynamic decoupling techniques \cite{simin2017locking}.  Numerous approaches to generate V\textsubscript{Si} have been studied, including irradiation using electrons \cite{widmann2015coherent, christle2015isolated}, neutrons \cite{fuchs2015engineering}, and protons \cite{kraus2017three}. In an effort to optimize V\textsubscript{Si} generation, the impact of different irradiation approaches on spin coherence has been systematically investigated \cite{kasper2020influence}. Direct laser writing \cite{chen2019laser}, ion implantation \cite{wang2017efficient}, and proton beam writing  \cite{kraus2017three} have been investigated for deterministic defect placement (Fig.~\ref{fig:color_centers}~(b,c)). Recently, the theoretically predicted \cite{soykal2016silicon} excited state fine structure of the V\textsubscript{Si} was experimentally confirmed in both inequivalent lattice sites in 4H-SiC \cite{banks2019resonant,nagy2019high}, enabling high-fidelity spin initialization \cite{nagy2019high} via the spin-selective intersystem crossing pathways \cite{soykal2016silicon, dong2019spin}.  The observation of narrow optical transitions \cite{nagy2019high, banks2019resonant, morioka2020spin}, as shown in Fig.~\ref{fig:color_centers}(d),  allowed for the demonstration of highly-indistinguishable photon emission with above-resonant driving \cite{morioka2020spin} (Fig.~\ref{fig:color_centers}(e)), an important step toward implementing cluster-state generation proposals \cite{economou2016spin, russo2018photonic}. The theoretically-predicted first-order DC Stark shift \cite{udvarhelyi2019spectrally} has been observed in ensembles \cite{ruhl2019stark} as well as in single defects \cite{lukin2020spectrally}. The demonstrated tuning range of 200~GHz \cite{lukin2020spectrally} is sufficient to overcome spectral inhomogeneity of defect ensembles \cite{nagy2019high}, a prerequisite for spin-spin entanglement of multiple V\textsubscript{Si} via photon interference. Furthermore, owing to the remarkable spectral stability of the V\textsubscript{Si} optical transitions, spectrally-reconfigurable single-photon emission from the V\textsubscript{Si} has been obtained via Floquet engineering \cite{lukin2020spectrally} (Fig.~\ref{fig:color_centers}(f)). Integration into semiconductor devices such as the p-i-n junction has enabled electrical readout of the spin-state \cite{niethammer2019coherent} and electrical control of the charge state \cite{widmann2019electrical} at room temperature. An essential next step toward realizing useful quantum photonic devices with the V\textsubscript{Si} will be the demonstration of spectrally-stable near-transform-limited defects in nanostructures.

\subsubsection*{The Divacancy}
The neutral divacancy is composed of adjacent silicon and carbon vacancies, denoted by V\textsubscript{Si}V$_\text{C}^0$. The combination of $C_{3v}$ symmetry, six active electrons, and spin-1 electronic structure render the optical and spin properties of the defect similar to the nitrogen vacancy (NV) center in diamond. However, the optical transitions are in the 1100~nm range, which is more favorable than the diamond NV center's 637~nm emission for optical communications and for integration into nanophotonic structures; The DWF of 0.07 \cite{christle2017isolated} is also an improvement over the NV center. Single divacancies with narrow optical linewidths and spin coherence up to 1~ms have been observed in SiC crystals without isotope purification \cite{christle2017isolated}. Recently, the discovery of dressed clock transitions have enabled the demonstration of divacancy coherence of 64~ms in material with natural isotope content \cite{miao2020universal}. Divacancy ensembles have been used to achieve a high degree of polarization of the SiC nuclear spin bath \cite{falk2015optical, ivady2016high}. Ensemble entanglement with nuclear spins at ambient conditions has been shown \cite{klimov2015quantum}, as well as the control of single divacancy-coupled nuclear spins \cite{bourassa2020entanglement}. Electrical and mechanical spin control of the divacancy have been demonstrated \cite{klimov2014electrically, falk2014electrically, whiteley2019spin}. In a major step toward wafer-scale optical and electrical integration of color centers, commercial \textit{p-i-n} junction SiC devices have been engineered to host individually-addressable divacancies with nearly lifetime-limited optical transitions, millisecond spin-coherence times, as well as optical and electrical charge control\cite{anderson2019electrical}. Furthermore, these devices can produce a Stark shift as large as 850~GHz (Fig.~\ref{fig:color_centers}(g,h)).  This is the first demonstration of such a combination of state-of-the-art optically-addressable spin-qubit properties in a single scalable semiconductor platform, opening opportunities for multi-qubit integration once combined with LDOS-enhancing photonic structures.

\subsubsection*{Emerging defects} 

Numerous other defects in SiC are currently under investigation for applications in quantum photonics. The nitrogen-vacancy center (N\textsubscript{C}V$_\text{Si}^-$) in SiC has been identified as a color center with a favorable emission frequency near the telecommunications S-band \cite{von2016nv, csore2017characterization, zargaleh2016evidence, zargaleh2018nitrogen}. Recently, coherent spin control has been observed in  N\textsubscript{C}V$_\text{Si}^-$ ensembles \cite{wang2020coherent}. Further studies are necessary to investigate the cause of the low brightness of the N\textsubscript{C}V$_\text{Si}^-$ as compared to its diamond counterpart; it may be due to a low quantum efficiency (i.e., a large percentage of the excited state decay is into the non-radiative intersystem crossing) or a long-lived metastable state. Another defect, the chromium ion (Cr$^{4+}$), has been identified as a promising quantum memory. Cr$^{4+}$ has an optical excited state lifetime of 155~\textmu s, and emits 75\% of photons into the ZPL in the near-IR (see Fig.~\ref{fig:color_centers}(i)) \cite{koehl2017resonant,diler2020coherent}. This suggests that integration of the Cr$^{4+}$ into nanophotonic structures may enable a large reduction of lifetime, which is essential for the efficient readout of single spins without a cycling transition \cite{raha2020optical}. However, the intrinsic optical and spin coherence of Cr$^{4+}$ defects remains an outstanding question; So far, only high density Cr$^{4+}$ ensembles have been studied, where the optical linewidths are $10^4$ times broader than lifetime-limited, and the measured spin-coherence time of 81~\textmu s is limited by spin-spin interactions in the ensemble \cite{diler2020coherent}. Finally, we highlight the vanadium impurity in SiC, V$^{4+}$, which is notable for its emission in the O-band and unusual optical lifetime of 108 and 167~ns (for the brighter inequivalent lattice sites) \cite{wolfowicz2019vanadium}. Particularly interesting is the strong sensitivity of the V$^{4+}$ optical transition to nearby nuclear spins, suggesting potential applications for optically-resolved nuclear spin registers \cite{wolfowicz2019vanadium}.

\subsection*{SiC Photonics}
Wafer-scale growth and processing of 4H and 6H polytypes of SiC was developed in the 1990's for applications in high-power electronics. Soon after, 4H- and 6H-SiC-on-insulator (SiCOI) were demonstrated \cite{di1997silicon} using the same ion-implantation (Smart-Cut) method that is used to produce silicon-on-insulator (SOI) wafers. This technology enabled the first demonstration of photonic crystal cavities (PhCs) in SiC \cite{song2011demonstration, yamada2014second}. As the development of photonics in Smart-Cut 4H-SiC continued \cite{cardenas2015optical} and Smart-Cut SiCOI became optimized on a wafer-scale \cite{yi2020wafer}, the intrinsic optical absorption of the SiC thin films was identified as the limiting factor for high-Q SiC photonics, limiting waveguide losses to $>5$~dB/cm \cite{zheng2019high}. Although further optimization of the implantation conditions may remedy the low material quality \cite{cardenas2015optical}, it is unclear whether the Smart-Cut method is suitable for producing films of SiC with the same nearly-pristine crystal quality as silicon-on-insulator. The difference between Smart-Cut SOI and SiCOI stems from the drastically different thermal properties of silicon and SiC: The lattice of silicon will soften and heal at the modest temperatures achievable in standard quartz furnaces. SiC, in turn, is one of the most refractory materials, subliming at 2700$^\circ$C. Repairing the lattice in post-processing without destroying the substrate is thus likely impossible.

Another approach to SiC photonics took advantage of the heteroepitaxial growth of 3C-SiC films on silicon. A variety of 3C-SiC-on-Si photonics devices have been demonstrated, including PhCs \cite{cardenas2013high, radulaski2013photonic} and whispering-gallery-mode resonators \cite{lu2014high}. However, this approach also suffers from substantial intrinsic material absorption, due to the high density of crystal defects near the growth interface caused by the Si-SiC lattice mismatch. Recently, a technique based on film transfer and back-side polishing introduced the 3C-SiC-on-insulator platform and enabled waveguides with losses down to 1.5~dB/cm, still likely limited by material absorption \cite{fan2018high, fan2020high}.

\begin{figure*}[tp]
\centering
\includegraphics[width=\textwidth]{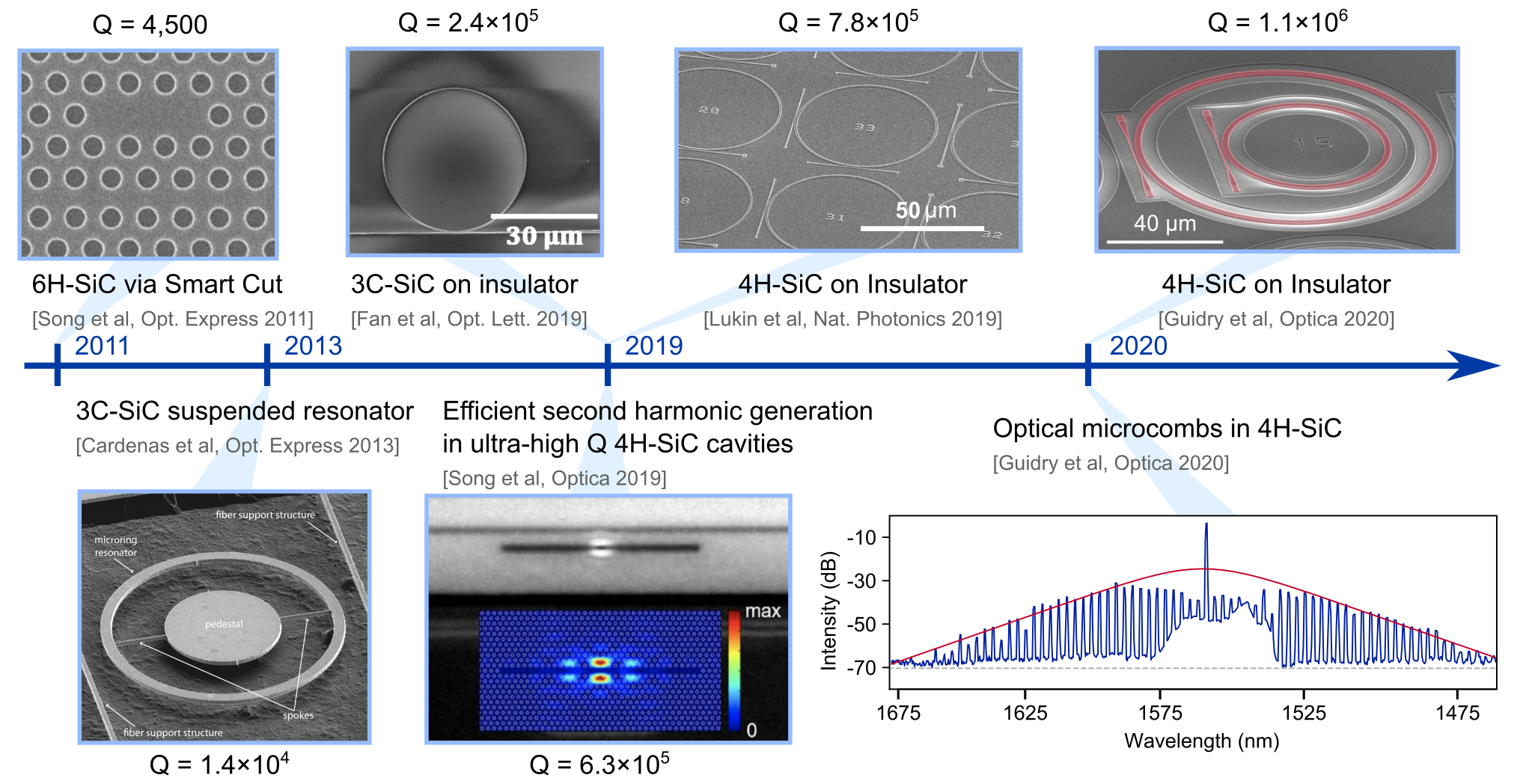}
\caption{Timeline of SiC photonics development. First demonstration of SiC photonic device using the Smart Cut approach with 6H-SiC \cite{song2011demonstration}. Soon after, suspended resonators in 3C-SiC-on-Si were demonstrated \cite{cardenas2013high}. Strong intrinsic absorption of low quality Smart Cut and heteroepitaxial 3C films was hypothesized to limit the achievable Q-factors. Using thicker 3C-SiC epilayers or thinning down bulk-crystal 4H-SiC, enabled record Q factors in 3C-SiC \cite{fan2018high, fan2020high}, ultra-high Q PhCs \cite{song2019ultrahigh}, and low-loss 4H-SiC-on-Insulator waveguides \cite{Lukin20194H}. Recently, devices with Q factors exceeding $10^6$ were shown, enabling the demonstration of optical parametric oscillation and microcomb formation \cite{Guidry2020Optical}.  Reproduced from \cite{song2011demonstration, cardenas2013high, fan2020high, song2019ultrahigh, Lukin20194H, Guidry2020Optical}}\label{fig:photonics}
\end{figure*}

\begin{figure*}[tp]
\centering
\includegraphics[width=0.5\textwidth]{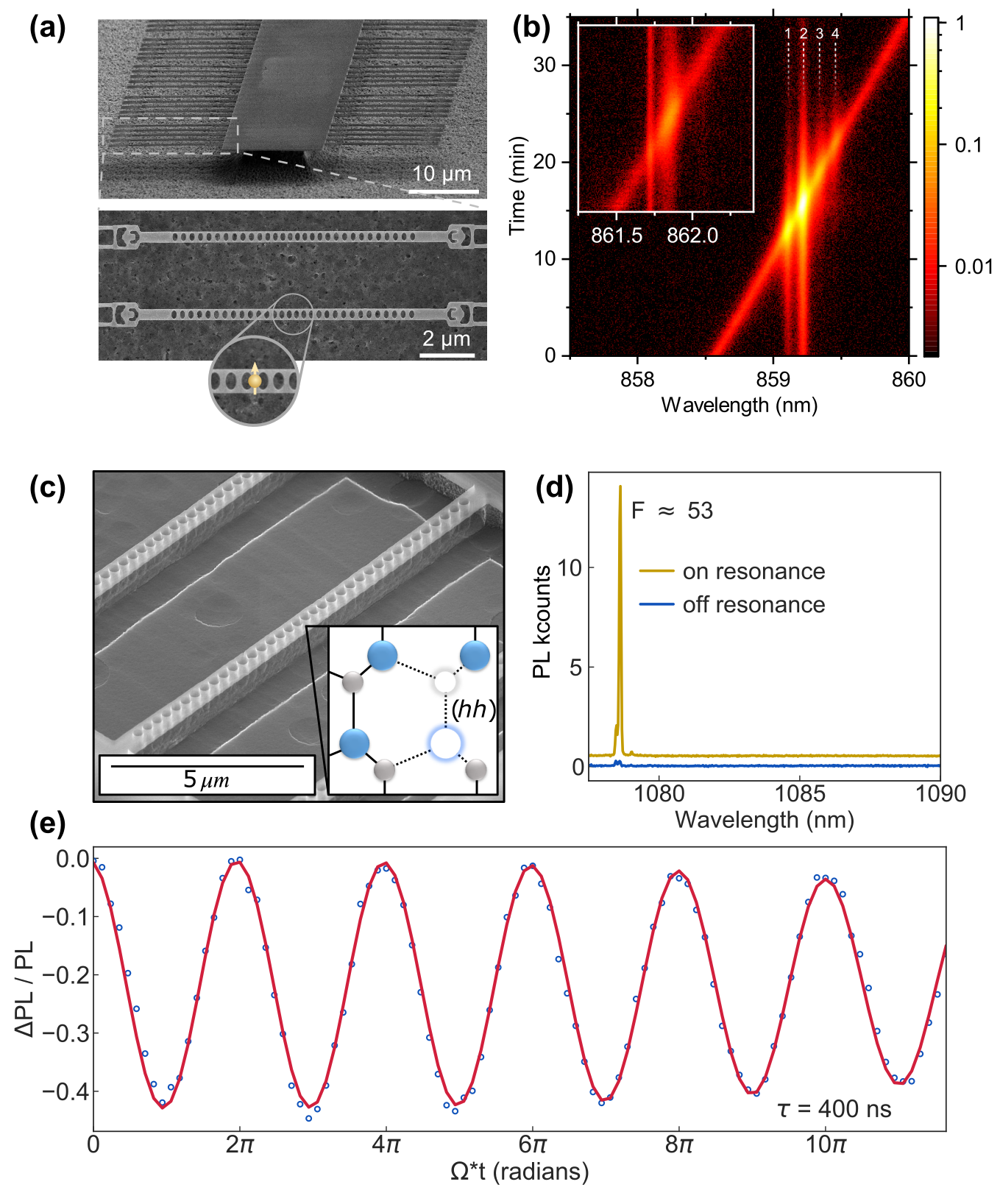}
\caption{SiC quantum photonics with defects. (a) An array of nanophotonic cavities fabricated in 4H-SiC-on-insulator, integrated with single V\textsubscript{Si} color centers \cite{Lukin20194H}. (b) A cavity is tuned via gas condensation through the color center resonance, thereby enhancing the optical transition by 120 times. (c) Nanophotonic cavities fabricated via the photoelectrochemical undercut technique \cite{crook2020purcell}, integrated with single divacancies. (d) Divacancy emission spectrum on and off resonance with the cavity, showing enhancement of 53 times. (e) Coherent spin control of the cavity-integrated divacancy. Reproduced from (a,b) \cite{Lukin20194H}, (c-e) \cite{crook2020purcell}.}\label{fig:q_photonics}
\end{figure*}

It is in the quantum context that the material quality of SiC thin films falls under the highest scrutiny: the coherence properties of color centers are highly sensitive even to low densities of unwanted defects. Indeed, in the first demonstration of SiC color centers coupled to a nanophotonic resonator --- using PhCs fabricated in 3C-SiC-on-Si --- the color center optical coherence was shown to be severely degraded by the lattice mismatch between Si and SiC  \cite{calusine2016cavity}. Similarly, color centers with good optical coherence have not been observed in Smart-Cut SiCOI, a consequence of the lattice damage induced by Smart-Cut ion-implantation: The dose required in the Smart-Cut process ($10^{16} - 10^{17}$~ions/cm\textsuperscript{2}) exceeds the dose used to generate spatially-resolvable single defects by four orders of magnitude. Incidentally, similar challenges arose in the development of thin-film diamond quantum photonics. In diamond, the absence of high quality thin films prompted the development of bulk-carving methods, such as angle-etching \cite{burek2014high}, for fabricating nanophotonic devices. The same technique has been recently demonstrated in SiC \cite{song2018high}. An alternative bulk-carving technique has been developed for 4H-SiC, relying on the advanced SiC homoepitaxy technology and doping-selective photoelectrochemical etching \cite{bracher2017selective}. This method enabled the first demonstration of Purcell enhancement of emission from V\textsubscript{Si} ensembles in 4H-SiC PhCs \cite{bracher2017selective}. Recently, using the same photonics platform, coherent spin control of a single divacancy integrated into PhCs was realized \cite{crook2020purcell} (Fig.~\ref{fig:q_photonics}~(c-e)).

Leveraging the wafer-scale production of 4H-SiC and the advanced grinding and polishing equipment developed for processing it, a method for fabricating ``quantum-grade'' 4H-SiC-on-insulator was recently introduced \cite{Lukin20194H}. This method enables 4H-SiCOI substrates with the same crystalline quality as bulk SiC crystal. Using 4H-SiCOI produced this way, ultra-high quality (Q) factor PhCs (Q $=6.3\times10^5$)  have been fabricated \cite{song2019ultrahigh}. 4H-SiCOI also enabled quantum photonic devices with single color centers in a CMOS compatible architecture  \cite{Lukin20194H} (Fig.~\ref{fig:q_photonics}~(a-b)).  Unfettered by the material absorption limit of previous approaches, integrated SiC photonics with propagation loss below 0.5~dB/cm (corresponding to a Q factor exceeding 1 million) have become possible. Low-loss microring resonators were used to demonstrate optical parametric oscillation and microcomb formation \cite{Guidry2020Optical}, establishing SiC as a promising material for integrated nonlinear photonics. Recently, the intrinsic absorption of 4H-SiC was measured to be as low as 0.02~dB/cm in an unoptimized sublimation-grown sample \cite{Guidry2020Optical}, suggesting that integrated photonics in SiC with Q factors of at least $10^7$ are possible.

\dl{
\subsection*{Comparison with diamond quantum photonics}
The key color center quantum photonics demonstrations (such as single-shot readout of spin \cite{robledo2011high, sukachev2017silicon}, cavity integration of emitters with narrow optical transitions \cite{burek2017fiber, bhaskar2020experimental, machielse2019quantum}, cavity-mediated spin-spin interactions \cite{evans2018photon}, and nuclear spin quantum register \cite{bradley2019ten}) have so far been in the diamond platform, specifically with the NV and the Silicon Vacancy color centers. Both of these color centers possess at least one optical transition that does not suffer from non-radiative spin-flip processes. The cyclicity of such a transition was utilized for  single-shot readout of spin, even without photonic cavity integration \cite{robledo2011high, sukachev2017silicon}. A similar cycling transition has not yet been definitively demonstrated in a SiC color center, although a potential candidate in the $kh$ divacancy has been identified \cite{miao2019electrically}. However, a cycling transition is not required for single-shot readout if cyclicity can be enhanced with a photonic cavity via the Purcell effect \cite{raha2020optical}.  So far, only centrosymmetric color centers in diamond have been integrated into nanophotonic structures while retaining narrow linewidths \cite{machielse2019quantum, wan2020large, rugar2020narrow}. Although inversion symmetry is not in principle a prerequisite for defect insensitivity to electric fields \cite{udvarhelyi2019spectrally}, integration of a noncentrosymmetric defect into a nanostructure while preserving its spectral stability is yet to be demonstrated. It should be noted that crystals without inversion symmetry such as SiC do not host inversion-symmetric defects. 

With regard to photonic devices, state-of-the-art microring resonators in SiC and diamond are currently comparable \cite{Guidry2020Optical, hausmann2014diamond}, whereas the photonic crystal nanocavities in SiC are superior due to the wider range of device designs accessible in the thin film platform \cite{burek2014high, song2019ultrahigh}. There are some key differences between diamond and SiC with regard to the intrinsic material properties: The larger bandgap of diamond results in a wider transparency window into the ultraviolet range. Although the SiC bandgap is narrower, it strikes a balance between optical transparency (which spans the visible frequencies) and ease of doping, enabling semiconductor structures such as \textit{p-i-n} junctions. The SiC $\chi^{(3)}$ nonlinearity is an order of magnitude stronger than that of diamond, lowering the power requirement for generating optical parametric oscillation and optical frequency combs \cite{Guidry2020Optical, hausmann2014diamond}. SiC also possesses a strong $\chi^{(2)}$ nonlinearity of 12~pm/V \cite{sato2009accurate}, which is absent in diamond due to the inversion symmetry of its lattice. }

\section*{Toward multi-node quantum photonics with defects}

\subsection*{Optimizing the single quantum node}

\subsubsection*{Emitter-cavity coupling}
As motivated in the introduction, the density of states of a defect's electromagnetic environment must be modified in order to enhance and direct its photon emission, suggesting a nanophotonic cavity coupled to an optically-active spin qubit as the basic building block of a defect-based integrated quantum photonic device. Despite remarkable progress in the development of the cavity-defect node, there remains a large performance gap between integrated quantum photonics devices and their classical photonics counterparts. To date, the state-of-the-art photonic crystal cavities in silicon \cite{asano2017photonic} and whispering gallery resonators in silica \cite{lee2012chemically} have Q factors of $10^7$ and $10^9$ ($10^8$ for a fully-integrated device \cite{yang2018bridging}), respectively. In contrast, photonic devices integrated with spin defects have so far demonstrated Q factors in the $10^3-10^4$ range. It is not surprising that spin-defect photonics are orders of magnitude behind, since the quantum material platforms (i.e., diamond, YSO, YVO, YAG, and SiC) are relatively new to the photonics scene. From this perspective, SiC-on-insulator is uniquely suited for bridging the classical-quantum photonics  gap.  SiCOI is already in the top three platforms for demonstrating high $Q/V$ photonic resonators, after silicon on insulator \cite{asano2017photonic} and lithium niobate on insulator \cite{li2019photon}. The high refractive index of SiC ($n~2.6$) allows for the fabrication of resonators based on 2D-photonic crystal cavities, the class of nanophotonic devices with the highest $Q/V$ ratio to-date \cite{asano2017photonic}. The impact of bridging the classical-quantum photonics gap will be an orders-of-magnitude improvement in readout fidelity and entanglement generation rates over current state-of-the-art spin-qubit photonics demonstrations. Furthermore, it will unlock new regimes of spin-qubit operation, including generation of transform-limited (and thus indistinguishable) photons from rare earth ions \cite{kindem2020control} and strong coupling of a color center to a cavity \cite{bhaskar2020experimental}. In SiC, these advances could likely be implemented with defects such as the Cr\textsuperscript{4+} ion, the divacancy, and the silicon vacancy.

\dl{
What $Q/V$ is sufficient, then? Naturally, the answer will depend on the defect used and the application. A defect with a lower branching ratio into the enhanced transition (for example, due to a small DWF or low quantum efficiency) will require greater $Q/V$ to achieve the same Purcell enhancement. From the photonics perspective, the emitter-cavity node is sufficiently optimized once the fraction of emitter excitations that do not result in a photon emitted into the cavity becomes negligible compared to other system losses, \textit{and} the spin state readout fidelity exceeds the fidelity of other single-register qubit operations.  For readout, the presence of a cycling transition would relax the $Q/V$ requirements \cite{raha2020optical, bhaskar2020experimental}. In selecting a promising quantum emitter, the brightness of a defect plays a role: High brightness can be an indicator of high quantum efficiency and the presence of cycling transitions. Low brightness, however, does not necessarily mean the contrary, since the presence of a long-lived metastable state can result in low count rates even if the non-radiative decay rate into the metastable states is slow. Thus, in contrast with single photon-source applications, the defect brightness is not the key metric in spin-based quantum technologies. }


\subsubsection*{Mitigating emitter degradation in nanostructures}
Another challenge for defect-based quantum photonics is mitigating the degradation of optical properties of defects in nanostructures. Often, defects with narrow, stable transitions in bulk material degrade severely when a material interface is nearby. The common understanding is that the linewidth degradation is caused by spectral diffusion (i.e., rapid temporal fluctuations in the defect's optical transitions due to fluctuating charges on the nearby surfaces). So far, the greatest success in nanophotonic integration has been had with centrosymmetric defects in diamond such as the silicon vacancy \cite{machielse2019quantum, bhaskar2020experimental, wan2020large}, the germanium vacancy \cite{wan2020large}, and the tin vacancy \cite{rugar2020narrow}. And yet, even these defects which should be maximally insensitive to environmental fluctuations display spectral diffusion that by several times exceeds the transform-limited linewidth, likely as a result of strain introduced during growth and fabrication, which lifts the symmetry of these defects and renders them sensitive to electric field to first order \cite{machielse2019quantum, wan2020large, rugar2020narrow}. The problem of spectral stability must be solved before the spin-defect quantum photonics technology becomes scalable.

Unlike many intrinsic properties of defects over which the experimentalist has no control, spectral diffusion in nanostructures is an extrinsic property and is amenable to systematic material science engineering. Surface passivation, either chemical or plasma-assisted, is one technique that has been successfully used to increase the photon or phonon lifetime in nanoresonators \cite{maccabe2019phononic, kuruma2020surface}, but to our knowledge has not yet been explored for defect stabilization. A comprehensive study that seeks to understand rather than simply eliminate spectral diffusion is needed to resolve this problem globally. For example, a comparison of the spectral diffusion  of different defect types in the same environment can help elucidate the degradation mechanisms.

The approach toward understanding spectral diffusion and resolving it will likely be specific to each quantum material platform. In the case of SiC, surfaces may potentially be passivated with a graphene layer, which can be readily grown on hexagonal SiC \cite{mishra2016graphene}. Although the strong optical absorption of graphene renders this an impractical solution when combined with photonics, it would constitute a valuable proof-of-concept demonstration of passivation-enabled compensation of spectral diffusion. An entirely different method to achieve near-surface emitter stabilization in SiC may take advantage of charge depletion using advanced doping epitaxy available in SiC \cite{anderson2019electrical}; implemented successfully in bulk material, the charge depletion technique has yet to be investigated in nanostructures. \dl{We note here that charge depletion can be directly integrated with existing nanophotonic architectures, as has been done in other platforms \cite{ellis2011ultralow}. Although the high temperature required for dopant activation in SiC \cite{rao1999donor} would necessitate definition of diode structure prior to the fabrication of the SiCOI material stack\cite{Lukin20194H}, it is a technologically straightforward process.} Another active approach to mitigating the effects of spectral diffusion may be via optimized optical excitation of the emitter, either via time-dependent drive \cite{fotso2016suppressing}  analogous to radiofrequency modulation demonstrations to extend defect spin coherence \cite{bluvstein2019extending}, or via optimized steady-state illumination \cite{tran2019suppression}. Fortunately, the challenge of stabilization of defects in nanostrucutres is as formidable as the possible strategies for overcoming it are numerous.

\subsection*{Scaling-up quantum photonic processors}

When designing a nanophotonic defect-based quantum node, it is crucial to look ahead toward multi-node scalability. This introduces two additional single-node system requirements. First, since the nodes must be spectrally identical during operation, the single node must be spectrally tunable to overcome inherent variations in resonator frequencies and the inhomogeneous broadening of defects. Second, the node must have an efficient waveguide interface, in order to transfer the emitted photons into the inter-node link with very high efficiency. The exact efficiency requirements will depend strongly on the application. Quantum communication and simulation will likely place a less stringent requirement than fault tolerant quantum computation, where proposals require no more than 10\% cumulative loss at all stages of the circuit \cite{nickerson2013topological, nickerson2014freely}.

The spectral tuning of cavities and color centers has seen excellent progress in a variety of platforms. For cavity tuning, the primary technique has been cryogenic gas condensation \cite{evans2018photon}: a heated gas tube delivers argon or xenon gas to the sample which then condenses on the cold sample surface, red-shifting the cavity frequency. Since condensed gas can be selectively and gradually removed via heating by a milliwatt laser, one can tune individual cavities onto resonance by applying the appropriate laser pulse. Numerous other techniques have also been employed, including atomic layer deposition \cite{yang2007digital}, laser-assisted oxidation \cite{kiravittaya2011tuning}, and index-shifting materials \cite{faraon2008local}. In all, cavity tuning is amenable to further optimization and numerous routes to multi-node scalability exist. \dl{ We note that although in principle the electrooptic effect allows fast modulation of cavity resonance, the Stark effect in emitters is typically much stronger and thus if rapid modulation is desired, it will likely be advantageous to modulate the emitter with respect to the cavity rather than vice versa.  Defect tuning has been demonstrated using both Stark shift \cite{anderson2019electrical, Lukin20194H} and strain \cite{meesala2018strain}, with  several demonstrations of the tuning range far exceeding the inhomogeneous broadening. Fast, high amplitude spectral modulation of quantum emitters has been shown \cite{miao2019electrically, lukin2020spectrally}.} One caveat, however, is that current tuning demonstrations are limited to compensating one degree of freedom, whereas spectral inhomogeneity in defects is higher-dimensional in nature (vector for electric field and tensor for strain). Thus, for most defects (an exception are defects with degeneracy broken by spin-spin interactions only, like the V\textsubscript{Si} \cite{soykal2016silicon}), a single tuning degree of freedom is only sufficient to make one optical transition degenerate\cite{meesala2018strain}. Thus, if a protocol makes use of multiple optical transitions of a single defect,  each additional transition must be controlled with an independent tunable laser, which is not scalable \cite{machielse2019quantum}. Cavity-assisted Raman emission can allow to overcome this limitation \cite{sun2018cavity}. Overall, while engineering challenges are still ahead, there is already a framework for building a fully-tunable quantum node and a clear path toward scalability.

In addition to spectral tuning, an efficient cavity-waveguide interface is the second essential requirement for multi-node devices. To achieve this, the cavity must be designed to be significantly over-coupled to the waveguide (i.e., cavity loss into the waveguide must dominate over other loss channels), implying that ultimately the actual Q factor will be much lower than the highest Q attainable in the platform. Consequently, this places an even greater requirement on the intrinsic cavity Q/V to maintain the same Purcell enhancement.  Furthermore, once the photon has entered the waveguide, great attention must be paid to any losses that the photon may experience as it propagates between nodes.

\begin{figure*}[tp]
\centering
\includegraphics[width=1\textwidth]{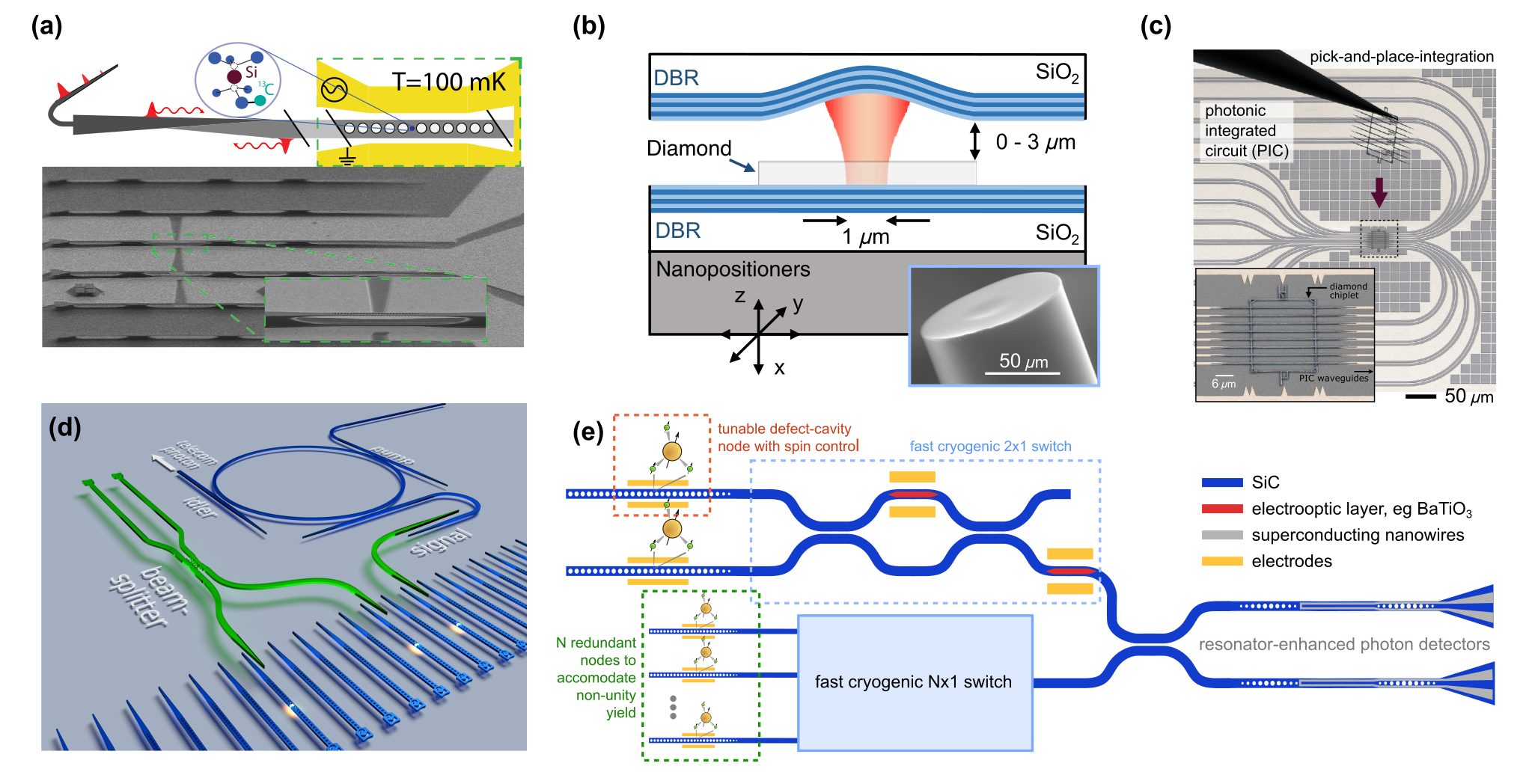}
\caption{ \dl{Approaches to scaling-up spin-based quantum photonic technologies (a) Diamond nanophotonic cavity with a single silicon vacancy defect and an adiabatically-coupled fiber interface. \cite{nguyen2019quantum} (b) Free-space coupled Fabry-P\'{e}rot microcavity enhancing the emission of an NV center in diamond \cite{riedel2017deterministic}. Inset: the concave mirror can also be fabricated directly on the tip of a fiber \cite{hunger2010fiber}. (c) Pick-and-place heterogeneous on-chip integration of diamond microchiplets containing silicon and germanium vacancy centers on top of aluminum nitride photonic waveguides \cite{wan2020large} (image courtesy of Noel Wan). (d) A heterogeneous approach without pick-and-place can be realized by using a secondary layer of photonic interconnects to post-select working quantum nodes \cite{Lukin20194H}. (e) A conceptual diagram demonstrating how the example photonic network shown in Fig.~\ref{fig:intro}b could be realized in a fully monolithic platform. In order to account for non-unity fabrication yield, $N$ redundant nodes are fabricated in the place of one node, and a Nx1 switch (composed of cascaded 2x1 switches) selects one working node. Reproduced from: (a) \cite{nguyen2019quantum}, (b) \cite{riedel2017deterministic, hunger2010fiber} (d) \cite{Lukin20194H}.}\label{fig:scale_up} }
\end{figure*}

At this point, it is important to note a distinction between the principles of scalability for classical computers and quantum photonic processors. Classical technologies based on semiconductor transistors progressed through ``scaling down'', where the individual node size has been reduced further and further to accommodate greater computing power, until the limiting factor has become the wires rather than the transistors themselves. In contrast, a defect-based quantum photonic computer does not fundamentally enjoy gains from small overall size (as long as the photon emission-enhancing component maximizes large Purcell factor through high Q/V). Compared to the resistive losses that limit classical computers, short-distance optical communications (fibers or waveguides) are effectively lossless. The speed of a defect-based quantum computer will not be limited by the internode-link communication rates (photon transit time), but by the (much slower) physics of the quantum node, namely, the spin manipulation and readout of the defect-cavity system. This difference in paradigms is especially relevant in context of the distinct challenge of device yield present in defect-based quantum photonics, since the technological complexity of the single node precludes fabricating quantum nodes with a yield exceeding 99\%. This suggests that scaling-up will require a degree of device post-selection and reconfigurability, as is already done on other platforms such as trapped atoms \cite{barredo2016atom}. In light of the above considerations, integration via off-chip fiber interconnects should not apriori be excluded in the near future (before the increasing number of quantum nodes makes it impractical).

In this context, we comment on the current leading approaches toward multi-node scalability:

\begin{enumerate}
    \item \textbf{Integrated nanophotonic devices with a fiber interface.} This approach relies on a nanophotonic cavity for enhancing the light-matter interaction, but routes the photons from the cavity directly into a fiber for off-chip processing (Fig.~\ref{fig:scale_up}a). \cite{burek2017fiber, machielse2019quantum, bhaskar2020experimental}  This approach naturally enables 100\% device yield within a quantum network via post-selection: individual devices are characterized and working devices are integrated together via (low loss) fiber interconnects. Because the nanophotonic waveguide and optical fibers are effectively lossless at the relevant length scales, photon loss is incurred exclusively at the fiber-waveguide interface, with demonstrated efficiencies as high as 96\% \cite{burek2017fiber}. 
    
    \item \textbf{Fabry-P\'{e}rot microcavities.} Recently, photonic resonators based on concave dielectric mirrors have enabled breakthrough demonstrations of cavity-integrated light-matter interactions \cite{najer2019gated}. In this approach, a concave mirror is fabricated \cite{hunger2012laser} either in bulk silica \cite{riedel2017deterministic} or, notably, directly on the tip of a fiber \cite{hunger2010fiber}, and forms a distributed Bragg reflector microcavity with the buried Bragg mirror beneath the active quantum medium, incorporating the defect in-between (Fig.~\ref{fig:scale_up}b). As with fiber-coupled on-chip nanophotonic cavities, unity yield is achieved by post-selection; The fiber-tip scans the surface to isolate a suitable defect, and the cavity resonance is tuned by controlling the fiber height piezoelectrically. Notably, this technique can in principle be applied to any defect that can be integrated into a smooth, thin membrane. \cite{haussler2019diamond, merkel2020coherent, riedel2017deterministic} The fiber Fabry-P\'{e}rot microcavity is in a sense a distillation of the fiber-coupled nanophotonic cavity approach, as the cavity output mode is Gaussian and can efficiently be coupled into single mode fibers. The efficiency is limited by losses in the optical components required for coupling the photons into the single-mode fiber, reflection losses at surfaces, and a slight modal mismatch between the cavity and fiber modes. Recently, a total coupling efficiency of 68\% into the fiber mode has been demonstrated \cite{tomm2020bright}. 
    
    \item \textbf{Pick-and-place heterogeneous on-chip integration.}  This approach aims to address the concern of yield in a fully integrated fashion, by transferring quantum nodes onto an integrated photonic circuit after characterization. Pick-and-place is a particularly promising technique in material platforms not suitable for standard photonics processing: For instance, it has recently enabled large-scale on-chip integration of diamond color centers with aluminum nitride interconnects (Fig.~\ref{fig:scale_up}c) \cite{wan2020large}. Although promising for fully chip-scale integration,  pick-and-place currently suffers from high experimental losses in the adiabatic transfer of photons from the transferred quantum material to the integrated photonic circuit (transmission of 34\% \cite{wan2020large}) suggesting that the transfer efficiency to the inter-node link may be a serious impediment for scalability. Although highly efficient interlayer adiabatic transfer is possible (approaching 99\% \cite{singaravelu2019low}), it relies on long, well-aligned (200~\textmu m) tapers that are difficult to achieve using a pick-and-place technique (but are possible with heterogeneous integration of multiple thin-film layers (Fig.~\ref{fig:scale_up}d)). Thus, we see an efficient waveguide interface as the key challenge in developing pick-and-place as a method for multi-node QIP.
\end{enumerate}

\section*{Perspective on fully-integrated quantum photonics}

Most advanced spin-defect experiments to-date rely on the strategy of coupling emitted photons into an optical fiber as soon as possible (approaches 1 and 2 above). However, for applications other than fiber network communications, a photon in a fiber is not an advantage over a photon in an integrated waveguide: Practical realization of the key components for quantum networks identified in Fig.~\ref{fig:intro} actually plays to the strengths of integrated photonics rather than fiber optics. Integrated photonics have already achieved system complexity beyond what can be practically realized in macroscopic fiber-based devices, integrating hundreds of elements with mean fidelities of linear components exceeding 99.9\% \cite{harris2017quantum}. On-chip integration of single-photon detectors has seen remarkable progress in recent years \cite{pernice2012high, najafi2015chip, esmaeil2020efficient}, and integration with photonic resonators will likely enable narrowband integrated single photon detection with efficiencies exceeding 99\% in the near future. In our view, the chip-integrated approach is the most promising for large-scale quantum systems. This architecture is illustrated in Fig.~\ref{fig:scale_up}(e), using the example photonic network introduced in Fig.~\ref{fig:intro}(b). 

In this fully-monolithic realization of chip-integrated spin-based quantum technologies, the photon never leaves the chip, never couples into a fiber or passes through lossy bulk active elements, and is not subject to system fluctuations inherent in a large-scale macroscopic system. In fact, the photon never leaves the waveguide into which the quantum node emitted it, because switching, interference, and detection can all be realized in a waveguide geometry \cite{najafi2015chip,harris2017quantum}. Compact and efficient on-chip photon detectors can be placed anywhere in the integrated circuit to convert a (fragile) quantum signal into a (robust) classical signal which can be routed off-chip via standard CMOS electronics such as vias and buried electrical layers, aiding in the realization of circuit connectivity. The relative simplicity of the integrated approach is a source of optimism for satisfying the extremely low loss requirements of useful quantum photonic computation. 

SiC-on-insulator is a promising candidate to realize a fully-integrated defect-based quantum photonic processor, using high Q/V photonic crystal cavities, fast cryogenic optical modulators, and integrated detectors. In order to overcome the challenge of non-unity yield, each quantum node may consist of $N$ redundant cavity-coupled spin-defect elements coupled to a bus waveguide. Using this configuration, one can achieve post-selection without any additional cavity-waveguide losses, by tuning all but one working node away from the quantum network operation frequency. If necessary, a similar approach can be employed for post-selection of detectors (which are to be integrated with low-Q resonators or long waveguides to maximize photon absorption probability). Each node is electrically interfaced to tune the defect optical transition and to coherently manipulate the spin. Fast cryogenic modulators and switches based on a directional coupler or resonator drop-filter configuration can be integrated directly into the SiC platform, taking advantage of its electrooptic effect. Cryogenic integration based on this approach has only recently been demonstrated \cite{eltes2020integrated}. To increase the bandwidth and decrease the footprint, an additional electro-optically active layer, such as barium titanate (BaTiO\textsubscript{3}) \cite{eltes2020integrated}, can be sputtered and patterned in an adiabatic taper atop the SiC waveguide to minimize scattering loss. Finally, regardless of the optical frequency of operation of the quantum processor, efficient frequency conversion to the telecommunications band using the strong intrinsic second-order nonlinearity of SiC (12~pm/V) \cite{sato2009accurate} would prepare the optically-encoded quantum information for long-distance communication.

\begin{figure*}[tp]
\centering
\includegraphics[width=\textwidth]{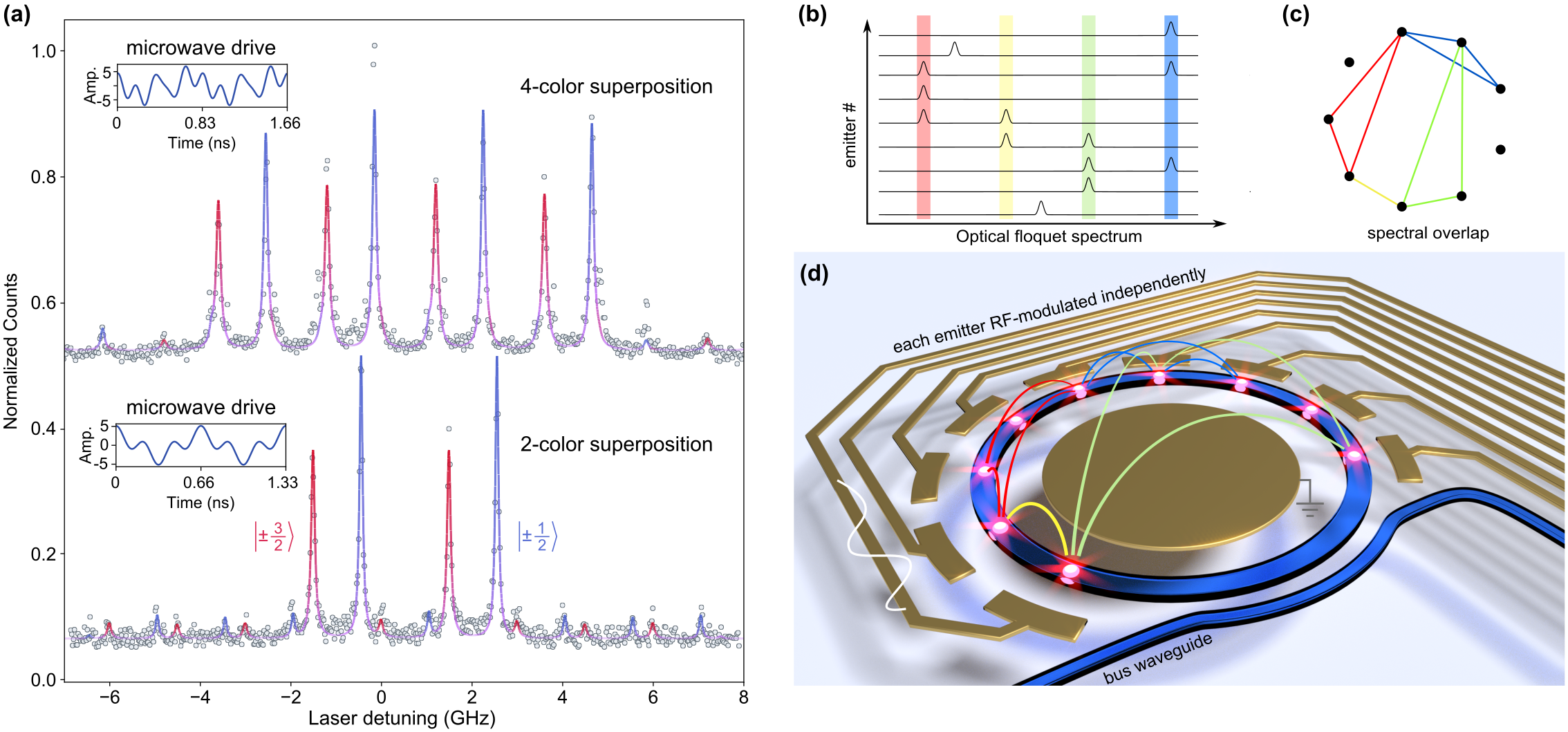}
\caption{Quantum information processing with spectrally-reconfigurable defects. (a) By rapidly modulating an optical transition, engineering of exotic photon emission spectra from color centers was demonstrated \cite{lukin2020spectrally}. Via optimized Stark modulation (insets), the V\textsubscript{Si} has been engineered to emit photons in a superposition of 4 (top) and 2 (bottom) colors. (Reproduced from \cite{lukin2020spectrally}). Based on this approach one may design nearly-arbitrary reconfigurable connectivity via spectral overlap, by independently controlling color centers interacting together via photon-photon interactions through the resonator.  This proposal is illustrated in (b,c) where nine emitters are prepared in Floquet states to create a specific connectivity via spectral overlap. (d) A concept figure of the possible experimental implementation of this approach.} \label{fig:floquet}
\end{figure*}

In the development of scalable quantum photonic circuits, an emerging technique to device engineering --- photonics inverse design \cite{piggott2015inverse, molesky2018inverse, su2020nanophotonic} --- will likely play an important role. The near-unity fidelities achieved with classical photonics design, as mentioned above, are limited to simple photonics building blocks such as a directional coupler. However, for more complex tasks, inverse design offers a novel approach for performing targeted optimization against multiple metrics, including robustness to fabrication imperfections (thus improving the yield) and fabrication constraints \cite{piggott2017fabrication}. For example, an classical adiabatic taper between photonic layers \cite{singaravelu2019low} is extremely sensitive to misalignment, precluding efficient interlayer coupling in pick-and-place; this limitation is an opportunity to apply inverse design. The powerful functionality of inverse design has been used to demonstrate efficient mode conversion \cite{vercruysse2019dispersion} and free-space coupling \cite{sapra2019inverse}, even under stringent fabrication constraints \cite{dory2019inverse}. The simultaneous inverse design of photonic structures and embedded color centers may enable a fully solid-state implementation of a quantum simulator recently proposed for trapped atoms \cite{gonzalez2015subwavelength}.

Finally, we briefly address some new approaches to QIP that may be enabled with recent advancements in color center control and integrated photonics. The recent proposals in quantum simulation based on Floquet state engineering for the superconducting qubit platform \cite{blais2020quantum, onodera2020quantum} may find similar implementation in a color-center photonics: By coupling an array of emitters to a single nanophotonic resonator, it may be possible to encode arbitrary couplings between them in the frequency basis, taking advantage of nontrivial, multi-peaked emission spectra to connect multiple emitters in nontrivial topologies, as shown in Fig.~\ref{fig:floquet}. A similar drive-optimization approach may also be used to overcome inhomogeneous broadening in emitter ensembles where individual emitters cannot be controlled but rather a single, global drive signal must be used \cite{lukin2020spectrally}. 

\dl{
\section*{Conclusion}

In summary, spin-based photonic technologies for quantum computing will likely operate in the network architecture (Fig.~\ref{fig:intro}) and will require the integration of spin-qubit registers with high quality photonic structures and efficient photon detectors to reduce the total photon loss below the demanding thresholds for quantum computing \cite{nickerson2014freely}. While there may be numerous approaches for achieving this goal, we believe that a fully chip-integrated quantum photonic platform holds the most promise, as this approach is most scalable and avoids additional coupling loss from waveguide interconnects. 
SiC has emerged as a promising platform for realizing this technology, with demonstrations of wafer-scale integration of high quality emitters into semiconductor junctions \cite{anderson2019electrical}, isotopic engineering for nuclear spin registers \cite{bourassa2020entanglement}, indistinguishable single photon emission \cite{morioka2020spin}, and quantum-grade SiC-on-Insulator platform for fabrication of photonic devices \cite{Lukin20194H}. However, several key results must be demonstrated to confirm its promise,  starting with the integration of large arrays of tunable of narrow-linewidth emitters into nanostructures.
}

\section*{Popular Summary}

Defects in crystals are usually undesirable imperfections that arise during crystal growth and processing. Some defects, however, have properties that make them useful for quantum computing. Defects like some color centers in silicon carbide and diamond have the ability to store quantum information for a long time in their ground states; furthermore, they can efficiently encode their spin state information onto photons in order to transmit the information over long distances. However, it is challenging to efficiently collect photons emitted by the defects, and nanoscale photonic engineering is necessary to strengthen the light-matter interaction to enable quantum computation. Silicon carbide is simultaneously host to high quality color centers and can be used to fabricate large-scale monolithic photonic device networks, which makes it a unique quantum material for scalable quantum photonic computation.

In this Perspective, we discuss the challenges and prospects of developing a fully-integrated color center-based quantum photonic computer. We begin with a brief review of the state-of-the-art of quantum photonics in silicon carbide, compare it to other leading solid state platforms (such as diamond), and outline the current leading approaches for implementing color center quantum technologies. Finally, we discuss how the industry-compatible silicon carbide-on-insulator platform can pioneer integrated color-center quantum technologies (including quantum simulation and quantum computation) in the next 5-10 years.

\vspace{1em}
\emph{Acknowledgements}: We thank Rahul Trivedi, Shuo Sun, Kiyoul Yang, Sophia Economou, Wenzheng Dong, Florian Kaiser, Oney O. Soykal, Shahriar Aghaeimeibodi, Daniel Riedel, Alexander D. White, Tian Zhong, Yizhong Huang, for insightful discussions. This work is funded in part by the U.S. Department of Energy, Office of Science, under Awards DE-SC0019174 and DE-Ac02-76SF00515; and the National Science Foundation under Awards 1839056 and EFRI-1741660; 
D.M.L. acknowledges the Fong Stanford Graduate Fellowship (SGF) and the National Defense Science and Engineering Graduate Fellowship.
M.A.G. acknowledges the Albion Hewlett SGF and the NSF Graduate Research Fellowship. 
\bibliography{library.bib}{}

\begin{thebibliography}{155}%
\makeatletter
\providecommand \@ifxundefined [1]{%
 \@ifx{#1\undefined}
}%
\providecommand \@ifnum [1]{%
 \ifnum #1\expandafter \@firstoftwo
 \else \expandafter \@secondoftwo
 \fi
}%
\providecommand \@ifx [1]{%
 \ifx #1\expandafter \@firstoftwo
 \else \expandafter \@secondoftwo
 \fi
}%
\providecommand \natexlab [1]{#1}%
\providecommand \enquote  [1]{``#1''}%
\providecommand \bibnamefont  [1]{#1}%
\providecommand \bibfnamefont [1]{#1}%
\providecommand \citenamefont [1]{#1}%
\providecommand \href@noop [0]{\@secondoftwo}%
\providecommand \href [0]{\begingroup \@sanitize@url \@href}%
\providecommand \@href[1]{\@@startlink{#1}\@@href}%
\providecommand \@@href[1]{\endgroup#1\@@endlink}%
\providecommand \@sanitize@url [0]{\catcode `\\12\catcode `\$12\catcode
  `\&12\catcode `\#12\catcode `\^12\catcode `\_12\catcode `\%12\relax}%
\providecommand \@@startlink[1]{}%
\providecommand \@@endlink[0]{}%
\providecommand \url  [0]{\begingroup\@sanitize@url \@url }%
\providecommand \@url [1]{\endgroup\@href {#1}{\urlprefix }}%
\providecommand \urlprefix  [0]{URL }%
\providecommand \Eprint [0]{\href }%
\providecommand \doibase [0]{http://dx.doi.org/}%
\providecommand \selectlanguage [0]{\@gobble}%
\providecommand \bibinfo  [0]{\@secondoftwo}%
\providecommand \bibfield  [0]{\@secondoftwo}%
\providecommand \translation [1]{[#1]}%
\providecommand \BibitemOpen [0]{}%
\providecommand \bibitemStop [0]{}%
\providecommand \bibitemNoStop [0]{.\EOS\space}%
\providecommand \EOS [0]{\spacefactor3000\relax}%
\providecommand \BibitemShut  [1]{\csname bibitem#1\endcsname}%
\let\auto@bib@innerbib\@empty
\bibitem [{\citenamefont {Arute}\ \emph {et~al.}(2019)\citenamefont {Arute},
  \citenamefont {Arya}, \citenamefont {Babbush}, \citenamefont {Bacon},
  \citenamefont {Bardin}, \citenamefont {Barends}, \citenamefont {Biswas},
  \citenamefont {Boixo}, \citenamefont {Brandao}, \citenamefont {Buell} \emph
  {et~al.}}]{arute2019quantum}%
  \BibitemOpen
  \bibfield  {author} {\bibinfo {author} {\bibfnamefont {F.}~\bibnamefont
  {Arute}}, \bibinfo {author} {\bibfnamefont {K.}~\bibnamefont {Arya}},
  \bibinfo {author} {\bibfnamefont {R.}~\bibnamefont {Babbush}}, \bibinfo
  {author} {\bibfnamefont {D.}~\bibnamefont {Bacon}}, \bibinfo {author}
  {\bibfnamefont {J.~C.}\ \bibnamefont {Bardin}}, \bibinfo {author}
  {\bibfnamefont {R.}~\bibnamefont {Barends}}, \bibinfo {author} {\bibfnamefont
  {R.}~\bibnamefont {Biswas}}, \bibinfo {author} {\bibfnamefont
  {S.}~\bibnamefont {Boixo}}, \bibinfo {author} {\bibfnamefont {F.~G.}\
  \bibnamefont {Brandao}}, \bibinfo {author} {\bibfnamefont {D.~A.}\
  \bibnamefont {Buell}},  \emph {et~al.},\ }\href@noop {} {\bibfield  {journal}
  {\bibinfo  {journal} {Nature}\ }\textbf {\bibinfo {volume} {574}},\ \bibinfo
  {pages} {505} (\bibinfo {year} {2019})}\BibitemShut {NoStop}%
\bibitem [{\citenamefont {Zhang}\ \emph {et~al.}(2017)\citenamefont {Zhang},
  \citenamefont {Pagano}, \citenamefont {Hess}, \citenamefont {Kyprianidis},
  \citenamefont {Becker}, \citenamefont {Kaplan}, \citenamefont {Gorshkov},
  \citenamefont {Gong},\ and\ \citenamefont {Monroe}}]{zhang2017observation}%
  \BibitemOpen
  \bibfield  {author} {\bibinfo {author} {\bibfnamefont {J.}~\bibnamefont
  {Zhang}}, \bibinfo {author} {\bibfnamefont {G.}~\bibnamefont {Pagano}},
  \bibinfo {author} {\bibfnamefont {P.~W.}\ \bibnamefont {Hess}}, \bibinfo
  {author} {\bibfnamefont {A.}~\bibnamefont {Kyprianidis}}, \bibinfo {author}
  {\bibfnamefont {P.}~\bibnamefont {Becker}}, \bibinfo {author} {\bibfnamefont
  {H.}~\bibnamefont {Kaplan}}, \bibinfo {author} {\bibfnamefont {A.~V.}\
  \bibnamefont {Gorshkov}}, \bibinfo {author} {\bibfnamefont {Z.-X.}\
  \bibnamefont {Gong}}, \ and\ \bibinfo {author} {\bibfnamefont
  {C.}~\bibnamefont {Monroe}},\ }\href@noop {} {\bibfield  {journal} {\bibinfo
  {journal} {Nature}\ }\textbf {\bibinfo {volume} {551}},\ \bibinfo {pages}
  {601} (\bibinfo {year} {2017})}\BibitemShut {NoStop}%
\bibitem [{\citenamefont {Bernien}\ \emph {et~al.}(2017)\citenamefont
  {Bernien}, \citenamefont {Schwartz}, \citenamefont {Keesling}, \citenamefont
  {Levine}, \citenamefont {Omran}, \citenamefont {Pichler}, \citenamefont
  {Choi}, \citenamefont {Zibrov}, \citenamefont {Endres}, \citenamefont
  {Greiner} \emph {et~al.}}]{bernien2017probing}%
  \BibitemOpen
  \bibfield  {author} {\bibinfo {author} {\bibfnamefont {H.}~\bibnamefont
  {Bernien}}, \bibinfo {author} {\bibfnamefont {S.}~\bibnamefont {Schwartz}},
  \bibinfo {author} {\bibfnamefont {A.}~\bibnamefont {Keesling}}, \bibinfo
  {author} {\bibfnamefont {H.}~\bibnamefont {Levine}}, \bibinfo {author}
  {\bibfnamefont {A.}~\bibnamefont {Omran}}, \bibinfo {author} {\bibfnamefont
  {H.}~\bibnamefont {Pichler}}, \bibinfo {author} {\bibfnamefont
  {S.}~\bibnamefont {Choi}}, \bibinfo {author} {\bibfnamefont {A.~S.}\
  \bibnamefont {Zibrov}}, \bibinfo {author} {\bibfnamefont {M.}~\bibnamefont
  {Endres}}, \bibinfo {author} {\bibfnamefont {M.}~\bibnamefont {Greiner}},
  \emph {et~al.},\ }\href@noop {} {\bibfield  {journal} {\bibinfo  {journal}
  {Nature}\ }\textbf {\bibinfo {volume} {551}},\ \bibinfo {pages} {579}
  (\bibinfo {year} {2017})}\BibitemShut {NoStop}%
\bibitem [{\citenamefont {Harris}\ \emph {et~al.}(2017)\citenamefont {Harris},
  \citenamefont {Steinbrecher}, \citenamefont {Prabhu}, \citenamefont {Lahini},
  \citenamefont {Mower}, \citenamefont {Bunandar}, \citenamefont {Chen},
  \citenamefont {Wong}, \citenamefont {Baehr-Jones}, \citenamefont {Hochberg}
  \emph {et~al.}}]{harris2017quantum}%
  \BibitemOpen
  \bibfield  {author} {\bibinfo {author} {\bibfnamefont {N.~C.}\ \bibnamefont
  {Harris}}, \bibinfo {author} {\bibfnamefont {G.~R.}\ \bibnamefont
  {Steinbrecher}}, \bibinfo {author} {\bibfnamefont {M.}~\bibnamefont
  {Prabhu}}, \bibinfo {author} {\bibfnamefont {Y.}~\bibnamefont {Lahini}},
  \bibinfo {author} {\bibfnamefont {J.}~\bibnamefont {Mower}}, \bibinfo
  {author} {\bibfnamefont {D.}~\bibnamefont {Bunandar}}, \bibinfo {author}
  {\bibfnamefont {C.}~\bibnamefont {Chen}}, \bibinfo {author} {\bibfnamefont
  {F.~N.}\ \bibnamefont {Wong}}, \bibinfo {author} {\bibfnamefont
  {T.}~\bibnamefont {Baehr-Jones}}, \bibinfo {author} {\bibfnamefont
  {M.}~\bibnamefont {Hochberg}},  \emph {et~al.},\ }\href@noop {} {\bibfield
  {journal} {\bibinfo  {journal} {Nature Photonics}\ }\textbf {\bibinfo
  {volume} {11}},\ \bibinfo {pages} {447} (\bibinfo {year} {2017})}\BibitemShut
  {NoStop}%
\bibitem [{\citenamefont {Atat{\"u}re}\ \emph {et~al.}(2018)\citenamefont
  {Atat{\"u}re}, \citenamefont {Englund}, \citenamefont {Vamivakas},
  \citenamefont {Lee},\ and\ \citenamefont {Wrachtrup}}]{atature2018material}%
  \BibitemOpen
  \bibfield  {author} {\bibinfo {author} {\bibfnamefont {M.}~\bibnamefont
  {Atat{\"u}re}}, \bibinfo {author} {\bibfnamefont {D.}~\bibnamefont
  {Englund}}, \bibinfo {author} {\bibfnamefont {N.}~\bibnamefont {Vamivakas}},
  \bibinfo {author} {\bibfnamefont {S.-Y.}\ \bibnamefont {Lee}}, \ and\
  \bibinfo {author} {\bibfnamefont {J.}~\bibnamefont {Wrachtrup}},\ }\href@noop
  {} {\bibfield  {journal} {\bibinfo  {journal} {Nature Reviews Materials}\
  }\textbf {\bibinfo {volume} {3}},\ \bibinfo {pages} {38} (\bibinfo {year}
  {2018})}\BibitemShut {NoStop}%
\bibitem [{\citenamefont {Awschalom}\ \emph {et~al.}(2018)\citenamefont
  {Awschalom}, \citenamefont {Hanson}, \citenamefont {Wrachtrup},\ and\
  \citenamefont {Zhou}}]{awschalom2018quantum}%
  \BibitemOpen
  \bibfield  {author} {\bibinfo {author} {\bibfnamefont {D.~D.}\ \bibnamefont
  {Awschalom}}, \bibinfo {author} {\bibfnamefont {R.}~\bibnamefont {Hanson}},
  \bibinfo {author} {\bibfnamefont {J.}~\bibnamefont {Wrachtrup}}, \ and\
  \bibinfo {author} {\bibfnamefont {B.~B.}\ \bibnamefont {Zhou}},\ }\href@noop
  {} {\bibfield  {journal} {\bibinfo  {journal} {Nature Photonics}\ }\textbf
  {\bibinfo {volume} {12}},\ \bibinfo {pages} {516} (\bibinfo {year}
  {2018})}\BibitemShut {NoStop}%
\bibitem [{\citenamefont {Wang}\ \emph {et~al.}(2019)\citenamefont {Wang},
  \citenamefont {Sciarrino}, \citenamefont {Laing},\ and\ \citenamefont
  {Thompson}}]{wang2019integrated}%
  \BibitemOpen
  \bibfield  {author} {\bibinfo {author} {\bibfnamefont {J.}~\bibnamefont
  {Wang}}, \bibinfo {author} {\bibfnamefont {F.}~\bibnamefont {Sciarrino}},
  \bibinfo {author} {\bibfnamefont {A.}~\bibnamefont {Laing}}, \ and\ \bibinfo
  {author} {\bibfnamefont {M.~G.}\ \bibnamefont {Thompson}},\ }\href@noop {}
  {\bibfield  {journal} {\bibinfo  {journal} {Nature Photonics}\ ,\ \bibinfo
  {pages} {1}} (\bibinfo {year} {2019})}\BibitemShut {NoStop}%
\bibitem [{\citenamefont {Elshaari}\ \emph {et~al.}(2020)\citenamefont
  {Elshaari}, \citenamefont {Pernice}, \citenamefont {Srinivasan},
  \citenamefont {Benson},\ and\ \citenamefont {Zwiller}}]{elshaari2020hybrid}%
  \BibitemOpen
  \bibfield  {author} {\bibinfo {author} {\bibfnamefont {A.~W.}\ \bibnamefont
  {Elshaari}}, \bibinfo {author} {\bibfnamefont {W.}~\bibnamefont {Pernice}},
  \bibinfo {author} {\bibfnamefont {K.}~\bibnamefont {Srinivasan}}, \bibinfo
  {author} {\bibfnamefont {O.}~\bibnamefont {Benson}}, \ and\ \bibinfo {author}
  {\bibfnamefont {V.}~\bibnamefont {Zwiller}},\ }\href@noop {} {\bibfield
  {journal} {\bibinfo  {journal} {Nature Photonics}\ ,\ \bibinfo {pages} {1}}
  (\bibinfo {year} {2020})}\BibitemShut {NoStop}%
\bibitem [{\citenamefont {Zhong}\ \emph {et~al.}(2015)\citenamefont {Zhong},
  \citenamefont {Hedges}, \citenamefont {Ahlefeldt}, \citenamefont
  {Bartholomew}, \citenamefont {Beavan}, \citenamefont {Wittig}, \citenamefont
  {Longdell},\ and\ \citenamefont {Sellars}}]{zhong2015optically}%
  \BibitemOpen
  \bibfield  {author} {\bibinfo {author} {\bibfnamefont {M.}~\bibnamefont
  {Zhong}}, \bibinfo {author} {\bibfnamefont {M.~P.}\ \bibnamefont {Hedges}},
  \bibinfo {author} {\bibfnamefont {R.~L.}\ \bibnamefont {Ahlefeldt}}, \bibinfo
  {author} {\bibfnamefont {J.~G.}\ \bibnamefont {Bartholomew}}, \bibinfo
  {author} {\bibfnamefont {S.~E.}\ \bibnamefont {Beavan}}, \bibinfo {author}
  {\bibfnamefont {S.~M.}\ \bibnamefont {Wittig}}, \bibinfo {author}
  {\bibfnamefont {J.~J.}\ \bibnamefont {Longdell}}, \ and\ \bibinfo {author}
  {\bibfnamefont {M.~J.}\ \bibnamefont {Sellars}},\ }\href@noop {} {\bibfield
  {journal} {\bibinfo  {journal} {Nature}\ }\textbf {\bibinfo {volume} {517}},\
  \bibinfo {pages} {177} (\bibinfo {year} {2015})}\BibitemShut {NoStop}%
\bibitem [{\citenamefont {Bradley}\ \emph {et~al.}(2019)\citenamefont
  {Bradley}, \citenamefont {Randall}, \citenamefont {Abobeih}, \citenamefont
  {Berrevoets}, \citenamefont {Degen}, \citenamefont {Bakker}, \citenamefont
  {Markham}, \citenamefont {Twitchen},\ and\ \citenamefont
  {Taminiau}}]{bradley2019ten}%
  \BibitemOpen
  \bibfield  {author} {\bibinfo {author} {\bibfnamefont {C.}~\bibnamefont
  {Bradley}}, \bibinfo {author} {\bibfnamefont {J.}~\bibnamefont {Randall}},
  \bibinfo {author} {\bibfnamefont {M.}~\bibnamefont {Abobeih}}, \bibinfo
  {author} {\bibfnamefont {R.}~\bibnamefont {Berrevoets}}, \bibinfo {author}
  {\bibfnamefont {M.}~\bibnamefont {Degen}}, \bibinfo {author} {\bibfnamefont
  {M.}~\bibnamefont {Bakker}}, \bibinfo {author} {\bibfnamefont
  {M.}~\bibnamefont {Markham}}, \bibinfo {author} {\bibfnamefont
  {D.}~\bibnamefont {Twitchen}}, \ and\ \bibinfo {author} {\bibfnamefont
  {T.}~\bibnamefont {Taminiau}},\ }\href@noop {} {\bibfield  {journal}
  {\bibinfo  {journal} {Physical Review X}\ }\textbf {\bibinfo {volume} {9}},\
  \bibinfo {pages} {031045} (\bibinfo {year} {2019})}\BibitemShut {NoStop}%
\bibitem [{\citenamefont {Miao}\ \emph {et~al.}(2020)\citenamefont {Miao},
  \citenamefont {Blanton}, \citenamefont {Anderson}, \citenamefont {Bourassa},
  \citenamefont {Crook}, \citenamefont {Wolfowicz}, \citenamefont {Abe},
  \citenamefont {Ohshima},\ and\ \citenamefont
  {Awschalom}}]{miao2020universal}%
  \BibitemOpen
  \bibfield  {author} {\bibinfo {author} {\bibfnamefont {K.~C.}\ \bibnamefont
  {Miao}}, \bibinfo {author} {\bibfnamefont {J.~P.}\ \bibnamefont {Blanton}},
  \bibinfo {author} {\bibfnamefont {C.~P.}\ \bibnamefont {Anderson}}, \bibinfo
  {author} {\bibfnamefont {A.}~\bibnamefont {Bourassa}}, \bibinfo {author}
  {\bibfnamefont {A.~L.}\ \bibnamefont {Crook}}, \bibinfo {author}
  {\bibfnamefont {G.}~\bibnamefont {Wolfowicz}}, \bibinfo {author}
  {\bibfnamefont {H.}~\bibnamefont {Abe}}, \bibinfo {author} {\bibfnamefont
  {T.}~\bibnamefont {Ohshima}}, \ and\ \bibinfo {author} {\bibfnamefont
  {D.~D.}\ \bibnamefont {Awschalom}},\ }\href@noop {} {\bibfield  {journal}
  {\bibinfo  {journal} {Science}\ }\textbf {\bibinfo {volume} {369}},\ \bibinfo
  {pages} {1493} (\bibinfo {year} {2020})}\BibitemShut {NoStop}%
\bibitem [{\citenamefont {Hensen}\ \emph {et~al.}(2015)\citenamefont {Hensen},
  \citenamefont {Bernien}, \citenamefont {Dr{\'e}au}, \citenamefont {Reiserer},
  \citenamefont {Kalb}, \citenamefont {Blok}, \citenamefont {Ruitenberg},
  \citenamefont {Vermeulen}, \citenamefont {Schouten}, \citenamefont
  {Abell{\'a}n} \emph {et~al.}}]{hensen2015loophole}%
  \BibitemOpen
  \bibfield  {author} {\bibinfo {author} {\bibfnamefont {B.}~\bibnamefont
  {Hensen}}, \bibinfo {author} {\bibfnamefont {H.}~\bibnamefont {Bernien}},
  \bibinfo {author} {\bibfnamefont {A.~E.}\ \bibnamefont {Dr{\'e}au}}, \bibinfo
  {author} {\bibfnamefont {A.}~\bibnamefont {Reiserer}}, \bibinfo {author}
  {\bibfnamefont {N.}~\bibnamefont {Kalb}}, \bibinfo {author} {\bibfnamefont
  {M.~S.}\ \bibnamefont {Blok}}, \bibinfo {author} {\bibfnamefont
  {J.}~\bibnamefont {Ruitenberg}}, \bibinfo {author} {\bibfnamefont {R.~F.}\
  \bibnamefont {Vermeulen}}, \bibinfo {author} {\bibfnamefont {R.~N.}\
  \bibnamefont {Schouten}}, \bibinfo {author} {\bibfnamefont {C.}~\bibnamefont
  {Abell{\'a}n}},  \emph {et~al.},\ }\href@noop {} {\bibfield  {journal}
  {\bibinfo  {journal} {Nature}\ }\textbf {\bibinfo {volume} {526}},\ \bibinfo
  {pages} {682} (\bibinfo {year} {2015})}\BibitemShut {NoStop}%
\bibitem [{\citenamefont {Nemoto}\ \emph {et~al.}(2014)\citenamefont {Nemoto},
  \citenamefont {Trupke}, \citenamefont {Devitt}, \citenamefont {Stephens},
  \citenamefont {Scharfenberger}, \citenamefont {Buczak}, \citenamefont
  {N{\"o}bauer}, \citenamefont {Everitt}, \citenamefont {Schmiedmayer},\ and\
  \citenamefont {Munro}}]{nemoto2014photonic}%
  \BibitemOpen
  \bibfield  {author} {\bibinfo {author} {\bibfnamefont {K.}~\bibnamefont
  {Nemoto}}, \bibinfo {author} {\bibfnamefont {M.}~\bibnamefont {Trupke}},
  \bibinfo {author} {\bibfnamefont {S.~J.}\ \bibnamefont {Devitt}}, \bibinfo
  {author} {\bibfnamefont {A.~M.}\ \bibnamefont {Stephens}}, \bibinfo {author}
  {\bibfnamefont {B.}~\bibnamefont {Scharfenberger}}, \bibinfo {author}
  {\bibfnamefont {K.}~\bibnamefont {Buczak}}, \bibinfo {author} {\bibfnamefont
  {T.}~\bibnamefont {N{\"o}bauer}}, \bibinfo {author} {\bibfnamefont {M.~S.}\
  \bibnamefont {Everitt}}, \bibinfo {author} {\bibfnamefont {J.}~\bibnamefont
  {Schmiedmayer}}, \ and\ \bibinfo {author} {\bibfnamefont {W.~J.}\
  \bibnamefont {Munro}},\ }\href@noop {} {\bibfield  {journal} {\bibinfo
  {journal} {Physical Review X}\ }\textbf {\bibinfo {volume} {4}},\ \bibinfo
  {pages} {031022} (\bibinfo {year} {2014})}\BibitemShut {NoStop}%
\bibitem [{\citenamefont {Nickerson}\ \emph {et~al.}(2013)\citenamefont
  {Nickerson}, \citenamefont {Li},\ and\ \citenamefont
  {Benjamin}}]{nickerson2013topological}%
  \BibitemOpen
  \bibfield  {author} {\bibinfo {author} {\bibfnamefont {N.~H.}\ \bibnamefont
  {Nickerson}}, \bibinfo {author} {\bibfnamefont {Y.}~\bibnamefont {Li}}, \
  and\ \bibinfo {author} {\bibfnamefont {S.~C.}\ \bibnamefont {Benjamin}},\
  }\href@noop {} {\bibfield  {journal} {\bibinfo  {journal} {Nature
  communications}\ }\textbf {\bibinfo {volume} {4}},\ \bibinfo {pages} {1}
  (\bibinfo {year} {2013})}\BibitemShut {NoStop}%
\bibitem [{\citenamefont {Nickerson}\ \emph {et~al.}(2014)\citenamefont
  {Nickerson}, \citenamefont {Fitzsimons},\ and\ \citenamefont
  {Benjamin}}]{nickerson2014freely}%
  \BibitemOpen
  \bibfield  {author} {\bibinfo {author} {\bibfnamefont {N.~H.}\ \bibnamefont
  {Nickerson}}, \bibinfo {author} {\bibfnamefont {J.~F.}\ \bibnamefont
  {Fitzsimons}}, \ and\ \bibinfo {author} {\bibfnamefont {S.~C.}\ \bibnamefont
  {Benjamin}},\ }\href@noop {} {\bibfield  {journal} {\bibinfo  {journal}
  {Physical Review X}\ }\textbf {\bibinfo {volume} {4}},\ \bibinfo {pages}
  {041041} (\bibinfo {year} {2014})}\BibitemShut {NoStop}%
\bibitem [{\citenamefont {Buterakos}\ \emph {et~al.}(2017)\citenamefont
  {Buterakos}, \citenamefont {Barnes},\ and\ \citenamefont
  {Economou}}]{buterakos2017deterministic}%
  \BibitemOpen
  \bibfield  {author} {\bibinfo {author} {\bibfnamefont {D.}~\bibnamefont
  {Buterakos}}, \bibinfo {author} {\bibfnamefont {E.}~\bibnamefont {Barnes}}, \
  and\ \bibinfo {author} {\bibfnamefont {S.~E.}\ \bibnamefont {Economou}},\
  }\href@noop {} {\bibfield  {journal} {\bibinfo  {journal} {Physical Review
  X}\ }\textbf {\bibinfo {volume} {7}},\ \bibinfo {pages} {041023} (\bibinfo
  {year} {2017})}\BibitemShut {NoStop}%
\bibitem [{\citenamefont {Russo}\ \emph {et~al.}(2018)\citenamefont {Russo},
  \citenamefont {Barnes},\ and\ \citenamefont {Economou}}]{russo2018photonic}%
  \BibitemOpen
  \bibfield  {author} {\bibinfo {author} {\bibfnamefont {A.}~\bibnamefont
  {Russo}}, \bibinfo {author} {\bibfnamefont {E.}~\bibnamefont {Barnes}}, \
  and\ \bibinfo {author} {\bibfnamefont {S.~E.}\ \bibnamefont {Economou}},\
  }\href@noop {} {\bibfield  {journal} {\bibinfo  {journal} {Physical Review
  B}\ }\textbf {\bibinfo {volume} {98}},\ \bibinfo {pages} {085303} (\bibinfo
  {year} {2018})}\BibitemShut {NoStop}%
\bibitem [{\citenamefont {Schwartz}\ \emph {et~al.}(2016)\citenamefont
  {Schwartz}, \citenamefont {Cogan}, \citenamefont {Schmidgall}, \citenamefont
  {Don}, \citenamefont {Gantz}, \citenamefont {Kenneth}, \citenamefont
  {Lindner},\ and\ \citenamefont {Gershoni}}]{schwartz2016deterministic}%
  \BibitemOpen
  \bibfield  {author} {\bibinfo {author} {\bibfnamefont {I.}~\bibnamefont
  {Schwartz}}, \bibinfo {author} {\bibfnamefont {D.}~\bibnamefont {Cogan}},
  \bibinfo {author} {\bibfnamefont {E.~R.}\ \bibnamefont {Schmidgall}},
  \bibinfo {author} {\bibfnamefont {Y.}~\bibnamefont {Don}}, \bibinfo {author}
  {\bibfnamefont {L.}~\bibnamefont {Gantz}}, \bibinfo {author} {\bibfnamefont
  {O.}~\bibnamefont {Kenneth}}, \bibinfo {author} {\bibfnamefont {N.~H.}\
  \bibnamefont {Lindner}}, \ and\ \bibinfo {author} {\bibfnamefont
  {D.}~\bibnamefont {Gershoni}},\ }\href@noop {} {\bibfield  {journal}
  {\bibinfo  {journal} {Science}\ }\textbf {\bibinfo {volume} {354}},\ \bibinfo
  {pages} {434} (\bibinfo {year} {2016})}\BibitemShut {NoStop}%
\bibitem [{\citenamefont {Muralidharan}\ \emph {et~al.}(2016)\citenamefont
  {Muralidharan}, \citenamefont {Li}, \citenamefont {Kim}, \citenamefont
  {L{\"u}tkenhaus}, \citenamefont {Lukin},\ and\ \citenamefont
  {Jiang}}]{muralidharan2016optimal}%
  \BibitemOpen
  \bibfield  {author} {\bibinfo {author} {\bibfnamefont {S.}~\bibnamefont
  {Muralidharan}}, \bibinfo {author} {\bibfnamefont {L.}~\bibnamefont {Li}},
  \bibinfo {author} {\bibfnamefont {J.}~\bibnamefont {Kim}}, \bibinfo {author}
  {\bibfnamefont {N.}~\bibnamefont {L{\"u}tkenhaus}}, \bibinfo {author}
  {\bibfnamefont {M.~D.}\ \bibnamefont {Lukin}}, \ and\ \bibinfo {author}
  {\bibfnamefont {L.}~\bibnamefont {Jiang}},\ }\href@noop {} {\bibfield
  {journal} {\bibinfo  {journal} {Scientific reports}\ }\textbf {\bibinfo
  {volume} {6}},\ \bibinfo {pages} {20463} (\bibinfo {year}
  {2016})}\BibitemShut {NoStop}%
\bibitem [{\citenamefont {Borregaard}\ \emph {et~al.}(2020)\citenamefont
  {Borregaard}, \citenamefont {Pichler}, \citenamefont {Schr{\"o}der},
  \citenamefont {Lukin}, \citenamefont {Lodahl},\ and\ \citenamefont
  {S{\o}rensen}}]{borregaard2020one}%
  \BibitemOpen
  \bibfield  {author} {\bibinfo {author} {\bibfnamefont {J.}~\bibnamefont
  {Borregaard}}, \bibinfo {author} {\bibfnamefont {H.}~\bibnamefont {Pichler}},
  \bibinfo {author} {\bibfnamefont {T.}~\bibnamefont {Schr{\"o}der}}, \bibinfo
  {author} {\bibfnamefont {M.~D.}\ \bibnamefont {Lukin}}, \bibinfo {author}
  {\bibfnamefont {P.}~\bibnamefont {Lodahl}}, \ and\ \bibinfo {author}
  {\bibfnamefont {A.~S.}\ \bibnamefont {S{\o}rensen}},\ }\href@noop {}
  {\bibfield  {journal} {\bibinfo  {journal} {Physical Review X}\ }\textbf
  {\bibinfo {volume} {10}},\ \bibinfo {pages} {021071} (\bibinfo {year}
  {2020})}\BibitemShut {NoStop}%
\bibitem [{\citenamefont {Bhaskar}\ \emph {et~al.}(2020)\citenamefont
  {Bhaskar}, \citenamefont {Riedinger}, \citenamefont {Machielse},
  \citenamefont {Levonian}, \citenamefont {Nguyen}, \citenamefont {Knall},
  \citenamefont {Park}, \citenamefont {Englund}, \citenamefont {Lon{\v{c}}ar},
  \citenamefont {Sukachev} \emph {et~al.}}]{bhaskar2020experimental}%
  \BibitemOpen
  \bibfield  {author} {\bibinfo {author} {\bibfnamefont {M.~K.}\ \bibnamefont
  {Bhaskar}}, \bibinfo {author} {\bibfnamefont {R.}~\bibnamefont {Riedinger}},
  \bibinfo {author} {\bibfnamefont {B.}~\bibnamefont {Machielse}}, \bibinfo
  {author} {\bibfnamefont {D.~S.}\ \bibnamefont {Levonian}}, \bibinfo {author}
  {\bibfnamefont {C.~T.}\ \bibnamefont {Nguyen}}, \bibinfo {author}
  {\bibfnamefont {E.~N.}\ \bibnamefont {Knall}}, \bibinfo {author}
  {\bibfnamefont {H.}~\bibnamefont {Park}}, \bibinfo {author} {\bibfnamefont
  {D.}~\bibnamefont {Englund}}, \bibinfo {author} {\bibfnamefont
  {M.}~\bibnamefont {Lon{\v{c}}ar}}, \bibinfo {author} {\bibfnamefont {D.~D.}\
  \bibnamefont {Sukachev}},  \emph {et~al.},\ }\href@noop {} {\bibfield
  {journal} {\bibinfo  {journal} {Nature}\ }\textbf {\bibinfo {volume} {580}},\
  \bibinfo {pages} {60} (\bibinfo {year} {2020})}\BibitemShut {NoStop}%
\bibitem [{\citenamefont {Zhong}\ \emph {et~al.}(2017)\citenamefont {Zhong},
  \citenamefont {Kindem}, \citenamefont {Bartholomew}, \citenamefont {Rochman},
  \citenamefont {Craiciu}, \citenamefont {Miyazono}, \citenamefont
  {Bettinelli}, \citenamefont {Cavalli}, \citenamefont {Verma}, \citenamefont
  {Nam} \emph {et~al.}}]{zhong2017nanophotonic}%
  \BibitemOpen
  \bibfield  {author} {\bibinfo {author} {\bibfnamefont {T.}~\bibnamefont
  {Zhong}}, \bibinfo {author} {\bibfnamefont {J.~M.}\ \bibnamefont {Kindem}},
  \bibinfo {author} {\bibfnamefont {J.~G.}\ \bibnamefont {Bartholomew}},
  \bibinfo {author} {\bibfnamefont {J.}~\bibnamefont {Rochman}}, \bibinfo
  {author} {\bibfnamefont {I.}~\bibnamefont {Craiciu}}, \bibinfo {author}
  {\bibfnamefont {E.}~\bibnamefont {Miyazono}}, \bibinfo {author}
  {\bibfnamefont {M.}~\bibnamefont {Bettinelli}}, \bibinfo {author}
  {\bibfnamefont {E.}~\bibnamefont {Cavalli}}, \bibinfo {author} {\bibfnamefont
  {V.}~\bibnamefont {Verma}}, \bibinfo {author} {\bibfnamefont {S.~W.}\
  \bibnamefont {Nam}},  \emph {et~al.},\ }\href@noop {} {\bibfield  {journal}
  {\bibinfo  {journal} {Science}\ }\textbf {\bibinfo {volume} {357}},\ \bibinfo
  {pages} {1392} (\bibinfo {year} {2017})}\BibitemShut {NoStop}%
\bibitem [{\citenamefont {Kalb}\ \emph {et~al.}(2017)\citenamefont {Kalb},
  \citenamefont {Reiserer}, \citenamefont {Humphreys}, \citenamefont
  {Bakermans}, \citenamefont {Kamerling}, \citenamefont {Nickerson},
  \citenamefont {Benjamin}, \citenamefont {Twitchen}, \citenamefont {Markham},\
  and\ \citenamefont {Hanson}}]{kalb2017entanglement}%
  \BibitemOpen
  \bibfield  {author} {\bibinfo {author} {\bibfnamefont {N.}~\bibnamefont
  {Kalb}}, \bibinfo {author} {\bibfnamefont {A.~A.}\ \bibnamefont {Reiserer}},
  \bibinfo {author} {\bibfnamefont {P.~C.}\ \bibnamefont {Humphreys}}, \bibinfo
  {author} {\bibfnamefont {J.~J.}\ \bibnamefont {Bakermans}}, \bibinfo {author}
  {\bibfnamefont {S.~J.}\ \bibnamefont {Kamerling}}, \bibinfo {author}
  {\bibfnamefont {N.~H.}\ \bibnamefont {Nickerson}}, \bibinfo {author}
  {\bibfnamefont {S.~C.}\ \bibnamefont {Benjamin}}, \bibinfo {author}
  {\bibfnamefont {D.~J.}\ \bibnamefont {Twitchen}}, \bibinfo {author}
  {\bibfnamefont {M.}~\bibnamefont {Markham}}, \ and\ \bibinfo {author}
  {\bibfnamefont {R.}~\bibnamefont {Hanson}},\ }\href@noop {} {\bibfield
  {journal} {\bibinfo  {journal} {Science}\ }\textbf {\bibinfo {volume}
  {356}},\ \bibinfo {pages} {928} (\bibinfo {year} {2017})}\BibitemShut
  {NoStop}%
\bibitem [{\citenamefont {Evans}\ \emph {et~al.}(2018)\citenamefont {Evans},
  \citenamefont {Bhaskar}, \citenamefont {Sukachev}, \citenamefont {Nguyen},
  \citenamefont {Sipahigil}, \citenamefont {Burek}, \citenamefont {Machielse},
  \citenamefont {Zhang}, \citenamefont {Zibrov}, \citenamefont {Bielejec} \emph
  {et~al.}}]{evans2018photon}%
  \BibitemOpen
  \bibfield  {author} {\bibinfo {author} {\bibfnamefont {R.~E.}\ \bibnamefont
  {Evans}}, \bibinfo {author} {\bibfnamefont {M.~K.}\ \bibnamefont {Bhaskar}},
  \bibinfo {author} {\bibfnamefont {D.~D.}\ \bibnamefont {Sukachev}}, \bibinfo
  {author} {\bibfnamefont {C.~T.}\ \bibnamefont {Nguyen}}, \bibinfo {author}
  {\bibfnamefont {A.}~\bibnamefont {Sipahigil}}, \bibinfo {author}
  {\bibfnamefont {M.~J.}\ \bibnamefont {Burek}}, \bibinfo {author}
  {\bibfnamefont {B.}~\bibnamefont {Machielse}}, \bibinfo {author}
  {\bibfnamefont {G.~H.}\ \bibnamefont {Zhang}}, \bibinfo {author}
  {\bibfnamefont {A.~S.}\ \bibnamefont {Zibrov}}, \bibinfo {author}
  {\bibfnamefont {E.}~\bibnamefont {Bielejec}},  \emph {et~al.},\ }\href@noop
  {} {\bibfield  {journal} {\bibinfo  {journal} {Science}\ }\textbf {\bibinfo
  {volume} {362}},\ \bibinfo {pages} {662} (\bibinfo {year}
  {2018})}\BibitemShut {NoStop}%
\bibitem [{\citenamefont {Kindem}\ \emph {et~al.}(2020)\citenamefont {Kindem},
  \citenamefont {Ruskuc}, \citenamefont {Bartholomew}, \citenamefont {Rochman},
  \citenamefont {Huan},\ and\ \citenamefont {Faraon}}]{kindem2020control}%
  \BibitemOpen
  \bibfield  {author} {\bibinfo {author} {\bibfnamefont {J.~M.}\ \bibnamefont
  {Kindem}}, \bibinfo {author} {\bibfnamefont {A.}~\bibnamefont {Ruskuc}},
  \bibinfo {author} {\bibfnamefont {J.~G.}\ \bibnamefont {Bartholomew}},
  \bibinfo {author} {\bibfnamefont {J.}~\bibnamefont {Rochman}}, \bibinfo
  {author} {\bibfnamefont {Y.~Q.}\ \bibnamefont {Huan}}, \ and\ \bibinfo
  {author} {\bibfnamefont {A.}~\bibnamefont {Faraon}},\ }\href@noop {}
  {\bibfield  {journal} {\bibinfo  {journal} {Nature}\ }\textbf {\bibinfo
  {volume} {580}},\ \bibinfo {pages} {201} (\bibinfo {year}
  {2020})}\BibitemShut {NoStop}%
\bibitem [{\citenamefont {Raha}\ \emph {et~al.}(2020)\citenamefont {Raha},
  \citenamefont {Chen}, \citenamefont {Phenicie}, \citenamefont {Ourari},
  \citenamefont {Dibos},\ and\ \citenamefont {Thompson}}]{raha2020optical}%
  \BibitemOpen
  \bibfield  {author} {\bibinfo {author} {\bibfnamefont {M.}~\bibnamefont
  {Raha}}, \bibinfo {author} {\bibfnamefont {S.}~\bibnamefont {Chen}}, \bibinfo
  {author} {\bibfnamefont {C.~M.}\ \bibnamefont {Phenicie}}, \bibinfo {author}
  {\bibfnamefont {S.}~\bibnamefont {Ourari}}, \bibinfo {author} {\bibfnamefont
  {A.~M.}\ \bibnamefont {Dibos}}, \ and\ \bibinfo {author} {\bibfnamefont
  {J.~D.}\ \bibnamefont {Thompson}},\ }\href@noop {} {\bibfield  {journal}
  {\bibinfo  {journal} {Nature communications}\ }\textbf {\bibinfo {volume}
  {11}},\ \bibinfo {pages} {1} (\bibinfo {year} {2020})}\BibitemShut {NoStop}%
\bibitem [{\citenamefont {Wang}\ \emph {et~al.}(2018)\citenamefont {Wang},
  \citenamefont {Zhang}, \citenamefont {Chen}, \citenamefont {Bertrand},
  \citenamefont {Shams-Ansari}, \citenamefont {Chandrasekhar}, \citenamefont
  {Winzer},\ and\ \citenamefont {Lon{\v{c}}ar}}]{wang2018integrated}%
  \BibitemOpen
  \bibfield  {author} {\bibinfo {author} {\bibfnamefont {C.}~\bibnamefont
  {Wang}}, \bibinfo {author} {\bibfnamefont {M.}~\bibnamefont {Zhang}},
  \bibinfo {author} {\bibfnamefont {X.}~\bibnamefont {Chen}}, \bibinfo {author}
  {\bibfnamefont {M.}~\bibnamefont {Bertrand}}, \bibinfo {author}
  {\bibfnamefont {A.}~\bibnamefont {Shams-Ansari}}, \bibinfo {author}
  {\bibfnamefont {S.}~\bibnamefont {Chandrasekhar}}, \bibinfo {author}
  {\bibfnamefont {P.}~\bibnamefont {Winzer}}, \ and\ \bibinfo {author}
  {\bibfnamefont {M.}~\bibnamefont {Lon{\v{c}}ar}},\ }\href@noop {} {\bibfield
  {journal} {\bibinfo  {journal} {Nature}\ }\textbf {\bibinfo {volume} {562}},\
  \bibinfo {pages} {101} (\bibinfo {year} {2018})}\BibitemShut {NoStop}%
\bibitem [{\citenamefont {Heck}\ \emph {et~al.}(2014)\citenamefont {Heck},
  \citenamefont {Bauters}, \citenamefont {Davenport}, \citenamefont {Spencer},\
  and\ \citenamefont {Bowers}}]{heck2014ultra}%
  \BibitemOpen
  \bibfield  {author} {\bibinfo {author} {\bibfnamefont {M.~J.}\ \bibnamefont
  {Heck}}, \bibinfo {author} {\bibfnamefont {J.~F.}\ \bibnamefont {Bauters}},
  \bibinfo {author} {\bibfnamefont {M.~L.}\ \bibnamefont {Davenport}}, \bibinfo
  {author} {\bibfnamefont {D.~T.}\ \bibnamefont {Spencer}}, \ and\ \bibinfo
  {author} {\bibfnamefont {J.~E.}\ \bibnamefont {Bowers}},\ }\href@noop {}
  {\bibfield  {journal} {\bibinfo  {journal} {Laser \& Photonics Reviews}\
  }\textbf {\bibinfo {volume} {8}},\ \bibinfo {pages} {667} (\bibinfo {year}
  {2014})}\BibitemShut {NoStop}%
\bibitem [{\citenamefont {Liu}\ \emph {et~al.}(2020{\natexlab{a}})\citenamefont
  {Liu}, \citenamefont {Tian}, \citenamefont {Lucas}, \citenamefont {Raja},
  \citenamefont {Lihachev}, \citenamefont {Wang}, \citenamefont {He},
  \citenamefont {Liu}, \citenamefont {Anderson}, \citenamefont {Weng} \emph
  {et~al.}}]{liu2020monolithic}%
  \BibitemOpen
  \bibfield  {author} {\bibinfo {author} {\bibfnamefont {J.}~\bibnamefont
  {Liu}}, \bibinfo {author} {\bibfnamefont {H.}~\bibnamefont {Tian}}, \bibinfo
  {author} {\bibfnamefont {E.}~\bibnamefont {Lucas}}, \bibinfo {author}
  {\bibfnamefont {A.~S.}\ \bibnamefont {Raja}}, \bibinfo {author}
  {\bibfnamefont {G.}~\bibnamefont {Lihachev}}, \bibinfo {author}
  {\bibfnamefont {R.~N.}\ \bibnamefont {Wang}}, \bibinfo {author}
  {\bibfnamefont {J.}~\bibnamefont {He}}, \bibinfo {author} {\bibfnamefont
  {T.}~\bibnamefont {Liu}}, \bibinfo {author} {\bibfnamefont {M.~H.}\
  \bibnamefont {Anderson}}, \bibinfo {author} {\bibfnamefont {W.}~\bibnamefont
  {Weng}},  \emph {et~al.},\ }\href@noop {} {\bibfield  {journal} {\bibinfo
  {journal} {Nature}\ }\textbf {\bibinfo {volume} {583}},\ \bibinfo {pages}
  {385} (\bibinfo {year} {2020}{\natexlab{a}})}\BibitemShut {NoStop}%
\bibitem [{\citenamefont {Liu}\ \emph {et~al.}(2020{\natexlab{b}})\citenamefont
  {Liu}, \citenamefont {Huang}, \citenamefont {Wang}, \citenamefont {He},
  \citenamefont {Raja}, \citenamefont {Liu}, \citenamefont {Engelsen},\ and\
  \citenamefont {Kippenberg}}]{liu2020high}%
  \BibitemOpen
  \bibfield  {author} {\bibinfo {author} {\bibfnamefont {J.}~\bibnamefont
  {Liu}}, \bibinfo {author} {\bibfnamefont {G.}~\bibnamefont {Huang}}, \bibinfo
  {author} {\bibfnamefont {R.~N.}\ \bibnamefont {Wang}}, \bibinfo {author}
  {\bibfnamefont {J.}~\bibnamefont {He}}, \bibinfo {author} {\bibfnamefont
  {A.~S.}\ \bibnamefont {Raja}}, \bibinfo {author} {\bibfnamefont
  {T.}~\bibnamefont {Liu}}, \bibinfo {author} {\bibfnamefont {N.~J.}\
  \bibnamefont {Engelsen}}, \ and\ \bibinfo {author} {\bibfnamefont {T.~J.}\
  \bibnamefont {Kippenberg}},\ }\href@noop {} {\bibfield  {journal} {\bibinfo
  {journal} {arXiv preprint arXiv:2005.13949}\ } (\bibinfo {year}
  {2020}{\natexlab{b}})}\BibitemShut {NoStop}%
\bibitem [{\citenamefont {Castelletto}\ and\ \citenamefont
  {Boretti}(2020)}]{castelletto2020silicon}%
  \BibitemOpen
  \bibfield  {author} {\bibinfo {author} {\bibfnamefont {S.}~\bibnamefont
  {Castelletto}}\ and\ \bibinfo {author} {\bibfnamefont {A.}~\bibnamefont
  {Boretti}},\ }\href@noop {} {\bibfield  {journal} {\bibinfo  {journal}
  {Journal of Physics: Photonics}\ }\textbf {\bibinfo {volume} {2}},\ \bibinfo
  {pages} {022001} (\bibinfo {year} {2020})}\BibitemShut {NoStop}%
\bibitem [{\citenamefont {Lukin}\ \emph
  {et~al.}(2020{\natexlab{a}})\citenamefont {Lukin}, \citenamefont {Dory},
  \citenamefont {Guidry}, \citenamefont {Yang}, \citenamefont {Mishra},
  \citenamefont {Trivedi}, \citenamefont {Radulaski}, \citenamefont {Sun},
  \citenamefont {Vercruysse}, \citenamefont {Ahn} \emph
  {et~al.}}]{Lukin20194H}%
  \BibitemOpen
  \bibfield  {author} {\bibinfo {author} {\bibfnamefont {D.~M.}\ \bibnamefont
  {Lukin}}, \bibinfo {author} {\bibfnamefont {C.}~\bibnamefont {Dory}},
  \bibinfo {author} {\bibfnamefont {M.~A.}\ \bibnamefont {Guidry}}, \bibinfo
  {author} {\bibfnamefont {K.~Y.}\ \bibnamefont {Yang}}, \bibinfo {author}
  {\bibfnamefont {S.~D.}\ \bibnamefont {Mishra}}, \bibinfo {author}
  {\bibfnamefont {R.}~\bibnamefont {Trivedi}}, \bibinfo {author} {\bibfnamefont
  {M.}~\bibnamefont {Radulaski}}, \bibinfo {author} {\bibfnamefont
  {S.}~\bibnamefont {Sun}}, \bibinfo {author} {\bibfnamefont {D.}~\bibnamefont
  {Vercruysse}}, \bibinfo {author} {\bibfnamefont {G.~H.}\ \bibnamefont {Ahn}},
   \emph {et~al.},\ }\href@noop {} {\bibfield  {journal} {\bibinfo  {journal}
  {Nature Photonics}\ }\textbf {\bibinfo {volume} {14}},\ \bibinfo {pages}
  {330} (\bibinfo {year} {2020}{\natexlab{a}})}\BibitemShut {NoStop}%
\bibitem [{\citenamefont {Song}\ \emph {et~al.}(2019)\citenamefont {Song},
  \citenamefont {Asano}, \citenamefont {Jeon}, \citenamefont {Kim},
  \citenamefont {Chen}, \citenamefont {Kang},\ and\ \citenamefont
  {Noda}}]{song2019ultrahigh}%
  \BibitemOpen
  \bibfield  {author} {\bibinfo {author} {\bibfnamefont {B.-S.}\ \bibnamefont
  {Song}}, \bibinfo {author} {\bibfnamefont {T.}~\bibnamefont {Asano}},
  \bibinfo {author} {\bibfnamefont {S.}~\bibnamefont {Jeon}}, \bibinfo {author}
  {\bibfnamefont {H.}~\bibnamefont {Kim}}, \bibinfo {author} {\bibfnamefont
  {C.}~\bibnamefont {Chen}}, \bibinfo {author} {\bibfnamefont {D.~D.}\
  \bibnamefont {Kang}}, \ and\ \bibinfo {author} {\bibfnamefont
  {S.}~\bibnamefont {Noda}},\ }\href@noop {} {\bibfield  {journal} {\bibinfo
  {journal} {Optica}\ }\textbf {\bibinfo {volume} {6}},\ \bibinfo {pages} {991}
  (\bibinfo {year} {2019})}\BibitemShut {NoStop}%
\bibitem [{\citenamefont {Guidry}\ \emph {et~al.}(2020)\citenamefont {Guidry},
  \citenamefont {Yang}, \citenamefont {Lukin}, \citenamefont {Markosyan},
  \citenamefont {Yang}, \citenamefont {Fejer},\ and\ \citenamefont {Vu{\v
  c}kovi\'c}}]{Guidry2020Optical}%
  \BibitemOpen
  \bibfield  {author} {\bibinfo {author} {\bibfnamefont {M.~A.}\ \bibnamefont
  {Guidry}}, \bibinfo {author} {\bibfnamefont {K.~Y.}\ \bibnamefont {Yang}},
  \bibinfo {author} {\bibfnamefont {D.~M.}\ \bibnamefont {Lukin}}, \bibinfo
  {author} {\bibfnamefont {A.}~\bibnamefont {Markosyan}}, \bibinfo {author}
  {\bibfnamefont {J.}~\bibnamefont {Yang}}, \bibinfo {author} {\bibfnamefont
  {M.~M.}\ \bibnamefont {Fejer}}, \ and\ \bibinfo {author} {\bibfnamefont
  {J.}~\bibnamefont {Vu{\v c}kovi\'c}},\ }\href@noop {} {\bibfield  {journal}
  {\bibinfo  {journal} {Optica}\ }\textbf {\bibinfo {volume} {7}},\ \bibinfo
  {pages} {1139} (\bibinfo {year} {2020})}\BibitemShut {NoStop}%
\bibitem [{\citenamefont {Nagy}\ \emph {et~al.}(2019)\citenamefont {Nagy},
  \citenamefont {Niethammer}, \citenamefont {Widmann}, \citenamefont {Chen},
  \citenamefont {Udvarhelyi}, \citenamefont {Bonato}, \citenamefont {Hassan},
  \citenamefont {Karhu}, \citenamefont {Ivanov}, \citenamefont {Son} \emph
  {et~al.}}]{nagy2019high}%
  \BibitemOpen
  \bibfield  {author} {\bibinfo {author} {\bibfnamefont {R.}~\bibnamefont
  {Nagy}}, \bibinfo {author} {\bibfnamefont {M.}~\bibnamefont {Niethammer}},
  \bibinfo {author} {\bibfnamefont {M.}~\bibnamefont {Widmann}}, \bibinfo
  {author} {\bibfnamefont {Y.-C.}\ \bibnamefont {Chen}}, \bibinfo {author}
  {\bibfnamefont {P.}~\bibnamefont {Udvarhelyi}}, \bibinfo {author}
  {\bibfnamefont {C.}~\bibnamefont {Bonato}}, \bibinfo {author} {\bibfnamefont
  {J.~U.}\ \bibnamefont {Hassan}}, \bibinfo {author} {\bibfnamefont
  {R.}~\bibnamefont {Karhu}}, \bibinfo {author} {\bibfnamefont {I.~G.}\
  \bibnamefont {Ivanov}}, \bibinfo {author} {\bibfnamefont {N.~T.}\
  \bibnamefont {Son}},  \emph {et~al.},\ }\href@noop {} {\bibfield  {journal}
  {\bibinfo  {journal} {Nature communications}\ }\textbf {\bibinfo {volume}
  {10}},\ \bibinfo {pages} {1} (\bibinfo {year} {2019})}\BibitemShut {NoStop}%
\bibitem [{\citenamefont {Vasconcelos}\ \emph {et~al.}(2020)\citenamefont
  {Vasconcelos}, \citenamefont {Reisenbauer}, \citenamefont {Salter},
  \citenamefont {Wachter}, \citenamefont {Wirtitsch}, \citenamefont
  {Schmiedmayer}, \citenamefont {Walther},\ and\ \citenamefont
  {Trupke}}]{vasconcelos2020scalable}%
  \BibitemOpen
  \bibfield  {author} {\bibinfo {author} {\bibfnamefont {R.}~\bibnamefont
  {Vasconcelos}}, \bibinfo {author} {\bibfnamefont {S.}~\bibnamefont
  {Reisenbauer}}, \bibinfo {author} {\bibfnamefont {C.}~\bibnamefont {Salter}},
  \bibinfo {author} {\bibfnamefont {G.}~\bibnamefont {Wachter}}, \bibinfo
  {author} {\bibfnamefont {D.}~\bibnamefont {Wirtitsch}}, \bibinfo {author}
  {\bibfnamefont {J.}~\bibnamefont {Schmiedmayer}}, \bibinfo {author}
  {\bibfnamefont {P.}~\bibnamefont {Walther}}, \ and\ \bibinfo {author}
  {\bibfnamefont {M.}~\bibnamefont {Trupke}},\ }\href@noop {} {\bibfield
  {journal} {\bibinfo  {journal} {npj Quantum Information}\ }\textbf {\bibinfo
  {volume} {6}},\ \bibinfo {pages} {1} (\bibinfo {year} {2020})}\BibitemShut
  {NoStop}%
\bibitem [{\citenamefont {Son}\ \emph {et~al.}(2020)\citenamefont {Son},
  \citenamefont {Anderson}, \citenamefont {Bourassa}, \citenamefont {Miao},
  \citenamefont {Babin}, \citenamefont {Widmann}, \citenamefont {Niethammer},
  \citenamefont {Ul~Hassan}, \citenamefont {Morioka}, \citenamefont {Ivanov}
  \emph {et~al.}}]{son2020developing}%
  \BibitemOpen
  \bibfield  {author} {\bibinfo {author} {\bibfnamefont {N.~T.}\ \bibnamefont
  {Son}}, \bibinfo {author} {\bibfnamefont {C.~P.}\ \bibnamefont {Anderson}},
  \bibinfo {author} {\bibfnamefont {A.}~\bibnamefont {Bourassa}}, \bibinfo
  {author} {\bibfnamefont {K.~C.}\ \bibnamefont {Miao}}, \bibinfo {author}
  {\bibfnamefont {C.}~\bibnamefont {Babin}}, \bibinfo {author} {\bibfnamefont
  {M.}~\bibnamefont {Widmann}}, \bibinfo {author} {\bibfnamefont
  {M.}~\bibnamefont {Niethammer}}, \bibinfo {author} {\bibfnamefont
  {J.}~\bibnamefont {Ul~Hassan}}, \bibinfo {author} {\bibfnamefont
  {N.}~\bibnamefont {Morioka}}, \bibinfo {author} {\bibfnamefont {I.~G.}\
  \bibnamefont {Ivanov}},  \emph {et~al.},\ }\href@noop {} {\bibfield
  {journal} {\bibinfo  {journal} {Applied Physics Letters}\ }\textbf {\bibinfo
  {volume} {116}},\ \bibinfo {pages} {190501} (\bibinfo {year}
  {2020})}\BibitemShut {NoStop}%
\bibitem [{\citenamefont {Esmaeil~Zadeh}\ \emph {et~al.}(2020)\citenamefont
  {Esmaeil~Zadeh}, \citenamefont {Los}, \citenamefont {Gourgues}, \citenamefont
  {Chang}, \citenamefont {Elshaari}, \citenamefont {Zichi}, \citenamefont {van
  Staaden}, \citenamefont {Swens}, \citenamefont {Kalhor}, \citenamefont
  {Guardiani} \emph {et~al.}}]{esmaeil2020efficient}%
  \BibitemOpen
  \bibfield  {author} {\bibinfo {author} {\bibfnamefont {I.}~\bibnamefont
  {Esmaeil~Zadeh}}, \bibinfo {author} {\bibfnamefont {J.~W.}\ \bibnamefont
  {Los}}, \bibinfo {author} {\bibfnamefont {R.~B.}\ \bibnamefont {Gourgues}},
  \bibinfo {author} {\bibfnamefont {J.}~\bibnamefont {Chang}}, \bibinfo
  {author} {\bibfnamefont {A.~W.}\ \bibnamefont {Elshaari}}, \bibinfo {author}
  {\bibfnamefont {J.~R.}\ \bibnamefont {Zichi}}, \bibinfo {author}
  {\bibfnamefont {Y.~J.}\ \bibnamefont {van Staaden}}, \bibinfo {author}
  {\bibfnamefont {J.~P.}\ \bibnamefont {Swens}}, \bibinfo {author}
  {\bibfnamefont {N.}~\bibnamefont {Kalhor}}, \bibinfo {author} {\bibfnamefont
  {A.}~\bibnamefont {Guardiani}},  \emph {et~al.},\ }\href@noop {} {\bibfield
  {journal} {\bibinfo  {journal} {ACS Photonics}\ } (\bibinfo {year}
  {2020})}\BibitemShut {NoStop}%
\bibitem [{\citenamefont {Meade}\ \emph {et~al.}(1995)\citenamefont {Meade},
  \citenamefont {Winn},\ and\ \citenamefont
  {Joannopoulos}}]{meade1995photonic}%
  \BibitemOpen
  \bibfield  {author} {\bibinfo {author} {\bibfnamefont {R.}~\bibnamefont
  {Meade}}, \bibinfo {author} {\bibfnamefont {J.~N.}\ \bibnamefont {Winn}}, \
  and\ \bibinfo {author} {\bibfnamefont {J.}~\bibnamefont {Joannopoulos}},\
  }\href@noop {} {\enquote {\bibinfo {title} {Photonic crystals: Molding the
  flow of light},}\ } (\bibinfo {year} {1995})\BibitemShut {NoStop}%
\bibitem [{\citenamefont {Arcari}\ \emph {et~al.}(2014)\citenamefont {Arcari},
  \citenamefont {S{\"o}llner}, \citenamefont {Javadi}, \citenamefont {Hansen},
  \citenamefont {Mahmoodian}, \citenamefont {Liu}, \citenamefont {Thyrrestrup},
  \citenamefont {Lee}, \citenamefont {Song}, \citenamefont {Stobbe} \emph
  {et~al.}}]{arcari2014near}%
  \BibitemOpen
  \bibfield  {author} {\bibinfo {author} {\bibfnamefont {M.}~\bibnamefont
  {Arcari}}, \bibinfo {author} {\bibfnamefont {I.}~\bibnamefont {S{\"o}llner}},
  \bibinfo {author} {\bibfnamefont {A.}~\bibnamefont {Javadi}}, \bibinfo
  {author} {\bibfnamefont {S.~L.}\ \bibnamefont {Hansen}}, \bibinfo {author}
  {\bibfnamefont {S.}~\bibnamefont {Mahmoodian}}, \bibinfo {author}
  {\bibfnamefont {J.}~\bibnamefont {Liu}}, \bibinfo {author} {\bibfnamefont
  {H.}~\bibnamefont {Thyrrestrup}}, \bibinfo {author} {\bibfnamefont {E.~H.}\
  \bibnamefont {Lee}}, \bibinfo {author} {\bibfnamefont {J.~D.}\ \bibnamefont
  {Song}}, \bibinfo {author} {\bibfnamefont {S.}~\bibnamefont {Stobbe}},  \emph
  {et~al.},\ }\href@noop {} {\bibfield  {journal} {\bibinfo  {journal}
  {Physical review letters}\ }\textbf {\bibinfo {volume} {113}},\ \bibinfo
  {pages} {093603} (\bibinfo {year} {2014})}\BibitemShut {NoStop}%
\bibitem [{\citenamefont {Waks}\ and\ \citenamefont
  {Vuckovic}(2006)}]{waks2006dipole}%
  \BibitemOpen
  \bibfield  {author} {\bibinfo {author} {\bibfnamefont {E.}~\bibnamefont
  {Waks}}\ and\ \bibinfo {author} {\bibfnamefont {J.}~\bibnamefont
  {Vuckovic}},\ }\href@noop {} {\bibfield  {journal} {\bibinfo  {journal}
  {Physical review letters}\ }\textbf {\bibinfo {volume} {96}},\ \bibinfo
  {pages} {153601} (\bibinfo {year} {2006})}\BibitemShut {NoStop}%
\bibitem [{\citenamefont {Dong}\ \emph {et~al.}(2019)\citenamefont {Dong},
  \citenamefont {Doherty},\ and\ \citenamefont {Economou}}]{dong2019spin}%
  \BibitemOpen
  \bibfield  {author} {\bibinfo {author} {\bibfnamefont {W.}~\bibnamefont
  {Dong}}, \bibinfo {author} {\bibfnamefont {M.}~\bibnamefont {Doherty}}, \
  and\ \bibinfo {author} {\bibfnamefont {S.~E.}\ \bibnamefont {Economou}},\
  }\href@noop {} {\bibfield  {journal} {\bibinfo  {journal} {Physical Review
  B}\ }\textbf {\bibinfo {volume} {99}},\ \bibinfo {pages} {184102} (\bibinfo
  {year} {2019})}\BibitemShut {NoStop}%
\bibitem [{\citenamefont {Neumann}\ \emph {et~al.}(2010)\citenamefont
  {Neumann}, \citenamefont {Kolesov}, \citenamefont {Naydenov}, \citenamefont
  {Beck}, \citenamefont {Rempp}, \citenamefont {Steiner}, \citenamefont
  {Jacques}, \citenamefont {Balasubramanian}, \citenamefont {Markham},
  \citenamefont {Twitchen} \emph {et~al.}}]{neumann2010quantum}%
  \BibitemOpen
  \bibfield  {author} {\bibinfo {author} {\bibfnamefont {P.}~\bibnamefont
  {Neumann}}, \bibinfo {author} {\bibfnamefont {R.}~\bibnamefont {Kolesov}},
  \bibinfo {author} {\bibfnamefont {B.}~\bibnamefont {Naydenov}}, \bibinfo
  {author} {\bibfnamefont {J.}~\bibnamefont {Beck}}, \bibinfo {author}
  {\bibfnamefont {F.}~\bibnamefont {Rempp}}, \bibinfo {author} {\bibfnamefont
  {M.}~\bibnamefont {Steiner}}, \bibinfo {author} {\bibfnamefont
  {V.}~\bibnamefont {Jacques}}, \bibinfo {author} {\bibfnamefont
  {G.}~\bibnamefont {Balasubramanian}}, \bibinfo {author} {\bibfnamefont
  {M.}~\bibnamefont {Markham}}, \bibinfo {author} {\bibfnamefont
  {D.}~\bibnamefont {Twitchen}},  \emph {et~al.},\ }\href@noop {} {\bibfield
  {journal} {\bibinfo  {journal} {Nature Physics}\ }\textbf {\bibinfo {volume}
  {6}},\ \bibinfo {pages} {249} (\bibinfo {year} {2010})}\BibitemShut {NoStop}%
\bibitem [{\citenamefont {S{\"o}rman}\ \emph {et~al.}(2000)\citenamefont
  {S{\"o}rman}, \citenamefont {Son}, \citenamefont {Chen}, \citenamefont
  {Kordina}, \citenamefont {Hallin},\ and\ \citenamefont
  {Janz{\'e}n}}]{sorman2000silicon}%
  \BibitemOpen
  \bibfield  {author} {\bibinfo {author} {\bibfnamefont {E.}~\bibnamefont
  {S{\"o}rman}}, \bibinfo {author} {\bibfnamefont {N.}~\bibnamefont {Son}},
  \bibinfo {author} {\bibfnamefont {W.}~\bibnamefont {Chen}}, \bibinfo {author}
  {\bibfnamefont {O.}~\bibnamefont {Kordina}}, \bibinfo {author} {\bibfnamefont
  {C.}~\bibnamefont {Hallin}}, \ and\ \bibinfo {author} {\bibfnamefont
  {E.}~\bibnamefont {Janz{\'e}n}},\ }\href@noop {} {\bibfield  {journal}
  {\bibinfo  {journal} {Physical Review B}\ }\textbf {\bibinfo {volume} {61}},\
  \bibinfo {pages} {2613} (\bibinfo {year} {2000})}\BibitemShut {NoStop}%
\bibitem [{\citenamefont {Baranov}\ \emph {et~al.}(2011)\citenamefont
  {Baranov}, \citenamefont {Bundakova}, \citenamefont {Soltamova},
  \citenamefont {Orlinskii}, \citenamefont {Borovykh}, \citenamefont
  {Zondervan}, \citenamefont {Verberk},\ and\ \citenamefont
  {Schmidt}}]{baranov2011silicon}%
  \BibitemOpen
  \bibfield  {author} {\bibinfo {author} {\bibfnamefont {P.~G.}\ \bibnamefont
  {Baranov}}, \bibinfo {author} {\bibfnamefont {A.~P.}\ \bibnamefont
  {Bundakova}}, \bibinfo {author} {\bibfnamefont {A.~A.}\ \bibnamefont
  {Soltamova}}, \bibinfo {author} {\bibfnamefont {S.~B.}\ \bibnamefont
  {Orlinskii}}, \bibinfo {author} {\bibfnamefont {I.~V.}\ \bibnamefont
  {Borovykh}}, \bibinfo {author} {\bibfnamefont {R.}~\bibnamefont {Zondervan}},
  \bibinfo {author} {\bibfnamefont {R.}~\bibnamefont {Verberk}}, \ and\
  \bibinfo {author} {\bibfnamefont {J.}~\bibnamefont {Schmidt}},\ }\href@noop
  {} {\bibfield  {journal} {\bibinfo  {journal} {Physical Review B}\ }\textbf
  {\bibinfo {volume} {83}},\ \bibinfo {pages} {125203} (\bibinfo {year}
  {2011})}\BibitemShut {NoStop}%
\bibitem [{\citenamefont {Soltamov}\ \emph {et~al.}(2015)\citenamefont
  {Soltamov}, \citenamefont {Yavkin}, \citenamefont {Tolmachev}, \citenamefont
  {Babunts}, \citenamefont {Badalyan}, \citenamefont {Davydov}, \citenamefont
  {Mokhov}, \citenamefont {Proskuryakov}, \citenamefont {Orlinskii},\ and\
  \citenamefont {Baranov}}]{soltamov2015optically}%
  \BibitemOpen
  \bibfield  {author} {\bibinfo {author} {\bibfnamefont {V.}~\bibnamefont
  {Soltamov}}, \bibinfo {author} {\bibfnamefont {B.}~\bibnamefont {Yavkin}},
  \bibinfo {author} {\bibfnamefont {D.}~\bibnamefont {Tolmachev}}, \bibinfo
  {author} {\bibfnamefont {R.}~\bibnamefont {Babunts}}, \bibinfo {author}
  {\bibfnamefont {A.}~\bibnamefont {Badalyan}}, \bibinfo {author}
  {\bibfnamefont {V.~Y.}\ \bibnamefont {Davydov}}, \bibinfo {author}
  {\bibfnamefont {E.}~\bibnamefont {Mokhov}}, \bibinfo {author} {\bibfnamefont
  {I.}~\bibnamefont {Proskuryakov}}, \bibinfo {author} {\bibfnamefont
  {S.}~\bibnamefont {Orlinskii}}, \ and\ \bibinfo {author} {\bibfnamefont
  {P.}~\bibnamefont {Baranov}},\ }\href@noop {} {\bibfield  {journal} {\bibinfo
   {journal} {Physical review letters}\ }\textbf {\bibinfo {volume} {115}},\
  \bibinfo {pages} {247602} (\bibinfo {year} {2015})}\BibitemShut {NoStop}%
\bibitem [{\citenamefont {Simin}\ \emph {et~al.}(2017)\citenamefont {Simin},
  \citenamefont {Kraus}, \citenamefont {Sperlich}, \citenamefont {Ohshima},
  \citenamefont {Astakhov},\ and\ \citenamefont {Dyakonov}}]{simin2017locking}%
  \BibitemOpen
  \bibfield  {author} {\bibinfo {author} {\bibfnamefont {D.}~\bibnamefont
  {Simin}}, \bibinfo {author} {\bibfnamefont {H.}~\bibnamefont {Kraus}},
  \bibinfo {author} {\bibfnamefont {A.}~\bibnamefont {Sperlich}}, \bibinfo
  {author} {\bibfnamefont {T.}~\bibnamefont {Ohshima}}, \bibinfo {author}
  {\bibfnamefont {G.}~\bibnamefont {Astakhov}}, \ and\ \bibinfo {author}
  {\bibfnamefont {V.}~\bibnamefont {Dyakonov}},\ }\href@noop {} {\bibfield
  {journal} {\bibinfo  {journal} {Physical Review B}\ }\textbf {\bibinfo
  {volume} {95}},\ \bibinfo {pages} {161201} (\bibinfo {year}
  {2017})}\BibitemShut {NoStop}%
\bibitem [{\citenamefont {Banks}\ \emph {et~al.}(2019)\citenamefont {Banks},
  \citenamefont {Soykal}, \citenamefont {Myers-Ward}, \citenamefont {Gaskill},
  \citenamefont {Reinecke},\ and\ \citenamefont {Carter}}]{banks2019resonant}%
  \BibitemOpen
  \bibfield  {author} {\bibinfo {author} {\bibfnamefont {H.~B.}\ \bibnamefont
  {Banks}}, \bibinfo {author} {\bibfnamefont {{\"O}.~O.}\ \bibnamefont
  {Soykal}}, \bibinfo {author} {\bibfnamefont {R.~L.}\ \bibnamefont
  {Myers-Ward}}, \bibinfo {author} {\bibfnamefont {D.~K.}\ \bibnamefont
  {Gaskill}}, \bibinfo {author} {\bibfnamefont {T.}~\bibnamefont {Reinecke}}, \
  and\ \bibinfo {author} {\bibfnamefont {S.~G.}\ \bibnamefont {Carter}},\
  }\href@noop {} {\bibfield  {journal} {\bibinfo  {journal} {Physical Review
  Applied}\ }\textbf {\bibinfo {volume} {11}},\ \bibinfo {pages} {024013}
  (\bibinfo {year} {2019})}\BibitemShut {NoStop}%
\bibitem [{\citenamefont {Davidsson}\ \emph {et~al.}(2019)\citenamefont
  {Davidsson}, \citenamefont {Iv{\'a}dy}, \citenamefont {Armiento},
  \citenamefont {Ohshima}, \citenamefont {Son}, \citenamefont {Gali},\ and\
  \citenamefont {Abrikosov}}]{davidsson2019identification}%
  \BibitemOpen
  \bibfield  {author} {\bibinfo {author} {\bibfnamefont {J.}~\bibnamefont
  {Davidsson}}, \bibinfo {author} {\bibfnamefont {V.}~\bibnamefont
  {Iv{\'a}dy}}, \bibinfo {author} {\bibfnamefont {R.}~\bibnamefont {Armiento}},
  \bibinfo {author} {\bibfnamefont {T.}~\bibnamefont {Ohshima}}, \bibinfo
  {author} {\bibfnamefont {N.}~\bibnamefont {Son}}, \bibinfo {author}
  {\bibfnamefont {A.}~\bibnamefont {Gali}}, \ and\ \bibinfo {author}
  {\bibfnamefont {I.~A.}\ \bibnamefont {Abrikosov}},\ }\href@noop {} {\bibfield
   {journal} {\bibinfo  {journal} {Applied Physics Letters}\ }\textbf {\bibinfo
  {volume} {114}},\ \bibinfo {pages} {112107} (\bibinfo {year}
  {2019})}\BibitemShut {NoStop}%
\bibitem [{\citenamefont {Udvarhelyi}\ \emph {et~al.}(2020)\citenamefont
  {Udvarhelyi}, \citenamefont {Thiering}, \citenamefont {Morioka},
  \citenamefont {Babin}, \citenamefont {Kaiser}, \citenamefont {Lukin},
  \citenamefont {Ohshima}, \citenamefont {Ul-Hassan}, \citenamefont {Son},
  \citenamefont {Vu{\v{c}}kovi{\'c}} \emph {et~al.}}]{udvarhelyi2020vibronic}%
  \BibitemOpen
  \bibfield  {author} {\bibinfo {author} {\bibfnamefont {P.}~\bibnamefont
  {Udvarhelyi}}, \bibinfo {author} {\bibfnamefont {G.}~\bibnamefont
  {Thiering}}, \bibinfo {author} {\bibfnamefont {N.}~\bibnamefont {Morioka}},
  \bibinfo {author} {\bibfnamefont {C.}~\bibnamefont {Babin}}, \bibinfo
  {author} {\bibfnamefont {F.}~\bibnamefont {Kaiser}}, \bibinfo {author}
  {\bibfnamefont {D.}~\bibnamefont {Lukin}}, \bibinfo {author} {\bibfnamefont
  {T.}~\bibnamefont {Ohshima}}, \bibinfo {author} {\bibfnamefont
  {J.}~\bibnamefont {Ul-Hassan}}, \bibinfo {author} {\bibfnamefont {N.~T.}\
  \bibnamefont {Son}}, \bibinfo {author} {\bibfnamefont {J.}~\bibnamefont
  {Vu{\v{c}}kovi{\'c}}},  \emph {et~al.},\ }\href@noop {} {\bibfield  {journal}
  {\bibinfo  {journal} {Physical Review Applied}\ }\textbf {\bibinfo {volume}
  {13}},\ \bibinfo {pages} {054017} (\bibinfo {year} {2020})}\BibitemShut
  {NoStop}%
\bibitem [{\citenamefont {Shang}\ \emph {et~al.}(2020)\citenamefont {Shang},
  \citenamefont {Hashemi}, \citenamefont {Berenc{\'e}n}, \citenamefont {Komsa},
  \citenamefont {Erhart}, \citenamefont {Zhou}, \citenamefont {Helm},
  \citenamefont {Krasheninnikov},\ and\ \citenamefont
  {Astakhov}}]{shang2020local}%
  \BibitemOpen
  \bibfield  {author} {\bibinfo {author} {\bibfnamefont {Z.}~\bibnamefont
  {Shang}}, \bibinfo {author} {\bibfnamefont {A.}~\bibnamefont {Hashemi}},
  \bibinfo {author} {\bibfnamefont {Y.}~\bibnamefont {Berenc{\'e}n}}, \bibinfo
  {author} {\bibfnamefont {H.-P.}\ \bibnamefont {Komsa}}, \bibinfo {author}
  {\bibfnamefont {P.}~\bibnamefont {Erhart}}, \bibinfo {author} {\bibfnamefont
  {S.}~\bibnamefont {Zhou}}, \bibinfo {author} {\bibfnamefont {M.}~\bibnamefont
  {Helm}}, \bibinfo {author} {\bibfnamefont {A.}~\bibnamefont
  {Krasheninnikov}}, \ and\ \bibinfo {author} {\bibfnamefont {G.}~\bibnamefont
  {Astakhov}},\ }\href@noop {} {\bibfield  {journal} {\bibinfo  {journal}
  {Physical Review B}\ }\textbf {\bibinfo {volume} {101}},\ \bibinfo {pages}
  {144109} (\bibinfo {year} {2020})}\BibitemShut {NoStop}%
\bibitem [{\citenamefont {Morioka}\ \emph {et~al.}(2020)\citenamefont
  {Morioka}, \citenamefont {Babin}, \citenamefont {Nagy}, \citenamefont
  {Gediz}, \citenamefont {Hesselmeier}, \citenamefont {Liu}, \citenamefont
  {Joliffe}, \citenamefont {Niethammer}, \citenamefont {Dasari}, \citenamefont
  {Vorobyov} \emph {et~al.}}]{morioka2020spin}%
  \BibitemOpen
  \bibfield  {author} {\bibinfo {author} {\bibfnamefont {N.}~\bibnamefont
  {Morioka}}, \bibinfo {author} {\bibfnamefont {C.}~\bibnamefont {Babin}},
  \bibinfo {author} {\bibfnamefont {R.}~\bibnamefont {Nagy}}, \bibinfo {author}
  {\bibfnamefont {I.}~\bibnamefont {Gediz}}, \bibinfo {author} {\bibfnamefont
  {E.}~\bibnamefont {Hesselmeier}}, \bibinfo {author} {\bibfnamefont
  {D.}~\bibnamefont {Liu}}, \bibinfo {author} {\bibfnamefont {M.}~\bibnamefont
  {Joliffe}}, \bibinfo {author} {\bibfnamefont {M.}~\bibnamefont {Niethammer}},
  \bibinfo {author} {\bibfnamefont {D.}~\bibnamefont {Dasari}}, \bibinfo
  {author} {\bibfnamefont {V.}~\bibnamefont {Vorobyov}},  \emph {et~al.},\
  }\href@noop {} {\bibfield  {journal} {\bibinfo  {journal} {Nature
  communications}\ }\textbf {\bibinfo {volume} {11}},\ \bibinfo {pages} {1}
  (\bibinfo {year} {2020})}\BibitemShut {NoStop}%
\bibitem [{\citenamefont {R{\"u}hl}\ \emph {et~al.}(2019)\citenamefont
  {R{\"u}hl}, \citenamefont {Bergmann}, \citenamefont {Krieger},\ and\
  \citenamefont {Weber}}]{ruhl2019stark}%
  \BibitemOpen
  \bibfield  {author} {\bibinfo {author} {\bibfnamefont {M.}~\bibnamefont
  {R{\"u}hl}}, \bibinfo {author} {\bibfnamefont {L.}~\bibnamefont {Bergmann}},
  \bibinfo {author} {\bibfnamefont {M.}~\bibnamefont {Krieger}}, \ and\
  \bibinfo {author} {\bibfnamefont {H.~B.}\ \bibnamefont {Weber}},\ }\href@noop
  {} {\bibfield  {journal} {\bibinfo  {journal} {Nano Letters}\ } (\bibinfo
  {year} {2019})}\BibitemShut {NoStop}%
\bibitem [{\citenamefont {Lukin}\ \emph
  {et~al.}(2020{\natexlab{b}})\citenamefont {Lukin}, \citenamefont {White},
  \citenamefont {Guidry}, \citenamefont {Trivedi}, \citenamefont {Morioka},
  \citenamefont {Babin}, \citenamefont {Hassan}, \citenamefont {Son},
  \citenamefont {Ohshima}, \citenamefont {Vasireddy} \emph
  {et~al.}}]{lukin2020spectrally}%
  \BibitemOpen
  \bibfield  {author} {\bibinfo {author} {\bibfnamefont {D.~M.}\ \bibnamefont
  {Lukin}}, \bibinfo {author} {\bibfnamefont {A.~D.}\ \bibnamefont {White}},
  \bibinfo {author} {\bibfnamefont {M.~A.}\ \bibnamefont {Guidry}}, \bibinfo
  {author} {\bibfnamefont {R.}~\bibnamefont {Trivedi}}, \bibinfo {author}
  {\bibfnamefont {N.}~\bibnamefont {Morioka}}, \bibinfo {author} {\bibfnamefont
  {C.}~\bibnamefont {Babin}}, \bibinfo {author} {\bibfnamefont {J.~U.}\
  \bibnamefont {Hassan}}, \bibinfo {author} {\bibfnamefont {N.~T.}\
  \bibnamefont {Son}}, \bibinfo {author} {\bibfnamefont {T.}~\bibnamefont
  {Ohshima}}, \bibinfo {author} {\bibfnamefont {P.~K.}\ \bibnamefont
  {Vasireddy}},  \emph {et~al.},\ }\href@noop {} {\bibfield  {journal}
  {\bibinfo  {journal} {npj Quantum Info}\ }\textbf {\bibinfo {volume} {6}},\
  \bibinfo {pages} {80} (\bibinfo {year} {2020}{\natexlab{b}})}\BibitemShut
  {NoStop}%
\bibitem [{\citenamefont {Koehl}\ \emph {et~al.}(2011)\citenamefont {Koehl},
  \citenamefont {Buckley}, \citenamefont {Heremans}, \citenamefont {Calusine},\
  and\ \citenamefont {Awschalom}}]{koehl2011room}%
  \BibitemOpen
  \bibfield  {author} {\bibinfo {author} {\bibfnamefont {W.~F.}\ \bibnamefont
  {Koehl}}, \bibinfo {author} {\bibfnamefont {B.~B.}\ \bibnamefont {Buckley}},
  \bibinfo {author} {\bibfnamefont {F.~J.}\ \bibnamefont {Heremans}}, \bibinfo
  {author} {\bibfnamefont {G.}~\bibnamefont {Calusine}}, \ and\ \bibinfo
  {author} {\bibfnamefont {D.~D.}\ \bibnamefont {Awschalom}},\ }\href@noop {}
  {\bibfield  {journal} {\bibinfo  {journal} {Nature}\ }\textbf {\bibinfo
  {volume} {479}},\ \bibinfo {pages} {84} (\bibinfo {year} {2011})}\BibitemShut
  {NoStop}%
\bibitem [{\citenamefont {Falk}\ \emph {et~al.}(2013)\citenamefont {Falk},
  \citenamefont {Buckley}, \citenamefont {Calusine}, \citenamefont {Koehl},
  \citenamefont {Dobrovitski}, \citenamefont {Politi}, \citenamefont {Zorman},
  \citenamefont {Feng},\ and\ \citenamefont {Awschalom}}]{falk2013polytype}%
  \BibitemOpen
  \bibfield  {author} {\bibinfo {author} {\bibfnamefont {A.~L.}\ \bibnamefont
  {Falk}}, \bibinfo {author} {\bibfnamefont {B.~B.}\ \bibnamefont {Buckley}},
  \bibinfo {author} {\bibfnamefont {G.}~\bibnamefont {Calusine}}, \bibinfo
  {author} {\bibfnamefont {W.~F.}\ \bibnamefont {Koehl}}, \bibinfo {author}
  {\bibfnamefont {V.~V.}\ \bibnamefont {Dobrovitski}}, \bibinfo {author}
  {\bibfnamefont {A.}~\bibnamefont {Politi}}, \bibinfo {author} {\bibfnamefont
  {C.~A.}\ \bibnamefont {Zorman}}, \bibinfo {author} {\bibfnamefont {P.~X.-L.}\
  \bibnamefont {Feng}}, \ and\ \bibinfo {author} {\bibfnamefont {D.~D.}\
  \bibnamefont {Awschalom}},\ }\href@noop {} {\bibfield  {journal} {\bibinfo
  {journal} {Nature communications}\ }\textbf {\bibinfo {volume} {4}},\
  \bibinfo {pages} {1} (\bibinfo {year} {2013})}\BibitemShut {NoStop}%
\bibitem [{\citenamefont {Christle}\ \emph {et~al.}(2015)\citenamefont
  {Christle}, \citenamefont {Falk}, \citenamefont {Andrich}, \citenamefont
  {Klimov}, \citenamefont {Hassan}, \citenamefont {Son}, \citenamefont
  {Janz{\'e}n}, \citenamefont {Ohshima},\ and\ \citenamefont
  {Awschalom}}]{christle2015isolated}%
  \BibitemOpen
  \bibfield  {author} {\bibinfo {author} {\bibfnamefont {D.~J.}\ \bibnamefont
  {Christle}}, \bibinfo {author} {\bibfnamefont {A.~L.}\ \bibnamefont {Falk}},
  \bibinfo {author} {\bibfnamefont {P.}~\bibnamefont {Andrich}}, \bibinfo
  {author} {\bibfnamefont {P.~V.}\ \bibnamefont {Klimov}}, \bibinfo {author}
  {\bibfnamefont {J.~U.}\ \bibnamefont {Hassan}}, \bibinfo {author}
  {\bibfnamefont {N.~T.}\ \bibnamefont {Son}}, \bibinfo {author} {\bibfnamefont
  {E.}~\bibnamefont {Janz{\'e}n}}, \bibinfo {author} {\bibfnamefont
  {T.}~\bibnamefont {Ohshima}}, \ and\ \bibinfo {author} {\bibfnamefont
  {D.~D.}\ \bibnamefont {Awschalom}},\ }\href@noop {} {\bibfield  {journal}
  {\bibinfo  {journal} {Nature materials}\ }\textbf {\bibinfo {volume} {14}},\
  \bibinfo {pages} {160} (\bibinfo {year} {2015})}\BibitemShut {NoStop}%
\bibitem [{\citenamefont {Falk}\ \emph {et~al.}(2015)\citenamefont {Falk},
  \citenamefont {Klimov}, \citenamefont {Iv{\'a}dy}, \citenamefont {Sz{\'a}sz},
  \citenamefont {Christle}, \citenamefont {Koehl}, \citenamefont {Gali},\ and\
  \citenamefont {Awschalom}}]{falk2015optical}%
  \BibitemOpen
  \bibfield  {author} {\bibinfo {author} {\bibfnamefont {A.~L.}\ \bibnamefont
  {Falk}}, \bibinfo {author} {\bibfnamefont {P.~V.}\ \bibnamefont {Klimov}},
  \bibinfo {author} {\bibfnamefont {V.}~\bibnamefont {Iv{\'a}dy}}, \bibinfo
  {author} {\bibfnamefont {K.}~\bibnamefont {Sz{\'a}sz}}, \bibinfo {author}
  {\bibfnamefont {D.~J.}\ \bibnamefont {Christle}}, \bibinfo {author}
  {\bibfnamefont {W.~F.}\ \bibnamefont {Koehl}}, \bibinfo {author}
  {\bibfnamefont {{\'A}.}~\bibnamefont {Gali}}, \ and\ \bibinfo {author}
  {\bibfnamefont {D.~D.}\ \bibnamefont {Awschalom}},\ }\href@noop {} {\bibfield
   {journal} {\bibinfo  {journal} {Physical review letters}\ }\textbf {\bibinfo
  {volume} {114}},\ \bibinfo {pages} {247603} (\bibinfo {year}
  {2015})}\BibitemShut {NoStop}%
\bibitem [{\citenamefont {Klimov}\ \emph {et~al.}(2015)\citenamefont {Klimov},
  \citenamefont {Falk}, \citenamefont {Christle}, \citenamefont {Dobrovitski},\
  and\ \citenamefont {Awschalom}}]{klimov2015quantum}%
  \BibitemOpen
  \bibfield  {author} {\bibinfo {author} {\bibfnamefont {P.~V.}\ \bibnamefont
  {Klimov}}, \bibinfo {author} {\bibfnamefont {A.~L.}\ \bibnamefont {Falk}},
  \bibinfo {author} {\bibfnamefont {D.~J.}\ \bibnamefont {Christle}}, \bibinfo
  {author} {\bibfnamefont {V.~V.}\ \bibnamefont {Dobrovitski}}, \ and\ \bibinfo
  {author} {\bibfnamefont {D.~D.}\ \bibnamefont {Awschalom}},\ }\href@noop {}
  {\bibfield  {journal} {\bibinfo  {journal} {Science advances}\ }\textbf
  {\bibinfo {volume} {1}},\ \bibinfo {pages} {e1501015} (\bibinfo {year}
  {2015})}\BibitemShut {NoStop}%
\bibitem [{\citenamefont {Iv{\'a}dy}\ \emph {et~al.}(2016)\citenamefont
  {Iv{\'a}dy}, \citenamefont {Klimov}, \citenamefont {Miao}, \citenamefont
  {Falk}, \citenamefont {Christle}, \citenamefont {Sz{\'a}sz}, \citenamefont
  {Abrikosov}, \citenamefont {Awschalom},\ and\ \citenamefont
  {Gali}}]{ivady2016high}%
  \BibitemOpen
  \bibfield  {author} {\bibinfo {author} {\bibfnamefont {V.}~\bibnamefont
  {Iv{\'a}dy}}, \bibinfo {author} {\bibfnamefont {P.~V.}\ \bibnamefont
  {Klimov}}, \bibinfo {author} {\bibfnamefont {K.~C.}\ \bibnamefont {Miao}},
  \bibinfo {author} {\bibfnamefont {A.~L.}\ \bibnamefont {Falk}}, \bibinfo
  {author} {\bibfnamefont {D.~J.}\ \bibnamefont {Christle}}, \bibinfo {author}
  {\bibfnamefont {K.}~\bibnamefont {Sz{\'a}sz}}, \bibinfo {author}
  {\bibfnamefont {I.~A.}\ \bibnamefont {Abrikosov}}, \bibinfo {author}
  {\bibfnamefont {D.~D.}\ \bibnamefont {Awschalom}}, \ and\ \bibinfo {author}
  {\bibfnamefont {A.}~\bibnamefont {Gali}},\ }\href@noop {} {\bibfield
  {journal} {\bibinfo  {journal} {Physical review letters}\ }\textbf {\bibinfo
  {volume} {117}},\ \bibinfo {pages} {220503} (\bibinfo {year}
  {2016})}\BibitemShut {NoStop}%
\bibitem [{\citenamefont {Christle}\ \emph {et~al.}(2017)\citenamefont
  {Christle}, \citenamefont {Klimov}, \citenamefont {Charles}, \citenamefont
  {Sz{\'a}sz}, \citenamefont {Iv{\'a}dy}, \citenamefont {Jokubavicius},
  \citenamefont {Hassan}, \citenamefont {Syv{\"a}j{\"a}rvi}, \citenamefont
  {Koehl}, \citenamefont {Ohshima} \emph {et~al.}}]{christle2017isolated}%
  \BibitemOpen
  \bibfield  {author} {\bibinfo {author} {\bibfnamefont {D.~J.}\ \bibnamefont
  {Christle}}, \bibinfo {author} {\bibfnamefont {P.~V.}\ \bibnamefont
  {Klimov}}, \bibinfo {author} {\bibfnamefont {F.}~\bibnamefont {Charles}},
  \bibinfo {author} {\bibfnamefont {K.}~\bibnamefont {Sz{\'a}sz}}, \bibinfo
  {author} {\bibfnamefont {V.}~\bibnamefont {Iv{\'a}dy}}, \bibinfo {author}
  {\bibfnamefont {V.}~\bibnamefont {Jokubavicius}}, \bibinfo {author}
  {\bibfnamefont {J.~U.}\ \bibnamefont {Hassan}}, \bibinfo {author}
  {\bibfnamefont {M.}~\bibnamefont {Syv{\"a}j{\"a}rvi}}, \bibinfo {author}
  {\bibfnamefont {W.~F.}\ \bibnamefont {Koehl}}, \bibinfo {author}
  {\bibfnamefont {T.}~\bibnamefont {Ohshima}},  \emph {et~al.},\ }\href@noop {}
  {\bibfield  {journal} {\bibinfo  {journal} {Physical Review X}\ }\textbf
  {\bibinfo {volume} {7}},\ \bibinfo {pages} {021046} (\bibinfo {year}
  {2017})}\BibitemShut {NoStop}%
\bibitem [{\citenamefont {Miao}\ \emph {et~al.}(2019)\citenamefont {Miao},
  \citenamefont {Bourassa}, \citenamefont {Anderson}, \citenamefont {Whiteley},
  \citenamefont {Crook}, \citenamefont {Bayliss}, \citenamefont {Wolfowicz},
  \citenamefont {Thiering}, \citenamefont {Udvarhelyi}, \citenamefont
  {Iv{\'a}dy} \emph {et~al.}}]{miao2019electrically}%
  \BibitemOpen
  \bibfield  {author} {\bibinfo {author} {\bibfnamefont {K.~C.}\ \bibnamefont
  {Miao}}, \bibinfo {author} {\bibfnamefont {A.}~\bibnamefont {Bourassa}},
  \bibinfo {author} {\bibfnamefont {C.~P.}\ \bibnamefont {Anderson}}, \bibinfo
  {author} {\bibfnamefont {S.~J.}\ \bibnamefont {Whiteley}}, \bibinfo {author}
  {\bibfnamefont {A.~L.}\ \bibnamefont {Crook}}, \bibinfo {author}
  {\bibfnamefont {S.~L.}\ \bibnamefont {Bayliss}}, \bibinfo {author}
  {\bibfnamefont {G.}~\bibnamefont {Wolfowicz}}, \bibinfo {author}
  {\bibfnamefont {G.}~\bibnamefont {Thiering}}, \bibinfo {author}
  {\bibfnamefont {P.}~\bibnamefont {Udvarhelyi}}, \bibinfo {author}
  {\bibfnamefont {V.}~\bibnamefont {Iv{\'a}dy}},  \emph {et~al.},\ }\href@noop
  {} {\bibfield  {journal} {\bibinfo  {journal} {Science Advances}\ }\textbf
  {\bibinfo {volume} {5}},\ \bibinfo {pages} {eaay0527} (\bibinfo {year}
  {2019})}\BibitemShut {NoStop}%
\bibitem [{\citenamefont {Anderson}\ \emph {et~al.}(2019)\citenamefont
  {Anderson}, \citenamefont {Bourassa}, \citenamefont {Miao}, \citenamefont
  {Wolfowicz}, \citenamefont {Mintun}, \citenamefont {Crook}, \citenamefont
  {Abe}, \citenamefont {Hassan}, \citenamefont {Son}, \citenamefont {Ohshima}
  \emph {et~al.}}]{anderson2019electrical}%
  \BibitemOpen
  \bibfield  {author} {\bibinfo {author} {\bibfnamefont {C.~P.}\ \bibnamefont
  {Anderson}}, \bibinfo {author} {\bibfnamefont {A.}~\bibnamefont {Bourassa}},
  \bibinfo {author} {\bibfnamefont {K.~C.}\ \bibnamefont {Miao}}, \bibinfo
  {author} {\bibfnamefont {G.}~\bibnamefont {Wolfowicz}}, \bibinfo {author}
  {\bibfnamefont {P.~J.}\ \bibnamefont {Mintun}}, \bibinfo {author}
  {\bibfnamefont {A.~L.}\ \bibnamefont {Crook}}, \bibinfo {author}
  {\bibfnamefont {H.}~\bibnamefont {Abe}}, \bibinfo {author} {\bibfnamefont
  {J.~U.}\ \bibnamefont {Hassan}}, \bibinfo {author} {\bibfnamefont {N.~T.}\
  \bibnamefont {Son}}, \bibinfo {author} {\bibfnamefont {T.}~\bibnamefont
  {Ohshima}},  \emph {et~al.},\ }\href@noop {} {\bibfield  {journal} {\bibinfo
  {journal} {Science}\ }\textbf {\bibinfo {volume} {366}},\ \bibinfo {pages}
  {1225} (\bibinfo {year} {2019})}\BibitemShut {NoStop}%
\bibitem [{\citenamefont {Von~Bardeleben}\ \emph {et~al.}(2016)\citenamefont
  {Von~Bardeleben}, \citenamefont {Cantin}, \citenamefont {Cs{\'o}r{\'e}},
  \citenamefont {Gali}, \citenamefont {Rauls},\ and\ \citenamefont
  {Gerstmann}}]{von2016nv}%
  \BibitemOpen
  \bibfield  {author} {\bibinfo {author} {\bibfnamefont {H.}~\bibnamefont
  {Von~Bardeleben}}, \bibinfo {author} {\bibfnamefont {J.}~\bibnamefont
  {Cantin}}, \bibinfo {author} {\bibfnamefont {A.}~\bibnamefont
  {Cs{\'o}r{\'e}}}, \bibinfo {author} {\bibfnamefont {A.}~\bibnamefont {Gali}},
  \bibinfo {author} {\bibfnamefont {E.}~\bibnamefont {Rauls}}, \ and\ \bibinfo
  {author} {\bibfnamefont {U.}~\bibnamefont {Gerstmann}},\ }\href@noop {}
  {\bibfield  {journal} {\bibinfo  {journal} {Physical Review B}\ }\textbf
  {\bibinfo {volume} {94}},\ \bibinfo {pages} {121202} (\bibinfo {year}
  {2016})}\BibitemShut {NoStop}%
\bibitem [{\citenamefont {Zargaleh}\ \emph {et~al.}(2018)\citenamefont
  {Zargaleh}, \citenamefont {Hameau}, \citenamefont {Eble}, \citenamefont
  {Margaillan}, \citenamefont {von Bardeleben}, \citenamefont {Cantin},\ and\
  \citenamefont {Gao}}]{zargaleh2018nitrogen}%
  \BibitemOpen
  \bibfield  {author} {\bibinfo {author} {\bibfnamefont {S.~A.}\ \bibnamefont
  {Zargaleh}}, \bibinfo {author} {\bibfnamefont {S.}~\bibnamefont {Hameau}},
  \bibinfo {author} {\bibfnamefont {B.}~\bibnamefont {Eble}}, \bibinfo {author}
  {\bibfnamefont {F.}~\bibnamefont {Margaillan}}, \bibinfo {author}
  {\bibfnamefont {H.~J.}\ \bibnamefont {von Bardeleben}}, \bibinfo {author}
  {\bibfnamefont {J.-L.}\ \bibnamefont {Cantin}}, \ and\ \bibinfo {author}
  {\bibfnamefont {W.}~\bibnamefont {Gao}},\ }\href@noop {} {\bibfield
  {journal} {\bibinfo  {journal} {Physical Review B}\ }\textbf {\bibinfo
  {volume} {98}},\ \bibinfo {pages} {165203} (\bibinfo {year}
  {2018})}\BibitemShut {NoStop}%
\bibitem [{\citenamefont {Zargaleh}\ \emph {et~al.}(2016)\citenamefont
  {Zargaleh}, \citenamefont {Eble}, \citenamefont {Hameau}, \citenamefont
  {Cantin}, \citenamefont {Legrand}, \citenamefont {Bernard}, \citenamefont
  {Margaillan}, \citenamefont {Lauret}, \citenamefont {Roch}, \citenamefont
  {Von~Bardeleben} \emph {et~al.}}]{zargaleh2016evidence}%
  \BibitemOpen
  \bibfield  {author} {\bibinfo {author} {\bibfnamefont {S.}~\bibnamefont
  {Zargaleh}}, \bibinfo {author} {\bibfnamefont {B.}~\bibnamefont {Eble}},
  \bibinfo {author} {\bibfnamefont {S.}~\bibnamefont {Hameau}}, \bibinfo
  {author} {\bibfnamefont {J.-L.}\ \bibnamefont {Cantin}}, \bibinfo {author}
  {\bibfnamefont {L.}~\bibnamefont {Legrand}}, \bibinfo {author} {\bibfnamefont
  {M.}~\bibnamefont {Bernard}}, \bibinfo {author} {\bibfnamefont
  {F.}~\bibnamefont {Margaillan}}, \bibinfo {author} {\bibfnamefont {J.-S.}\
  \bibnamefont {Lauret}}, \bibinfo {author} {\bibfnamefont {J.-F.}\
  \bibnamefont {Roch}}, \bibinfo {author} {\bibfnamefont {H.}~\bibnamefont
  {Von~Bardeleben}},  \emph {et~al.},\ }\href@noop {} {\bibfield  {journal}
  {\bibinfo  {journal} {Physical Review B}\ }\textbf {\bibinfo {volume} {94}},\
  \bibinfo {pages} {060102} (\bibinfo {year} {2016})}\BibitemShut {NoStop}%
\bibitem [{\citenamefont {Cs{\'o}r{\'e}}\ \emph {et~al.}(2017)\citenamefont
  {Cs{\'o}r{\'e}}, \citenamefont {Von~Bardeleben}, \citenamefont {Cantin},\
  and\ \citenamefont {Gali}}]{csore2017characterization}%
  \BibitemOpen
  \bibfield  {author} {\bibinfo {author} {\bibfnamefont {A.}~\bibnamefont
  {Cs{\'o}r{\'e}}}, \bibinfo {author} {\bibfnamefont {H.}~\bibnamefont
  {Von~Bardeleben}}, \bibinfo {author} {\bibfnamefont {J.}~\bibnamefont
  {Cantin}}, \ and\ \bibinfo {author} {\bibfnamefont {A.}~\bibnamefont
  {Gali}},\ }\href@noop {} {\bibfield  {journal} {\bibinfo  {journal} {Physical
  Review B}\ }\textbf {\bibinfo {volume} {96}},\ \bibinfo {pages} {085204}
  (\bibinfo {year} {2017})}\BibitemShut {NoStop}%
\bibitem [{\citenamefont {Wang}\ \emph {et~al.}(2020)\citenamefont {Wang},
  \citenamefont {Yan}, \citenamefont {Li}, \citenamefont {Liu}, \citenamefont
  {Liu}, \citenamefont {Guo}, \citenamefont {Guo}, \citenamefont {Zhou},
  \citenamefont {Cui}, \citenamefont {Wang} \emph {et~al.}}]{wang2020coherent}%
  \BibitemOpen
  \bibfield  {author} {\bibinfo {author} {\bibfnamefont {J.-F.}\ \bibnamefont
  {Wang}}, \bibinfo {author} {\bibfnamefont {F.-F.}\ \bibnamefont {Yan}},
  \bibinfo {author} {\bibfnamefont {Q.}~\bibnamefont {Li}}, \bibinfo {author}
  {\bibfnamefont {Z.-H.}\ \bibnamefont {Liu}}, \bibinfo {author} {\bibfnamefont
  {H.}~\bibnamefont {Liu}}, \bibinfo {author} {\bibfnamefont {G.-P.}\
  \bibnamefont {Guo}}, \bibinfo {author} {\bibfnamefont {L.-P.}\ \bibnamefont
  {Guo}}, \bibinfo {author} {\bibfnamefont {X.}~\bibnamefont {Zhou}}, \bibinfo
  {author} {\bibfnamefont {J.-M.}\ \bibnamefont {Cui}}, \bibinfo {author}
  {\bibfnamefont {J.}~\bibnamefont {Wang}},  \emph {et~al.},\ }\href@noop {}
  {\bibfield  {journal} {\bibinfo  {journal} {Physical Review Letters}\
  }\textbf {\bibinfo {volume} {124}},\ \bibinfo {pages} {223601} (\bibinfo
  {year} {2020})}\BibitemShut {NoStop}%
\bibitem [{\citenamefont {Koehl}\ \emph {et~al.}(2017)\citenamefont {Koehl},
  \citenamefont {Diler}, \citenamefont {Whiteley}, \citenamefont {Bourassa},
  \citenamefont {Son}, \citenamefont {Janz{\'e}n},\ and\ \citenamefont
  {Awschalom}}]{koehl2017resonant}%
  \BibitemOpen
  \bibfield  {author} {\bibinfo {author} {\bibfnamefont {W.~F.}\ \bibnamefont
  {Koehl}}, \bibinfo {author} {\bibfnamefont {B.}~\bibnamefont {Diler}},
  \bibinfo {author} {\bibfnamefont {S.~J.}\ \bibnamefont {Whiteley}}, \bibinfo
  {author} {\bibfnamefont {A.}~\bibnamefont {Bourassa}}, \bibinfo {author}
  {\bibfnamefont {N.~T.}\ \bibnamefont {Son}}, \bibinfo {author} {\bibfnamefont
  {E.}~\bibnamefont {Janz{\'e}n}}, \ and\ \bibinfo {author} {\bibfnamefont
  {D.~D.}\ \bibnamefont {Awschalom}},\ }\href@noop {} {\bibfield  {journal}
  {\bibinfo  {journal} {Physical Review B}\ }\textbf {\bibinfo {volume} {95}},\
  \bibinfo {pages} {035207} (\bibinfo {year} {2017})}\BibitemShut {NoStop}%
\bibitem [{\citenamefont {Diler}\ \emph {et~al.}(2020)\citenamefont {Diler},
  \citenamefont {Whiteley}, \citenamefont {Anderson}, \citenamefont
  {Wolfowicz}, \citenamefont {Wesson}, \citenamefont {Bielejec}, \citenamefont
  {Heremans},\ and\ \citenamefont {Awschalom}}]{diler2020coherent}%
  \BibitemOpen
  \bibfield  {author} {\bibinfo {author} {\bibfnamefont {B.}~\bibnamefont
  {Diler}}, \bibinfo {author} {\bibfnamefont {S.~J.}\ \bibnamefont {Whiteley}},
  \bibinfo {author} {\bibfnamefont {C.~P.}\ \bibnamefont {Anderson}}, \bibinfo
  {author} {\bibfnamefont {G.}~\bibnamefont {Wolfowicz}}, \bibinfo {author}
  {\bibfnamefont {M.~E.}\ \bibnamefont {Wesson}}, \bibinfo {author}
  {\bibfnamefont {E.~S.}\ \bibnamefont {Bielejec}}, \bibinfo {author}
  {\bibfnamefont {F.~J.}\ \bibnamefont {Heremans}}, \ and\ \bibinfo {author}
  {\bibfnamefont {D.~D.}\ \bibnamefont {Awschalom}},\ }\href@noop {} {\bibfield
   {journal} {\bibinfo  {journal} {npj Quantum Information}\ }\textbf {\bibinfo
  {volume} {6}},\ \bibinfo {pages} {1} (\bibinfo {year} {2020})}\BibitemShut
  {NoStop}%
\bibitem [{\citenamefont {Spindlberger}\ \emph {et~al.}(2019)\citenamefont
  {Spindlberger}, \citenamefont {Cs{\'o}r{\'e}}, \citenamefont {Thiering},
  \citenamefont {Putz}, \citenamefont {Karhu}, \citenamefont {Hassan},
  \citenamefont {Son}, \citenamefont {Fromherz}, \citenamefont {Gali},\ and\
  \citenamefont {Trupke}}]{spindlberger2019optical}%
  \BibitemOpen
  \bibfield  {author} {\bibinfo {author} {\bibfnamefont {L.}~\bibnamefont
  {Spindlberger}}, \bibinfo {author} {\bibfnamefont {A.}~\bibnamefont
  {Cs{\'o}r{\'e}}}, \bibinfo {author} {\bibfnamefont {G.}~\bibnamefont
  {Thiering}}, \bibinfo {author} {\bibfnamefont {S.}~\bibnamefont {Putz}},
  \bibinfo {author} {\bibfnamefont {R.}~\bibnamefont {Karhu}}, \bibinfo
  {author} {\bibfnamefont {J.~U.}\ \bibnamefont {Hassan}}, \bibinfo {author}
  {\bibfnamefont {N.}~\bibnamefont {Son}}, \bibinfo {author} {\bibfnamefont
  {T.}~\bibnamefont {Fromherz}}, \bibinfo {author} {\bibfnamefont
  {A.}~\bibnamefont {Gali}}, \ and\ \bibinfo {author} {\bibfnamefont
  {M.}~\bibnamefont {Trupke}},\ }\href@noop {} {\bibfield  {journal} {\bibinfo
  {journal} {Physical Review Applied}\ }\textbf {\bibinfo {volume} {12}},\
  \bibinfo {pages} {014015} (\bibinfo {year} {2019})}\BibitemShut {NoStop}%
\bibitem [{\citenamefont {Wolfowicz}\ \emph {et~al.}(2019)\citenamefont
  {Wolfowicz}, \citenamefont {Anderson}, \citenamefont {Diler}, \citenamefont
  {Poluektov}, \citenamefont {Heremans},\ and\ \citenamefont
  {Awschalom}}]{wolfowicz2019vanadium}%
  \BibitemOpen
  \bibfield  {author} {\bibinfo {author} {\bibfnamefont {G.}~\bibnamefont
  {Wolfowicz}}, \bibinfo {author} {\bibfnamefont {C.~P.}\ \bibnamefont
  {Anderson}}, \bibinfo {author} {\bibfnamefont {B.}~\bibnamefont {Diler}},
  \bibinfo {author} {\bibfnamefont {O.~G.}\ \bibnamefont {Poluektov}}, \bibinfo
  {author} {\bibfnamefont {F.~J.}\ \bibnamefont {Heremans}}, \ and\ \bibinfo
  {author} {\bibfnamefont {D.~D.}\ \bibnamefont {Awschalom}},\ }\href@noop {}
  {\bibfield  {journal} {\bibinfo  {journal} {arXiv preprint arXiv:1908.09817}\
  } (\bibinfo {year} {2019})}\BibitemShut {NoStop}%
\bibitem [{\citenamefont {Kraus}\ \emph {et~al.}(2017)\citenamefont {Kraus},
  \citenamefont {Simin}, \citenamefont {Kasper}, \citenamefont {Suda},
  \citenamefont {Kawabata}, \citenamefont {Kada}, \citenamefont {Honda},
  \citenamefont {Hijikata}, \citenamefont {Ohshima}, \citenamefont {Dyakonov}
  \emph {et~al.}}]{kraus2017three}%
  \BibitemOpen
  \bibfield  {author} {\bibinfo {author} {\bibfnamefont {H.}~\bibnamefont
  {Kraus}}, \bibinfo {author} {\bibfnamefont {D.}~\bibnamefont {Simin}},
  \bibinfo {author} {\bibfnamefont {C.}~\bibnamefont {Kasper}}, \bibinfo
  {author} {\bibfnamefont {Y.}~\bibnamefont {Suda}}, \bibinfo {author}
  {\bibfnamefont {S.}~\bibnamefont {Kawabata}}, \bibinfo {author}
  {\bibfnamefont {W.}~\bibnamefont {Kada}}, \bibinfo {author} {\bibfnamefont
  {T.}~\bibnamefont {Honda}}, \bibinfo {author} {\bibfnamefont
  {Y.}~\bibnamefont {Hijikata}}, \bibinfo {author} {\bibfnamefont
  {T.}~\bibnamefont {Ohshima}}, \bibinfo {author} {\bibfnamefont
  {V.}~\bibnamefont {Dyakonov}},  \emph {et~al.},\ }\href@noop {} {\bibfield
  {journal} {\bibinfo  {journal} {Nano letters}\ }\textbf {\bibinfo {volume}
  {17}},\ \bibinfo {pages} {2865} (\bibinfo {year} {2017})}\BibitemShut
  {NoStop}%
\bibitem [{\citenamefont {Chen}\ \emph {et~al.}(2019)\citenamefont {Chen},
  \citenamefont {Salter}, \citenamefont {Niethammer}, \citenamefont {Widmann},
  \citenamefont {Kaiser}, \citenamefont {Nagy}, \citenamefont {Morioka},
  \citenamefont {Babin}, \citenamefont {Erlekampf}, \citenamefont {Berwian}
  \emph {et~al.}}]{chen2019laser}%
  \BibitemOpen
  \bibfield  {author} {\bibinfo {author} {\bibfnamefont {Y.-C.}\ \bibnamefont
  {Chen}}, \bibinfo {author} {\bibfnamefont {P.~S.}\ \bibnamefont {Salter}},
  \bibinfo {author} {\bibfnamefont {M.}~\bibnamefont {Niethammer}}, \bibinfo
  {author} {\bibfnamefont {M.}~\bibnamefont {Widmann}}, \bibinfo {author}
  {\bibfnamefont {F.}~\bibnamefont {Kaiser}}, \bibinfo {author} {\bibfnamefont
  {R.}~\bibnamefont {Nagy}}, \bibinfo {author} {\bibfnamefont {N.}~\bibnamefont
  {Morioka}}, \bibinfo {author} {\bibfnamefont {C.}~\bibnamefont {Babin}},
  \bibinfo {author} {\bibfnamefont {J.}~\bibnamefont {Erlekampf}}, \bibinfo
  {author} {\bibfnamefont {P.}~\bibnamefont {Berwian}},  \emph {et~al.},\
  }\href@noop {} {\bibfield  {journal} {\bibinfo  {journal} {Nano letters}\
  }\textbf {\bibinfo {volume} {19}},\ \bibinfo {pages} {2377} (\bibinfo {year}
  {2019})}\BibitemShut {NoStop}%
\bibitem [{\citenamefont {Kraus}\ \emph {et~al.}(2014)\citenamefont {Kraus},
  \citenamefont {Soltamov}, \citenamefont {Riedel}, \citenamefont {V{\"a}th},
  \citenamefont {Fuchs}, \citenamefont {Sperlich}, \citenamefont {Baranov},
  \citenamefont {Dyakonov},\ and\ \citenamefont {Astakhov}}]{kraus2014room}%
  \BibitemOpen
  \bibfield  {author} {\bibinfo {author} {\bibfnamefont {H.}~\bibnamefont
  {Kraus}}, \bibinfo {author} {\bibfnamefont {V.}~\bibnamefont {Soltamov}},
  \bibinfo {author} {\bibfnamefont {D.}~\bibnamefont {Riedel}}, \bibinfo
  {author} {\bibfnamefont {S.}~\bibnamefont {V{\"a}th}}, \bibinfo {author}
  {\bibfnamefont {F.}~\bibnamefont {Fuchs}}, \bibinfo {author} {\bibfnamefont
  {A.}~\bibnamefont {Sperlich}}, \bibinfo {author} {\bibfnamefont
  {P.}~\bibnamefont {Baranov}}, \bibinfo {author} {\bibfnamefont
  {V.}~\bibnamefont {Dyakonov}}, \ and\ \bibinfo {author} {\bibfnamefont
  {G.}~\bibnamefont {Astakhov}},\ }\href@noop {} {\bibfield  {journal}
  {\bibinfo  {journal} {Nature Physics}\ }\textbf {\bibinfo {volume} {10}},\
  \bibinfo {pages} {157} (\bibinfo {year} {2014})}\BibitemShut {NoStop}%
\bibitem [{\citenamefont {Simin}\ \emph {et~al.}(2015)\citenamefont {Simin},
  \citenamefont {Fuchs}, \citenamefont {Kraus}, \citenamefont {Sperlich},
  \citenamefont {Baranov}, \citenamefont {Astakhov},\ and\ \citenamefont
  {Dyakonov}}]{simin2015high}%
  \BibitemOpen
  \bibfield  {author} {\bibinfo {author} {\bibfnamefont {D.}~\bibnamefont
  {Simin}}, \bibinfo {author} {\bibfnamefont {F.}~\bibnamefont {Fuchs}},
  \bibinfo {author} {\bibfnamefont {H.}~\bibnamefont {Kraus}}, \bibinfo
  {author} {\bibfnamefont {A.}~\bibnamefont {Sperlich}}, \bibinfo {author}
  {\bibfnamefont {P.}~\bibnamefont {Baranov}}, \bibinfo {author} {\bibfnamefont
  {G.}~\bibnamefont {Astakhov}}, \ and\ \bibinfo {author} {\bibfnamefont
  {V.}~\bibnamefont {Dyakonov}},\ }\href@noop {} {\bibfield  {journal}
  {\bibinfo  {journal} {Physical Review Applied}\ }\textbf {\bibinfo {volume}
  {4}},\ \bibinfo {pages} {014009} (\bibinfo {year} {2015})}\BibitemShut
  {NoStop}%
\bibitem [{\citenamefont {Simin}\ \emph {et~al.}(2016)\citenamefont {Simin},
  \citenamefont {Soltamov}, \citenamefont {Poshakinskiy}, \citenamefont
  {Anisimov}, \citenamefont {Babunts}, \citenamefont {Tolmachev}, \citenamefont
  {Mokhov}, \citenamefont {Trupke}, \citenamefont {Tarasenko}, \citenamefont
  {Sperlich} \emph {et~al.}}]{simin2016all}%
  \BibitemOpen
  \bibfield  {author} {\bibinfo {author} {\bibfnamefont {D.}~\bibnamefont
  {Simin}}, \bibinfo {author} {\bibfnamefont {V.}~\bibnamefont {Soltamov}},
  \bibinfo {author} {\bibfnamefont {A.}~\bibnamefont {Poshakinskiy}}, \bibinfo
  {author} {\bibfnamefont {A.}~\bibnamefont {Anisimov}}, \bibinfo {author}
  {\bibfnamefont {R.}~\bibnamefont {Babunts}}, \bibinfo {author} {\bibfnamefont
  {D.}~\bibnamefont {Tolmachev}}, \bibinfo {author} {\bibfnamefont
  {E.}~\bibnamefont {Mokhov}}, \bibinfo {author} {\bibfnamefont
  {M.}~\bibnamefont {Trupke}}, \bibinfo {author} {\bibfnamefont
  {S.}~\bibnamefont {Tarasenko}}, \bibinfo {author} {\bibfnamefont
  {A.}~\bibnamefont {Sperlich}},  \emph {et~al.},\ }\href@noop {} {\bibfield
  {journal} {\bibinfo  {journal} {Physical Review X}\ }\textbf {\bibinfo
  {volume} {6}},\ \bibinfo {pages} {031014} (\bibinfo {year}
  {2016})}\BibitemShut {NoStop}%
\bibitem [{\citenamefont {Soltamov}\ \emph {et~al.}(2019)\citenamefont
  {Soltamov}, \citenamefont {Kasper}, \citenamefont {Poshakinskiy},
  \citenamefont {Anisimov}, \citenamefont {Mokhov}, \citenamefont {Sperlich},
  \citenamefont {Tarasenko}, \citenamefont {Baranov}, \citenamefont
  {Astakhov},\ and\ \citenamefont {Dyakonov}}]{soltamov2019excitation}%
  \BibitemOpen
  \bibfield  {author} {\bibinfo {author} {\bibfnamefont {V.}~\bibnamefont
  {Soltamov}}, \bibinfo {author} {\bibfnamefont {C.}~\bibnamefont {Kasper}},
  \bibinfo {author} {\bibfnamefont {A.}~\bibnamefont {Poshakinskiy}}, \bibinfo
  {author} {\bibfnamefont {A.}~\bibnamefont {Anisimov}}, \bibinfo {author}
  {\bibfnamefont {E.}~\bibnamefont {Mokhov}}, \bibinfo {author} {\bibfnamefont
  {A.}~\bibnamefont {Sperlich}}, \bibinfo {author} {\bibfnamefont
  {S.}~\bibnamefont {Tarasenko}}, \bibinfo {author} {\bibfnamefont
  {P.}~\bibnamefont {Baranov}}, \bibinfo {author} {\bibfnamefont
  {G.}~\bibnamefont {Astakhov}}, \ and\ \bibinfo {author} {\bibfnamefont
  {V.}~\bibnamefont {Dyakonov}},\ }\href@noop {} {\bibfield  {journal}
  {\bibinfo  {journal} {Nature communications}\ }\textbf {\bibinfo {volume}
  {10}},\ \bibinfo {pages} {1} (\bibinfo {year} {2019})}\BibitemShut {NoStop}%
\bibitem [{\citenamefont {Widmann}\ \emph {et~al.}(2015)\citenamefont
  {Widmann}, \citenamefont {Lee}, \citenamefont {Rendler}, \citenamefont {Son},
  \citenamefont {Fedder}, \citenamefont {Paik}, \citenamefont {Yang},
  \citenamefont {Zhao}, \citenamefont {Yang}, \citenamefont {Booker} \emph
  {et~al.}}]{widmann2015coherent}%
  \BibitemOpen
  \bibfield  {author} {\bibinfo {author} {\bibfnamefont {M.}~\bibnamefont
  {Widmann}}, \bibinfo {author} {\bibfnamefont {S.-Y.}\ \bibnamefont {Lee}},
  \bibinfo {author} {\bibfnamefont {T.}~\bibnamefont {Rendler}}, \bibinfo
  {author} {\bibfnamefont {N.~T.}\ \bibnamefont {Son}}, \bibinfo {author}
  {\bibfnamefont {H.}~\bibnamefont {Fedder}}, \bibinfo {author} {\bibfnamefont
  {S.}~\bibnamefont {Paik}}, \bibinfo {author} {\bibfnamefont {L.-P.}\
  \bibnamefont {Yang}}, \bibinfo {author} {\bibfnamefont {N.}~\bibnamefont
  {Zhao}}, \bibinfo {author} {\bibfnamefont {S.}~\bibnamefont {Yang}}, \bibinfo
  {author} {\bibfnamefont {I.}~\bibnamefont {Booker}},  \emph {et~al.},\
  }\href@noop {} {\bibfield  {journal} {\bibinfo  {journal} {Nature materials}\
  }\textbf {\bibinfo {volume} {14}},\ \bibinfo {pages} {164} (\bibinfo {year}
  {2015})}\BibitemShut {NoStop}%
\bibitem [{\citenamefont {Fuchs}\ \emph {et~al.}(2015)\citenamefont {Fuchs},
  \citenamefont {Stender}, \citenamefont {Trupke}, \citenamefont {Simin},
  \citenamefont {Pflaum}, \citenamefont {Dyakonov},\ and\ \citenamefont
  {Astakhov}}]{fuchs2015engineering}%
  \BibitemOpen
  \bibfield  {author} {\bibinfo {author} {\bibfnamefont {F.}~\bibnamefont
  {Fuchs}}, \bibinfo {author} {\bibfnamefont {B.}~\bibnamefont {Stender}},
  \bibinfo {author} {\bibfnamefont {M.}~\bibnamefont {Trupke}}, \bibinfo
  {author} {\bibfnamefont {D.}~\bibnamefont {Simin}}, \bibinfo {author}
  {\bibfnamefont {J.}~\bibnamefont {Pflaum}}, \bibinfo {author} {\bibfnamefont
  {V.}~\bibnamefont {Dyakonov}}, \ and\ \bibinfo {author} {\bibfnamefont
  {G.}~\bibnamefont {Astakhov}},\ }\href@noop {} {\bibfield  {journal}
  {\bibinfo  {journal} {Nature communications}\ }\textbf {\bibinfo {volume}
  {6}},\ \bibinfo {pages} {1} (\bibinfo {year} {2015})}\BibitemShut {NoStop}%
\bibitem [{\citenamefont {Kasper}\ \emph {et~al.}(2020)\citenamefont {Kasper},
  \citenamefont {Klenkert}, \citenamefont {Shang}, \citenamefont {Simin},
  \citenamefont {Gottscholl}, \citenamefont {Sperlich}, \citenamefont {Kraus},
  \citenamefont {Schneider}, \citenamefont {Zhou}, \citenamefont {Trupke} \emph
  {et~al.}}]{kasper2020influence}%
  \BibitemOpen
  \bibfield  {author} {\bibinfo {author} {\bibfnamefont {C.}~\bibnamefont
  {Kasper}}, \bibinfo {author} {\bibfnamefont {D.}~\bibnamefont {Klenkert}},
  \bibinfo {author} {\bibfnamefont {Z.}~\bibnamefont {Shang}}, \bibinfo
  {author} {\bibfnamefont {D.}~\bibnamefont {Simin}}, \bibinfo {author}
  {\bibfnamefont {A.}~\bibnamefont {Gottscholl}}, \bibinfo {author}
  {\bibfnamefont {A.}~\bibnamefont {Sperlich}}, \bibinfo {author}
  {\bibfnamefont {H.}~\bibnamefont {Kraus}}, \bibinfo {author} {\bibfnamefont
  {C.}~\bibnamefont {Schneider}}, \bibinfo {author} {\bibfnamefont
  {S.}~\bibnamefont {Zhou}}, \bibinfo {author} {\bibfnamefont {M.}~\bibnamefont
  {Trupke}},  \emph {et~al.},\ }\href@noop {} {\bibfield  {journal} {\bibinfo
  {journal} {Physical Review Applied}\ }\textbf {\bibinfo {volume} {13}},\
  \bibinfo {pages} {044054} (\bibinfo {year} {2020})}\BibitemShut {NoStop}%
\bibitem [{\citenamefont {Wang}\ \emph {et~al.}(2017)\citenamefont {Wang},
  \citenamefont {Zhou}, \citenamefont {Zhang}, \citenamefont {Liu},
  \citenamefont {Li}, \citenamefont {Li}, \citenamefont {Liu}, \citenamefont
  {Wang},\ and\ \citenamefont {Gao}}]{wang2017efficient}%
  \BibitemOpen
  \bibfield  {author} {\bibinfo {author} {\bibfnamefont {J.}~\bibnamefont
  {Wang}}, \bibinfo {author} {\bibfnamefont {Y.}~\bibnamefont {Zhou}}, \bibinfo
  {author} {\bibfnamefont {X.}~\bibnamefont {Zhang}}, \bibinfo {author}
  {\bibfnamefont {F.}~\bibnamefont {Liu}}, \bibinfo {author} {\bibfnamefont
  {Y.}~\bibnamefont {Li}}, \bibinfo {author} {\bibfnamefont {K.}~\bibnamefont
  {Li}}, \bibinfo {author} {\bibfnamefont {Z.}~\bibnamefont {Liu}}, \bibinfo
  {author} {\bibfnamefont {G.}~\bibnamefont {Wang}}, \ and\ \bibinfo {author}
  {\bibfnamefont {W.}~\bibnamefont {Gao}},\ }\href@noop {} {\bibfield
  {journal} {\bibinfo  {journal} {Physical Review Applied}\ }\textbf {\bibinfo
  {volume} {7}},\ \bibinfo {pages} {064021} (\bibinfo {year}
  {2017})}\BibitemShut {NoStop}%
\bibitem [{\citenamefont {Soykal}\ \emph {et~al.}(2016)\citenamefont {Soykal},
  \citenamefont {Dev},\ and\ \citenamefont {Economou}}]{soykal2016silicon}%
  \BibitemOpen
  \bibfield  {author} {\bibinfo {author} {\bibfnamefont {{\"O}.}~\bibnamefont
  {Soykal}}, \bibinfo {author} {\bibfnamefont {P.}~\bibnamefont {Dev}}, \ and\
  \bibinfo {author} {\bibfnamefont {S.~E.}\ \bibnamefont {Economou}},\
  }\href@noop {} {\bibfield  {journal} {\bibinfo  {journal} {Physical Review
  B}\ }\textbf {\bibinfo {volume} {93}},\ \bibinfo {pages} {081207} (\bibinfo
  {year} {2016})}\BibitemShut {NoStop}%
\bibitem [{\citenamefont {Economou}\ and\ \citenamefont
  {Dev}(2016)}]{economou2016spin}%
  \BibitemOpen
  \bibfield  {author} {\bibinfo {author} {\bibfnamefont {S.~E.}\ \bibnamefont
  {Economou}}\ and\ \bibinfo {author} {\bibfnamefont {P.}~\bibnamefont {Dev}},\
  }\href@noop {} {\bibfield  {journal} {\bibinfo  {journal} {Nanotechnology}\
  }\textbf {\bibinfo {volume} {27}},\ \bibinfo {pages} {504001} (\bibinfo
  {year} {2016})}\BibitemShut {NoStop}%
\bibitem [{\citenamefont {Udvarhelyi}\ \emph {et~al.}(2019)\citenamefont
  {Udvarhelyi}, \citenamefont {Nagy}, \citenamefont {Kaiser}, \citenamefont
  {Lee}, \citenamefont {Wrachtrup},\ and\ \citenamefont
  {Gali}}]{udvarhelyi2019spectrally}%
  \BibitemOpen
  \bibfield  {author} {\bibinfo {author} {\bibfnamefont {P.}~\bibnamefont
  {Udvarhelyi}}, \bibinfo {author} {\bibfnamefont {R.}~\bibnamefont {Nagy}},
  \bibinfo {author} {\bibfnamefont {F.}~\bibnamefont {Kaiser}}, \bibinfo
  {author} {\bibfnamefont {S.-Y.}\ \bibnamefont {Lee}}, \bibinfo {author}
  {\bibfnamefont {J.}~\bibnamefont {Wrachtrup}}, \ and\ \bibinfo {author}
  {\bibfnamefont {A.}~\bibnamefont {Gali}},\ }\href@noop {} {\bibfield
  {journal} {\bibinfo  {journal} {Physical Review Applied}\ }\textbf {\bibinfo
  {volume} {11}},\ \bibinfo {pages} {044022} (\bibinfo {year}
  {2019})}\BibitemShut {NoStop}%
\bibitem [{\citenamefont {Niethammer}\ \emph {et~al.}(2019)\citenamefont
  {Niethammer}, \citenamefont {Widmann}, \citenamefont {Rendler}, \citenamefont
  {Morioka}, \citenamefont {Chen}, \citenamefont {St{\"o}hr}, \citenamefont
  {Hassan}, \citenamefont {Onoda}, \citenamefont {Ohshima}, \citenamefont {Lee}
  \emph {et~al.}}]{niethammer2019coherent}%
  \BibitemOpen
  \bibfield  {author} {\bibinfo {author} {\bibfnamefont {M.}~\bibnamefont
  {Niethammer}}, \bibinfo {author} {\bibfnamefont {M.}~\bibnamefont {Widmann}},
  \bibinfo {author} {\bibfnamefont {T.}~\bibnamefont {Rendler}}, \bibinfo
  {author} {\bibfnamefont {N.}~\bibnamefont {Morioka}}, \bibinfo {author}
  {\bibfnamefont {Y.-C.}\ \bibnamefont {Chen}}, \bibinfo {author}
  {\bibfnamefont {R.}~\bibnamefont {St{\"o}hr}}, \bibinfo {author}
  {\bibfnamefont {J.~U.}\ \bibnamefont {Hassan}}, \bibinfo {author}
  {\bibfnamefont {S.}~\bibnamefont {Onoda}}, \bibinfo {author} {\bibfnamefont
  {T.}~\bibnamefont {Ohshima}}, \bibinfo {author} {\bibfnamefont {S.-Y.}\
  \bibnamefont {Lee}},  \emph {et~al.},\ }\href@noop {} {\bibfield  {journal}
  {\bibinfo  {journal} {Nature Communications}\ }\textbf {\bibinfo {volume}
  {10}},\ \bibinfo {pages} {1} (\bibinfo {year} {2019})}\BibitemShut {NoStop}%
\bibitem [{\citenamefont {Widmann}\ \emph {et~al.}(2019)\citenamefont
  {Widmann}, \citenamefont {Niethammer}, \citenamefont {Fedyanin},
  \citenamefont {Khramtsov}, \citenamefont {Rendler}, \citenamefont {Booker},
  \citenamefont {Ul~Hassan}, \citenamefont {Morioka}, \citenamefont {Chen},
  \citenamefont {Ivanov} \emph {et~al.}}]{widmann2019electrical}%
  \BibitemOpen
  \bibfield  {author} {\bibinfo {author} {\bibfnamefont {M.}~\bibnamefont
  {Widmann}}, \bibinfo {author} {\bibfnamefont {M.}~\bibnamefont {Niethammer}},
  \bibinfo {author} {\bibfnamefont {D.~Y.}\ \bibnamefont {Fedyanin}}, \bibinfo
  {author} {\bibfnamefont {I.~A.}\ \bibnamefont {Khramtsov}}, \bibinfo {author}
  {\bibfnamefont {T.}~\bibnamefont {Rendler}}, \bibinfo {author} {\bibfnamefont
  {I.~D.}\ \bibnamefont {Booker}}, \bibinfo {author} {\bibfnamefont
  {J.}~\bibnamefont {Ul~Hassan}}, \bibinfo {author} {\bibfnamefont
  {N.}~\bibnamefont {Morioka}}, \bibinfo {author} {\bibfnamefont {Y.-C.}\
  \bibnamefont {Chen}}, \bibinfo {author} {\bibfnamefont {I.~G.}\ \bibnamefont
  {Ivanov}},  \emph {et~al.},\ }\href@noop {} {\bibfield  {journal} {\bibinfo
  {journal} {Nano letters}\ }\textbf {\bibinfo {volume} {19}},\ \bibinfo
  {pages} {7173} (\bibinfo {year} {2019})}\BibitemShut {NoStop}%
\bibitem [{\citenamefont {Bourassa}\ \emph {et~al.}(2020)\citenamefont
  {Bourassa}, \citenamefont {Anderson}, \citenamefont {Miao}, \citenamefont
  {Onizhuk}, \citenamefont {Ma}, \citenamefont {Crook}, \citenamefont {Abe},
  \citenamefont {Ul-Hassan}, \citenamefont {Ohshima}, \citenamefont {Son} \emph
  {et~al.}}]{bourassa2020entanglement}%
  \BibitemOpen
  \bibfield  {author} {\bibinfo {author} {\bibfnamefont {A.}~\bibnamefont
  {Bourassa}}, \bibinfo {author} {\bibfnamefont {C.~P.}\ \bibnamefont
  {Anderson}}, \bibinfo {author} {\bibfnamefont {K.~C.}\ \bibnamefont {Miao}},
  \bibinfo {author} {\bibfnamefont {M.}~\bibnamefont {Onizhuk}}, \bibinfo
  {author} {\bibfnamefont {H.}~\bibnamefont {Ma}}, \bibinfo {author}
  {\bibfnamefont {A.~L.}\ \bibnamefont {Crook}}, \bibinfo {author}
  {\bibfnamefont {H.}~\bibnamefont {Abe}}, \bibinfo {author} {\bibfnamefont
  {J.}~\bibnamefont {Ul-Hassan}}, \bibinfo {author} {\bibfnamefont
  {T.}~\bibnamefont {Ohshima}}, \bibinfo {author} {\bibfnamefont {N.~T.}\
  \bibnamefont {Son}},  \emph {et~al.},\ }\href@noop {} {\bibfield  {journal}
  {\bibinfo  {journal} {Nature Materials}\ ,\ \bibinfo {pages} {1}} (\bibinfo
  {year} {2020})}\BibitemShut {NoStop}%
\bibitem [{\citenamefont {Klimov}\ \emph {et~al.}(2014)\citenamefont {Klimov},
  \citenamefont {Falk}, \citenamefont {Buckley},\ and\ \citenamefont
  {Awschalom}}]{klimov2014electrically}%
  \BibitemOpen
  \bibfield  {author} {\bibinfo {author} {\bibfnamefont {P.}~\bibnamefont
  {Klimov}}, \bibinfo {author} {\bibfnamefont {A.}~\bibnamefont {Falk}},
  \bibinfo {author} {\bibfnamefont {B.}~\bibnamefont {Buckley}}, \ and\
  \bibinfo {author} {\bibfnamefont {D.}~\bibnamefont {Awschalom}},\ }\href@noop
  {} {\bibfield  {journal} {\bibinfo  {journal} {Physical Review Letters}\
  }\textbf {\bibinfo {volume} {112}},\ \bibinfo {pages} {087601} (\bibinfo
  {year} {2014})}\BibitemShut {NoStop}%
\bibitem [{\citenamefont {Falk}\ \emph {et~al.}(2014)\citenamefont {Falk},
  \citenamefont {Klimov}, \citenamefont {Buckley}, \citenamefont {Iv{\'a}dy},
  \citenamefont {Abrikosov}, \citenamefont {Calusine}, \citenamefont {Koehl},
  \citenamefont {Gali},\ and\ \citenamefont
  {Awschalom}}]{falk2014electrically}%
  \BibitemOpen
  \bibfield  {author} {\bibinfo {author} {\bibfnamefont {A.~L.}\ \bibnamefont
  {Falk}}, \bibinfo {author} {\bibfnamefont {P.~V.}\ \bibnamefont {Klimov}},
  \bibinfo {author} {\bibfnamefont {B.~B.}\ \bibnamefont {Buckley}}, \bibinfo
  {author} {\bibfnamefont {V.}~\bibnamefont {Iv{\'a}dy}}, \bibinfo {author}
  {\bibfnamefont {I.~A.}\ \bibnamefont {Abrikosov}}, \bibinfo {author}
  {\bibfnamefont {G.}~\bibnamefont {Calusine}}, \bibinfo {author}
  {\bibfnamefont {W.~F.}\ \bibnamefont {Koehl}}, \bibinfo {author}
  {\bibfnamefont {{\'A}.}~\bibnamefont {Gali}}, \ and\ \bibinfo {author}
  {\bibfnamefont {D.~D.}\ \bibnamefont {Awschalom}},\ }\href@noop {} {\bibfield
   {journal} {\bibinfo  {journal} {Physical review letters}\ }\textbf {\bibinfo
  {volume} {112}},\ \bibinfo {pages} {187601} (\bibinfo {year}
  {2014})}\BibitemShut {NoStop}%
\bibitem [{\citenamefont {Whiteley}\ \emph {et~al.}(2019)\citenamefont
  {Whiteley}, \citenamefont {Wolfowicz}, \citenamefont {Anderson},
  \citenamefont {Bourassa}, \citenamefont {Ma}, \citenamefont {Ye},
  \citenamefont {Koolstra}, \citenamefont {Satzinger}, \citenamefont {Holt},
  \citenamefont {Heremans} \emph {et~al.}}]{whiteley2019spin}%
  \BibitemOpen
  \bibfield  {author} {\bibinfo {author} {\bibfnamefont {S.~J.}\ \bibnamefont
  {Whiteley}}, \bibinfo {author} {\bibfnamefont {G.}~\bibnamefont {Wolfowicz}},
  \bibinfo {author} {\bibfnamefont {C.~P.}\ \bibnamefont {Anderson}}, \bibinfo
  {author} {\bibfnamefont {A.}~\bibnamefont {Bourassa}}, \bibinfo {author}
  {\bibfnamefont {H.}~\bibnamefont {Ma}}, \bibinfo {author} {\bibfnamefont
  {M.}~\bibnamefont {Ye}}, \bibinfo {author} {\bibfnamefont {G.}~\bibnamefont
  {Koolstra}}, \bibinfo {author} {\bibfnamefont {K.~J.}\ \bibnamefont
  {Satzinger}}, \bibinfo {author} {\bibfnamefont {M.~V.}\ \bibnamefont {Holt}},
  \bibinfo {author} {\bibfnamefont {F.~J.}\ \bibnamefont {Heremans}},  \emph
  {et~al.},\ }\href@noop {} {\bibfield  {journal} {\bibinfo  {journal} {Nature
  Physics}\ }\textbf {\bibinfo {volume} {15}},\ \bibinfo {pages} {490}
  (\bibinfo {year} {2019})}\BibitemShut {NoStop}%
\bibitem [{\citenamefont {Di~Cioccio}\ \emph {et~al.}(1997)\citenamefont
  {Di~Cioccio}, \citenamefont {Letertre}, \citenamefont {Le~Tiec},
  \citenamefont {Papon}, \citenamefont {Jaussaud},\ and\ \citenamefont
  {Bruel}}]{di1997silicon}%
  \BibitemOpen
  \bibfield  {author} {\bibinfo {author} {\bibfnamefont {L.}~\bibnamefont
  {Di~Cioccio}}, \bibinfo {author} {\bibfnamefont {F.}~\bibnamefont
  {Letertre}}, \bibinfo {author} {\bibfnamefont {Y.}~\bibnamefont {Le~Tiec}},
  \bibinfo {author} {\bibfnamefont {A.}~\bibnamefont {Papon}}, \bibinfo
  {author} {\bibfnamefont {C.}~\bibnamefont {Jaussaud}}, \ and\ \bibinfo
  {author} {\bibfnamefont {M.}~\bibnamefont {Bruel}},\ }\href@noop {}
  {\bibfield  {journal} {\bibinfo  {journal} {Materials Science and
  Engineering: B}\ }\textbf {\bibinfo {volume} {46}},\ \bibinfo {pages} {349}
  (\bibinfo {year} {1997})}\BibitemShut {NoStop}%
\bibitem [{\citenamefont {Song}\ \emph {et~al.}(2011)\citenamefont {Song},
  \citenamefont {Yamada}, \citenamefont {Asano},\ and\ \citenamefont
  {Noda}}]{song2011demonstration}%
  \BibitemOpen
  \bibfield  {author} {\bibinfo {author} {\bibfnamefont {B.-S.}\ \bibnamefont
  {Song}}, \bibinfo {author} {\bibfnamefont {S.}~\bibnamefont {Yamada}},
  \bibinfo {author} {\bibfnamefont {T.}~\bibnamefont {Asano}}, \ and\ \bibinfo
  {author} {\bibfnamefont {S.}~\bibnamefont {Noda}},\ }\href@noop {} {\bibfield
   {journal} {\bibinfo  {journal} {Optics express}\ }\textbf {\bibinfo {volume}
  {19}},\ \bibinfo {pages} {11084} (\bibinfo {year} {2011})}\BibitemShut
  {NoStop}%
\bibitem [{\citenamefont {Yamada}\ \emph {et~al.}(2014)\citenamefont {Yamada},
  \citenamefont {Song}, \citenamefont {Jeon}, \citenamefont {Upham},
  \citenamefont {Tanaka}, \citenamefont {Asano},\ and\ \citenamefont
  {Noda}}]{yamada2014second}%
  \BibitemOpen
  \bibfield  {author} {\bibinfo {author} {\bibfnamefont {S.}~\bibnamefont
  {Yamada}}, \bibinfo {author} {\bibfnamefont {B.-S.}\ \bibnamefont {Song}},
  \bibinfo {author} {\bibfnamefont {S.}~\bibnamefont {Jeon}}, \bibinfo {author}
  {\bibfnamefont {J.}~\bibnamefont {Upham}}, \bibinfo {author} {\bibfnamefont
  {Y.}~\bibnamefont {Tanaka}}, \bibinfo {author} {\bibfnamefont
  {T.}~\bibnamefont {Asano}}, \ and\ \bibinfo {author} {\bibfnamefont
  {S.}~\bibnamefont {Noda}},\ }\href@noop {} {\bibfield  {journal} {\bibinfo
  {journal} {Optics letters}\ }\textbf {\bibinfo {volume} {39}},\ \bibinfo
  {pages} {1768} (\bibinfo {year} {2014})}\BibitemShut {NoStop}%
\bibitem [{\citenamefont {Cardenas}\ \emph {et~al.}(2015)\citenamefont
  {Cardenas}, \citenamefont {Yu}, \citenamefont {Okawachi}, \citenamefont
  {Poitras}, \citenamefont {Lau}, \citenamefont {Dutt}, \citenamefont {Gaeta},\
  and\ \citenamefont {Lipson}}]{cardenas2015optical}%
  \BibitemOpen
  \bibfield  {author} {\bibinfo {author} {\bibfnamefont {J.}~\bibnamefont
  {Cardenas}}, \bibinfo {author} {\bibfnamefont {M.}~\bibnamefont {Yu}},
  \bibinfo {author} {\bibfnamefont {Y.}~\bibnamefont {Okawachi}}, \bibinfo
  {author} {\bibfnamefont {C.~B.}\ \bibnamefont {Poitras}}, \bibinfo {author}
  {\bibfnamefont {R.~K.}\ \bibnamefont {Lau}}, \bibinfo {author} {\bibfnamefont
  {A.}~\bibnamefont {Dutt}}, \bibinfo {author} {\bibfnamefont {A.~L.}\
  \bibnamefont {Gaeta}}, \ and\ \bibinfo {author} {\bibfnamefont
  {M.}~\bibnamefont {Lipson}},\ }\href@noop {} {\bibfield  {journal} {\bibinfo
  {journal} {Optics letters}\ }\textbf {\bibinfo {volume} {40}},\ \bibinfo
  {pages} {4138} (\bibinfo {year} {2015})}\BibitemShut {NoStop}%
\bibitem [{\citenamefont {Yi}\ \emph {et~al.}(2020)\citenamefont {Yi},
  \citenamefont {Zheng}, \citenamefont {Huang}, \citenamefont {Lin},
  \citenamefont {Yan}, \citenamefont {You}, \citenamefont {Huang},
  \citenamefont {Zhang}, \citenamefont {Shen}, \citenamefont {Zhou} \emph
  {et~al.}}]{yi2020wafer}%
  \BibitemOpen
  \bibfield  {author} {\bibinfo {author} {\bibfnamefont {A.}~\bibnamefont
  {Yi}}, \bibinfo {author} {\bibfnamefont {Y.}~\bibnamefont {Zheng}}, \bibinfo
  {author} {\bibfnamefont {H.}~\bibnamefont {Huang}}, \bibinfo {author}
  {\bibfnamefont {J.}~\bibnamefont {Lin}}, \bibinfo {author} {\bibfnamefont
  {Y.}~\bibnamefont {Yan}}, \bibinfo {author} {\bibfnamefont {T.}~\bibnamefont
  {You}}, \bibinfo {author} {\bibfnamefont {K.}~\bibnamefont {Huang}}, \bibinfo
  {author} {\bibfnamefont {S.}~\bibnamefont {Zhang}}, \bibinfo {author}
  {\bibfnamefont {C.}~\bibnamefont {Shen}}, \bibinfo {author} {\bibfnamefont
  {M.}~\bibnamefont {Zhou}},  \emph {et~al.},\ }\href@noop {} {\bibfield
  {journal} {\bibinfo  {journal} {Optical Materials}\ ,\ \bibinfo {pages}
  {109990}} (\bibinfo {year} {2020})}\BibitemShut {NoStop}%
\bibitem [{\citenamefont {Zheng}\ \emph {et~al.}(2019)\citenamefont {Zheng},
  \citenamefont {Pu}, \citenamefont {Yi}, \citenamefont {Chang}, \citenamefont
  {You}, \citenamefont {Huang}, \citenamefont {Kamel}, \citenamefont
  {Henriksen}, \citenamefont {J{\o}rgensen}, \citenamefont {Ou} \emph
  {et~al.}}]{zheng2019high}%
  \BibitemOpen
  \bibfield  {author} {\bibinfo {author} {\bibfnamefont {Y.}~\bibnamefont
  {Zheng}}, \bibinfo {author} {\bibfnamefont {M.}~\bibnamefont {Pu}}, \bibinfo
  {author} {\bibfnamefont {A.}~\bibnamefont {Yi}}, \bibinfo {author}
  {\bibfnamefont {B.}~\bibnamefont {Chang}}, \bibinfo {author} {\bibfnamefont
  {T.}~\bibnamefont {You}}, \bibinfo {author} {\bibfnamefont {K.}~\bibnamefont
  {Huang}}, \bibinfo {author} {\bibfnamefont {A.~N.}\ \bibnamefont {Kamel}},
  \bibinfo {author} {\bibfnamefont {M.~R.}\ \bibnamefont {Henriksen}}, \bibinfo
  {author} {\bibfnamefont {A.~A.}\ \bibnamefont {J{\o}rgensen}}, \bibinfo
  {author} {\bibfnamefont {X.}~\bibnamefont {Ou}},  \emph {et~al.},\
  }\href@noop {} {\bibfield  {journal} {\bibinfo  {journal} {Optics express}\
  }\textbf {\bibinfo {volume} {27}},\ \bibinfo {pages} {13053} (\bibinfo {year}
  {2019})}\BibitemShut {NoStop}%
\bibitem [{\citenamefont {Cardenas}\ \emph {et~al.}(2013)\citenamefont
  {Cardenas}, \citenamefont {Zhang}, \citenamefont {Phare}, \citenamefont
  {Shah}, \citenamefont {Poitras}, \citenamefont {Guha},\ and\ \citenamefont
  {Lipson}}]{cardenas2013high}%
  \BibitemOpen
  \bibfield  {author} {\bibinfo {author} {\bibfnamefont {J.}~\bibnamefont
  {Cardenas}}, \bibinfo {author} {\bibfnamefont {M.}~\bibnamefont {Zhang}},
  \bibinfo {author} {\bibfnamefont {C.~T.}\ \bibnamefont {Phare}}, \bibinfo
  {author} {\bibfnamefont {S.~Y.}\ \bibnamefont {Shah}}, \bibinfo {author}
  {\bibfnamefont {C.~B.}\ \bibnamefont {Poitras}}, \bibinfo {author}
  {\bibfnamefont {B.}~\bibnamefont {Guha}}, \ and\ \bibinfo {author}
  {\bibfnamefont {M.}~\bibnamefont {Lipson}},\ }\href@noop {} {\bibfield
  {journal} {\bibinfo  {journal} {Optics express}\ }\textbf {\bibinfo {volume}
  {21}},\ \bibinfo {pages} {16882} (\bibinfo {year} {2013})}\BibitemShut
  {NoStop}%
\bibitem [{\citenamefont {Radulaski}\ \emph {et~al.}(2013)\citenamefont
  {Radulaski}, \citenamefont {Babinec}, \citenamefont {Buckley}, \citenamefont
  {Rundquist}, \citenamefont {Provine}, \citenamefont {Alassaad}, \citenamefont
  {Ferro},\ and\ \citenamefont {Vu{\v{c}}kovi{\'c}}}]{radulaski2013photonic}%
  \BibitemOpen
  \bibfield  {author} {\bibinfo {author} {\bibfnamefont {M.}~\bibnamefont
  {Radulaski}}, \bibinfo {author} {\bibfnamefont {T.~M.}\ \bibnamefont
  {Babinec}}, \bibinfo {author} {\bibfnamefont {S.}~\bibnamefont {Buckley}},
  \bibinfo {author} {\bibfnamefont {A.}~\bibnamefont {Rundquist}}, \bibinfo
  {author} {\bibfnamefont {J.}~\bibnamefont {Provine}}, \bibinfo {author}
  {\bibfnamefont {K.}~\bibnamefont {Alassaad}}, \bibinfo {author}
  {\bibfnamefont {G.}~\bibnamefont {Ferro}}, \ and\ \bibinfo {author}
  {\bibfnamefont {J.}~\bibnamefont {Vu{\v{c}}kovi{\'c}}},\ }\href@noop {}
  {\bibfield  {journal} {\bibinfo  {journal} {Optics express}\ }\textbf
  {\bibinfo {volume} {21}},\ \bibinfo {pages} {32623} (\bibinfo {year}
  {2013})}\BibitemShut {NoStop}%
\bibitem [{\citenamefont {Lu}\ \emph {et~al.}(2014)\citenamefont {Lu},
  \citenamefont {Lee}, \citenamefont {Feng},\ and\ \citenamefont
  {Lin}}]{lu2014high}%
  \BibitemOpen
  \bibfield  {author} {\bibinfo {author} {\bibfnamefont {X.}~\bibnamefont
  {Lu}}, \bibinfo {author} {\bibfnamefont {J.~Y.}\ \bibnamefont {Lee}},
  \bibinfo {author} {\bibfnamefont {P.~X.-L.}\ \bibnamefont {Feng}}, \ and\
  \bibinfo {author} {\bibfnamefont {Q.}~\bibnamefont {Lin}},\ }\href@noop {}
  {\bibfield  {journal} {\bibinfo  {journal} {Applied Physics Letters}\
  }\textbf {\bibinfo {volume} {104}},\ \bibinfo {pages} {181103} (\bibinfo
  {year} {2014})}\BibitemShut {NoStop}%
\bibitem [{\citenamefont {Fan}\ \emph {et~al.}(2018)\citenamefont {Fan},
  \citenamefont {Moradinejad}, \citenamefont {Wu}, \citenamefont {Eftekhar},\
  and\ \citenamefont {Adibi}}]{fan2018high}%
  \BibitemOpen
  \bibfield  {author} {\bibinfo {author} {\bibfnamefont {T.}~\bibnamefont
  {Fan}}, \bibinfo {author} {\bibfnamefont {H.}~\bibnamefont {Moradinejad}},
  \bibinfo {author} {\bibfnamefont {X.}~\bibnamefont {Wu}}, \bibinfo {author}
  {\bibfnamefont {A.~A.}\ \bibnamefont {Eftekhar}}, \ and\ \bibinfo {author}
  {\bibfnamefont {A.}~\bibnamefont {Adibi}},\ }\href@noop {} {\bibfield
  {journal} {\bibinfo  {journal} {Optics express}\ }\textbf {\bibinfo {volume}
  {26}},\ \bibinfo {pages} {25814} (\bibinfo {year} {2018})}\BibitemShut
  {NoStop}%
\bibitem [{\citenamefont {Fan}\ \emph {et~al.}(2020)\citenamefont {Fan},
  \citenamefont {Wu}, \citenamefont {Eftekhar}, \citenamefont {Bosi},
  \citenamefont {Moradinejad}, \citenamefont {Woods},\ and\ \citenamefont
  {Adibi}}]{fan2020high}%
  \BibitemOpen
  \bibfield  {author} {\bibinfo {author} {\bibfnamefont {T.}~\bibnamefont
  {Fan}}, \bibinfo {author} {\bibfnamefont {X.}~\bibnamefont {Wu}}, \bibinfo
  {author} {\bibfnamefont {A.~A.}\ \bibnamefont {Eftekhar}}, \bibinfo {author}
  {\bibfnamefont {M.}~\bibnamefont {Bosi}}, \bibinfo {author} {\bibfnamefont
  {H.}~\bibnamefont {Moradinejad}}, \bibinfo {author} {\bibfnamefont {E.~V.}\
  \bibnamefont {Woods}}, \ and\ \bibinfo {author} {\bibfnamefont
  {A.}~\bibnamefont {Adibi}},\ }\href@noop {} {\bibfield  {journal} {\bibinfo
  {journal} {Optics Letters}\ }\textbf {\bibinfo {volume} {45}},\ \bibinfo
  {pages} {153} (\bibinfo {year} {2020})}\BibitemShut {NoStop}%
\bibitem [{\citenamefont {Calusine}\ \emph {et~al.}(2016)\citenamefont
  {Calusine}, \citenamefont {Politi},\ and\ \citenamefont
  {Awschalom}}]{calusine2016cavity}%
  \BibitemOpen
  \bibfield  {author} {\bibinfo {author} {\bibfnamefont {G.}~\bibnamefont
  {Calusine}}, \bibinfo {author} {\bibfnamefont {A.}~\bibnamefont {Politi}}, \
  and\ \bibinfo {author} {\bibfnamefont {D.~D.}\ \bibnamefont {Awschalom}},\
  }\href@noop {} {\bibfield  {journal} {\bibinfo  {journal} {Physical Review
  Applied}\ }\textbf {\bibinfo {volume} {6}},\ \bibinfo {pages} {014019}
  (\bibinfo {year} {2016})}\BibitemShut {NoStop}%
\bibitem [{\citenamefont {Burek}\ \emph {et~al.}(2014)\citenamefont {Burek},
  \citenamefont {Chu}, \citenamefont {Liddy}, \citenamefont {Patel},
  \citenamefont {Rochman}, \citenamefont {Meesala}, \citenamefont {Hong},
  \citenamefont {Quan}, \citenamefont {Lukin},\ and\ \citenamefont
  {Lon{\v{c}}ar}}]{burek2014high}%
  \BibitemOpen
  \bibfield  {author} {\bibinfo {author} {\bibfnamefont {M.~J.}\ \bibnamefont
  {Burek}}, \bibinfo {author} {\bibfnamefont {Y.}~\bibnamefont {Chu}}, \bibinfo
  {author} {\bibfnamefont {M.~S.}\ \bibnamefont {Liddy}}, \bibinfo {author}
  {\bibfnamefont {P.}~\bibnamefont {Patel}}, \bibinfo {author} {\bibfnamefont
  {J.}~\bibnamefont {Rochman}}, \bibinfo {author} {\bibfnamefont
  {S.}~\bibnamefont {Meesala}}, \bibinfo {author} {\bibfnamefont
  {W.}~\bibnamefont {Hong}}, \bibinfo {author} {\bibfnamefont {Q.}~\bibnamefont
  {Quan}}, \bibinfo {author} {\bibfnamefont {M.~D.}\ \bibnamefont {Lukin}}, \
  and\ \bibinfo {author} {\bibfnamefont {M.}~\bibnamefont {Lon{\v{c}}ar}},\
  }\href@noop {} {\bibfield  {journal} {\bibinfo  {journal} {Nature
  communications}\ }\textbf {\bibinfo {volume} {5}},\ \bibinfo {pages} {1}
  (\bibinfo {year} {2014})}\BibitemShut {NoStop}%
\bibitem [{\citenamefont {Song}\ \emph {et~al.}(2018)\citenamefont {Song},
  \citenamefont {Jeon}, \citenamefont {Kim}, \citenamefont {Kang},
  \citenamefont {Asano},\ and\ \citenamefont {Noda}}]{song2018high}%
  \BibitemOpen
  \bibfield  {author} {\bibinfo {author} {\bibfnamefont {B.-S.}\ \bibnamefont
  {Song}}, \bibinfo {author} {\bibfnamefont {S.}~\bibnamefont {Jeon}}, \bibinfo
  {author} {\bibfnamefont {H.}~\bibnamefont {Kim}}, \bibinfo {author}
  {\bibfnamefont {D.~D.}\ \bibnamefont {Kang}}, \bibinfo {author}
  {\bibfnamefont {T.}~\bibnamefont {Asano}}, \ and\ \bibinfo {author}
  {\bibfnamefont {S.}~\bibnamefont {Noda}},\ }\href@noop {} {\bibfield
  {journal} {\bibinfo  {journal} {Applied Physics Letters}\ }\textbf {\bibinfo
  {volume} {113}},\ \bibinfo {pages} {231106} (\bibinfo {year}
  {2018})}\BibitemShut {NoStop}%
\bibitem [{\citenamefont {Bracher}\ \emph {et~al.}(2017)\citenamefont
  {Bracher}, \citenamefont {Zhang},\ and\ \citenamefont
  {Hu}}]{bracher2017selective}%
  \BibitemOpen
  \bibfield  {author} {\bibinfo {author} {\bibfnamefont {D.~O.}\ \bibnamefont
  {Bracher}}, \bibinfo {author} {\bibfnamefont {X.}~\bibnamefont {Zhang}}, \
  and\ \bibinfo {author} {\bibfnamefont {E.~L.}\ \bibnamefont {Hu}},\
  }\href@noop {} {\bibfield  {journal} {\bibinfo  {journal} {Proceedings of the
  National Academy of Sciences}\ }\textbf {\bibinfo {volume} {114}},\ \bibinfo
  {pages} {4060} (\bibinfo {year} {2017})}\BibitemShut {NoStop}%
\bibitem [{\citenamefont {Crook}\ \emph {et~al.}(2020)\citenamefont {Crook},
  \citenamefont {Anderson}, \citenamefont {Miao}, \citenamefont {Bourassa},
  \citenamefont {Lee}, \citenamefont {Bayliss}, \citenamefont {Bracher},
  \citenamefont {Zhang}, \citenamefont {Abe}, \citenamefont {Ohshima} \emph
  {et~al.}}]{crook2020purcell}%
  \BibitemOpen
  \bibfield  {author} {\bibinfo {author} {\bibfnamefont {A.~L.}\ \bibnamefont
  {Crook}}, \bibinfo {author} {\bibfnamefont {C.~P.}\ \bibnamefont {Anderson}},
  \bibinfo {author} {\bibfnamefont {K.~C.}\ \bibnamefont {Miao}}, \bibinfo
  {author} {\bibfnamefont {A.}~\bibnamefont {Bourassa}}, \bibinfo {author}
  {\bibfnamefont {H.}~\bibnamefont {Lee}}, \bibinfo {author} {\bibfnamefont
  {S.~L.}\ \bibnamefont {Bayliss}}, \bibinfo {author} {\bibfnamefont {D.~O.}\
  \bibnamefont {Bracher}}, \bibinfo {author} {\bibfnamefont {X.}~\bibnamefont
  {Zhang}}, \bibinfo {author} {\bibfnamefont {H.}~\bibnamefont {Abe}}, \bibinfo
  {author} {\bibfnamefont {T.}~\bibnamefont {Ohshima}},  \emph {et~al.},\
  }\href@noop {} {\bibfield  {journal} {\bibinfo  {journal} {Nano Letters}\ }
  (\bibinfo {year} {2020})}\BibitemShut {NoStop}%
\bibitem [{\citenamefont {Robledo}\ \emph {et~al.}(2011)\citenamefont
  {Robledo}, \citenamefont {Childress}, \citenamefont {Bernien}, \citenamefont
  {Hensen}, \citenamefont {Alkemade},\ and\ \citenamefont
  {Hanson}}]{robledo2011high}%
  \BibitemOpen
  \bibfield  {author} {\bibinfo {author} {\bibfnamefont {L.}~\bibnamefont
  {Robledo}}, \bibinfo {author} {\bibfnamefont {L.}~\bibnamefont {Childress}},
  \bibinfo {author} {\bibfnamefont {H.}~\bibnamefont {Bernien}}, \bibinfo
  {author} {\bibfnamefont {B.}~\bibnamefont {Hensen}}, \bibinfo {author}
  {\bibfnamefont {P.~F.}\ \bibnamefont {Alkemade}}, \ and\ \bibinfo {author}
  {\bibfnamefont {R.}~\bibnamefont {Hanson}},\ }\href@noop {} {\bibfield
  {journal} {\bibinfo  {journal} {Nature}\ }\textbf {\bibinfo {volume} {477}},\
  \bibinfo {pages} {574} (\bibinfo {year} {2011})}\BibitemShut {NoStop}%
\bibitem [{\citenamefont {Sukachev}\ \emph {et~al.}(2017)\citenamefont
  {Sukachev}, \citenamefont {Sipahigil}, \citenamefont {Nguyen}, \citenamefont
  {Bhaskar}, \citenamefont {Evans}, \citenamefont {Jelezko},\ and\
  \citenamefont {Lukin}}]{sukachev2017silicon}%
  \BibitemOpen
  \bibfield  {author} {\bibinfo {author} {\bibfnamefont {D.~D.}\ \bibnamefont
  {Sukachev}}, \bibinfo {author} {\bibfnamefont {A.}~\bibnamefont {Sipahigil}},
  \bibinfo {author} {\bibfnamefont {C.~T.}\ \bibnamefont {Nguyen}}, \bibinfo
  {author} {\bibfnamefont {M.~K.}\ \bibnamefont {Bhaskar}}, \bibinfo {author}
  {\bibfnamefont {R.~E.}\ \bibnamefont {Evans}}, \bibinfo {author}
  {\bibfnamefont {F.}~\bibnamefont {Jelezko}}, \ and\ \bibinfo {author}
  {\bibfnamefont {M.~D.}\ \bibnamefont {Lukin}},\ }\href@noop {} {\bibfield
  {journal} {\bibinfo  {journal} {Physical review letters}\ }\textbf {\bibinfo
  {volume} {119}},\ \bibinfo {pages} {223602} (\bibinfo {year}
  {2017})}\BibitemShut {NoStop}%
\bibitem [{\citenamefont {Burek}\ \emph {et~al.}(2017)\citenamefont {Burek},
  \citenamefont {Meuwly}, \citenamefont {Evans}, \citenamefont {Bhaskar},
  \citenamefont {Sipahigil}, \citenamefont {Meesala}, \citenamefont
  {Machielse}, \citenamefont {Sukachev}, \citenamefont {Nguyen}, \citenamefont
  {Pacheco} \emph {et~al.}}]{burek2017fiber}%
  \BibitemOpen
  \bibfield  {author} {\bibinfo {author} {\bibfnamefont {M.~J.}\ \bibnamefont
  {Burek}}, \bibinfo {author} {\bibfnamefont {C.}~\bibnamefont {Meuwly}},
  \bibinfo {author} {\bibfnamefont {R.~E.}\ \bibnamefont {Evans}}, \bibinfo
  {author} {\bibfnamefont {M.~K.}\ \bibnamefont {Bhaskar}}, \bibinfo {author}
  {\bibfnamefont {A.}~\bibnamefont {Sipahigil}}, \bibinfo {author}
  {\bibfnamefont {S.}~\bibnamefont {Meesala}}, \bibinfo {author} {\bibfnamefont
  {B.}~\bibnamefont {Machielse}}, \bibinfo {author} {\bibfnamefont {D.~D.}\
  \bibnamefont {Sukachev}}, \bibinfo {author} {\bibfnamefont {C.~T.}\
  \bibnamefont {Nguyen}}, \bibinfo {author} {\bibfnamefont {J.~L.}\
  \bibnamefont {Pacheco}},  \emph {et~al.},\ }\href@noop {} {\bibfield
  {journal} {\bibinfo  {journal} {Physical Review Applied}\ }\textbf {\bibinfo
  {volume} {8}},\ \bibinfo {pages} {024026} (\bibinfo {year}
  {2017})}\BibitemShut {NoStop}%
\bibitem [{\citenamefont {Machielse}\ \emph {et~al.}(2019)\citenamefont
  {Machielse}, \citenamefont {Bogdanovic}, \citenamefont {Meesala},
  \citenamefont {Gauthier}, \citenamefont {Burek}, \citenamefont {Joe},
  \citenamefont {Chalupnik}, \citenamefont {Sohn}, \citenamefont {Holzgrafe},
  \citenamefont {Evans} \emph {et~al.}}]{machielse2019quantum}%
  \BibitemOpen
  \bibfield  {author} {\bibinfo {author} {\bibfnamefont {B.}~\bibnamefont
  {Machielse}}, \bibinfo {author} {\bibfnamefont {S.}~\bibnamefont
  {Bogdanovic}}, \bibinfo {author} {\bibfnamefont {S.}~\bibnamefont {Meesala}},
  \bibinfo {author} {\bibfnamefont {S.}~\bibnamefont {Gauthier}}, \bibinfo
  {author} {\bibfnamefont {M.~J.}\ \bibnamefont {Burek}}, \bibinfo {author}
  {\bibfnamefont {G.}~\bibnamefont {Joe}}, \bibinfo {author} {\bibfnamefont
  {M.}~\bibnamefont {Chalupnik}}, \bibinfo {author} {\bibfnamefont {Y.-I.}\
  \bibnamefont {Sohn}}, \bibinfo {author} {\bibfnamefont {J.}~\bibnamefont
  {Holzgrafe}}, \bibinfo {author} {\bibfnamefont {R.~E.}\ \bibnamefont
  {Evans}},  \emph {et~al.},\ }\href@noop {} {\bibfield  {journal} {\bibinfo
  {journal} {Physical Review X}\ }\textbf {\bibinfo {volume} {9}},\ \bibinfo
  {pages} {031022} (\bibinfo {year} {2019})}\BibitemShut {NoStop}%
\bibitem [{\citenamefont {Wan}\ \emph {et~al.}(2020)\citenamefont {Wan},
  \citenamefont {Lu}, \citenamefont {Chen}, \citenamefont {Walsh},
  \citenamefont {Trusheim}, \citenamefont {De~Santis}, \citenamefont {Bersin},
  \citenamefont {Harris}, \citenamefont {Mouradian}, \citenamefont {Christen}
  \emph {et~al.}}]{wan2020large}%
  \BibitemOpen
  \bibfield  {author} {\bibinfo {author} {\bibfnamefont {N.~H.}\ \bibnamefont
  {Wan}}, \bibinfo {author} {\bibfnamefont {T.-J.}\ \bibnamefont {Lu}},
  \bibinfo {author} {\bibfnamefont {K.~C.}\ \bibnamefont {Chen}}, \bibinfo
  {author} {\bibfnamefont {M.~P.}\ \bibnamefont {Walsh}}, \bibinfo {author}
  {\bibfnamefont {M.~E.}\ \bibnamefont {Trusheim}}, \bibinfo {author}
  {\bibfnamefont {L.}~\bibnamefont {De~Santis}}, \bibinfo {author}
  {\bibfnamefont {E.~A.}\ \bibnamefont {Bersin}}, \bibinfo {author}
  {\bibfnamefont {I.~B.}\ \bibnamefont {Harris}}, \bibinfo {author}
  {\bibfnamefont {S.~L.}\ \bibnamefont {Mouradian}}, \bibinfo {author}
  {\bibfnamefont {I.~R.}\ \bibnamefont {Christen}},  \emph {et~al.},\
  }\href@noop {} {\bibfield  {journal} {\bibinfo  {journal} {Nature}\ }\textbf
  {\bibinfo {volume} {583}},\ \bibinfo {pages} {226} (\bibinfo {year}
  {2020})}\BibitemShut {NoStop}%
\bibitem [{\citenamefont {Rugar}\ \emph {et~al.}(2020)\citenamefont {Rugar},
  \citenamefont {Dory}, \citenamefont {Aghaeimeibodi}, \citenamefont {Lu},
  \citenamefont {Sun}, \citenamefont {Mishra}, \citenamefont {Shen},
  \citenamefont {Melosh},\ and\ \citenamefont
  {Vu{\v{c}}kovi{\'c}}}]{rugar2020narrow}%
  \BibitemOpen
  \bibfield  {author} {\bibinfo {author} {\bibfnamefont {A.~E.}\ \bibnamefont
  {Rugar}}, \bibinfo {author} {\bibfnamefont {C.}~\bibnamefont {Dory}},
  \bibinfo {author} {\bibfnamefont {S.}~\bibnamefont {Aghaeimeibodi}}, \bibinfo
  {author} {\bibfnamefont {H.}~\bibnamefont {Lu}}, \bibinfo {author}
  {\bibfnamefont {S.}~\bibnamefont {Sun}}, \bibinfo {author} {\bibfnamefont
  {S.~D.}\ \bibnamefont {Mishra}}, \bibinfo {author} {\bibfnamefont {Z.-X.}\
  \bibnamefont {Shen}}, \bibinfo {author} {\bibfnamefont {N.~A.}\ \bibnamefont
  {Melosh}}, \ and\ \bibinfo {author} {\bibfnamefont {J.}~\bibnamefont
  {Vu{\v{c}}kovi{\'c}}},\ }\href@noop {} {\bibfield  {journal} {\bibinfo
  {journal} {ACS Photonics}\ }\textbf {\bibinfo {volume} {7}},\ \bibinfo
  {pages} {2356–2361} (\bibinfo {year} {2020})}\BibitemShut {NoStop}%
\bibitem [{\citenamefont {Hausmann}\ \emph {et~al.}(2014)\citenamefont
  {Hausmann}, \citenamefont {Bulu}, \citenamefont {Venkataraman}, \citenamefont
  {Deotare},\ and\ \citenamefont {Lon{\v{c}}ar}}]{hausmann2014diamond}%
  \BibitemOpen
  \bibfield  {author} {\bibinfo {author} {\bibfnamefont {B.}~\bibnamefont
  {Hausmann}}, \bibinfo {author} {\bibfnamefont {I.}~\bibnamefont {Bulu}},
  \bibinfo {author} {\bibfnamefont {V.}~\bibnamefont {Venkataraman}}, \bibinfo
  {author} {\bibfnamefont {P.}~\bibnamefont {Deotare}}, \ and\ \bibinfo
  {author} {\bibfnamefont {M.}~\bibnamefont {Lon{\v{c}}ar}},\ }\href@noop {}
  {\bibfield  {journal} {\bibinfo  {journal} {Nature Photonics}\ }\textbf
  {\bibinfo {volume} {8}},\ \bibinfo {pages} {369} (\bibinfo {year}
  {2014})}\BibitemShut {NoStop}%
\bibitem [{\citenamefont {Sato}\ \emph {et~al.}(2009)\citenamefont {Sato},
  \citenamefont {Abe}, \citenamefont {Shoji}, \citenamefont {Suda},\ and\
  \citenamefont {Kondo}}]{sato2009accurate}%
  \BibitemOpen
  \bibfield  {author} {\bibinfo {author} {\bibfnamefont {H.}~\bibnamefont
  {Sato}}, \bibinfo {author} {\bibfnamefont {M.}~\bibnamefont {Abe}}, \bibinfo
  {author} {\bibfnamefont {I.}~\bibnamefont {Shoji}}, \bibinfo {author}
  {\bibfnamefont {J.}~\bibnamefont {Suda}}, \ and\ \bibinfo {author}
  {\bibfnamefont {T.}~\bibnamefont {Kondo}},\ }\href@noop {} {\bibfield
  {journal} {\bibinfo  {journal} {JOSA B}\ }\textbf {\bibinfo {volume} {26}},\
  \bibinfo {pages} {1892} (\bibinfo {year} {2009})}\BibitemShut {NoStop}%
\bibitem [{\citenamefont {Asano}\ \emph {et~al.}(2017)\citenamefont {Asano},
  \citenamefont {Ochi}, \citenamefont {Takahashi}, \citenamefont {Kishimoto},\
  and\ \citenamefont {Noda}}]{asano2017photonic}%
  \BibitemOpen
  \bibfield  {author} {\bibinfo {author} {\bibfnamefont {T.}~\bibnamefont
  {Asano}}, \bibinfo {author} {\bibfnamefont {Y.}~\bibnamefont {Ochi}},
  \bibinfo {author} {\bibfnamefont {Y.}~\bibnamefont {Takahashi}}, \bibinfo
  {author} {\bibfnamefont {K.}~\bibnamefont {Kishimoto}}, \ and\ \bibinfo
  {author} {\bibfnamefont {S.}~\bibnamefont {Noda}},\ }\href@noop {} {\bibfield
   {journal} {\bibinfo  {journal} {Optics express}\ }\textbf {\bibinfo {volume}
  {25}},\ \bibinfo {pages} {1769} (\bibinfo {year} {2017})}\BibitemShut
  {NoStop}%
\bibitem [{\citenamefont {Lee}\ \emph {et~al.}(2012)\citenamefont {Lee},
  \citenamefont {Chen}, \citenamefont {Li}, \citenamefont {Yang}, \citenamefont
  {Jeon}, \citenamefont {Painter},\ and\ \citenamefont
  {Vahala}}]{lee2012chemically}%
  \BibitemOpen
  \bibfield  {author} {\bibinfo {author} {\bibfnamefont {H.}~\bibnamefont
  {Lee}}, \bibinfo {author} {\bibfnamefont {T.}~\bibnamefont {Chen}}, \bibinfo
  {author} {\bibfnamefont {J.}~\bibnamefont {Li}}, \bibinfo {author}
  {\bibfnamefont {K.~Y.}\ \bibnamefont {Yang}}, \bibinfo {author}
  {\bibfnamefont {S.}~\bibnamefont {Jeon}}, \bibinfo {author} {\bibfnamefont
  {O.}~\bibnamefont {Painter}}, \ and\ \bibinfo {author} {\bibfnamefont
  {K.~J.}\ \bibnamefont {Vahala}},\ }\href@noop {} {\bibfield  {journal}
  {\bibinfo  {journal} {Nature Photonics}\ }\textbf {\bibinfo {volume} {6}},\
  \bibinfo {pages} {369} (\bibinfo {year} {2012})}\BibitemShut {NoStop}%
\bibitem [{\citenamefont {Yang}\ \emph {et~al.}(2018)\citenamefont {Yang},
  \citenamefont {Oh}, \citenamefont {Lee}, \citenamefont {Yang}, \citenamefont
  {Yi}, \citenamefont {Shen}, \citenamefont {Wang},\ and\ \citenamefont
  {Vahala}}]{yang2018bridging}%
  \BibitemOpen
  \bibfield  {author} {\bibinfo {author} {\bibfnamefont {K.~Y.}\ \bibnamefont
  {Yang}}, \bibinfo {author} {\bibfnamefont {D.~Y.}\ \bibnamefont {Oh}},
  \bibinfo {author} {\bibfnamefont {S.~H.}\ \bibnamefont {Lee}}, \bibinfo
  {author} {\bibfnamefont {Q.-F.}\ \bibnamefont {Yang}}, \bibinfo {author}
  {\bibfnamefont {X.}~\bibnamefont {Yi}}, \bibinfo {author} {\bibfnamefont
  {B.}~\bibnamefont {Shen}}, \bibinfo {author} {\bibfnamefont {H.}~\bibnamefont
  {Wang}}, \ and\ \bibinfo {author} {\bibfnamefont {K.}~\bibnamefont
  {Vahala}},\ }\href@noop {} {\bibfield  {journal} {\bibinfo  {journal} {Nature
  Photonics}\ }\textbf {\bibinfo {volume} {12}},\ \bibinfo {pages} {297}
  (\bibinfo {year} {2018})}\BibitemShut {NoStop}%
\bibitem [{\citenamefont {Li}\ \emph {et~al.}(2019)\citenamefont {Li},
  \citenamefont {Liang}, \citenamefont {Luo}, \citenamefont {He}, \citenamefont
  {Ling},\ and\ \citenamefont {Lin}}]{li2019photon}%
  \BibitemOpen
  \bibfield  {author} {\bibinfo {author} {\bibfnamefont {M.}~\bibnamefont
  {Li}}, \bibinfo {author} {\bibfnamefont {H.}~\bibnamefont {Liang}}, \bibinfo
  {author} {\bibfnamefont {R.}~\bibnamefont {Luo}}, \bibinfo {author}
  {\bibfnamefont {Y.}~\bibnamefont {He}}, \bibinfo {author} {\bibfnamefont
  {J.}~\bibnamefont {Ling}}, \ and\ \bibinfo {author} {\bibfnamefont
  {Q.}~\bibnamefont {Lin}},\ }\href@noop {} {\bibfield  {journal} {\bibinfo
  {journal} {Optica}\ }\textbf {\bibinfo {volume} {6}},\ \bibinfo {pages} {860}
  (\bibinfo {year} {2019})}\BibitemShut {NoStop}%
\bibitem [{\citenamefont {MacCabe}\ \emph {et~al.}(2019)\citenamefont
  {MacCabe}, \citenamefont {Ren}, \citenamefont {Luo}, \citenamefont {Cohen},
  \citenamefont {Zhou}, \citenamefont {Sipahigil}, \citenamefont
  {Mirhosseini},\ and\ \citenamefont {Painter}}]{maccabe2019phononic}%
  \BibitemOpen
  \bibfield  {author} {\bibinfo {author} {\bibfnamefont {G.~S.}\ \bibnamefont
  {MacCabe}}, \bibinfo {author} {\bibfnamefont {H.}~\bibnamefont {Ren}},
  \bibinfo {author} {\bibfnamefont {J.}~\bibnamefont {Luo}}, \bibinfo {author}
  {\bibfnamefont {J.~D.}\ \bibnamefont {Cohen}}, \bibinfo {author}
  {\bibfnamefont {H.}~\bibnamefont {Zhou}}, \bibinfo {author} {\bibfnamefont
  {A.}~\bibnamefont {Sipahigil}}, \bibinfo {author} {\bibfnamefont
  {M.}~\bibnamefont {Mirhosseini}}, \ and\ \bibinfo {author} {\bibfnamefont
  {O.}~\bibnamefont {Painter}},\ }\href@noop {} {\bibfield  {journal} {\bibinfo
   {journal} {arXiv preprint arXiv:1901.04129}\ } (\bibinfo {year}
  {2019})}\BibitemShut {NoStop}%
\bibitem [{\citenamefont {Kuruma}\ \emph {et~al.}(2020)\citenamefont {Kuruma},
  \citenamefont {Ota}, \citenamefont {Kakuda}, \citenamefont {Iwamoto},\ and\
  \citenamefont {Arakawa}}]{kuruma2020surface}%
  \BibitemOpen
  \bibfield  {author} {\bibinfo {author} {\bibfnamefont {K.}~\bibnamefont
  {Kuruma}}, \bibinfo {author} {\bibfnamefont {Y.}~\bibnamefont {Ota}},
  \bibinfo {author} {\bibfnamefont {M.}~\bibnamefont {Kakuda}}, \bibinfo
  {author} {\bibfnamefont {S.}~\bibnamefont {Iwamoto}}, \ and\ \bibinfo
  {author} {\bibfnamefont {Y.}~\bibnamefont {Arakawa}},\ }\href@noop {}
  {\bibfield  {journal} {\bibinfo  {journal} {APL Photonics}\ }\textbf
  {\bibinfo {volume} {5}},\ \bibinfo {pages} {046106} (\bibinfo {year}
  {2020})}\BibitemShut {NoStop}%
\bibitem [{\citenamefont {Mishra}\ \emph {et~al.}(2016)\citenamefont {Mishra},
  \citenamefont {Boeckl}, \citenamefont {Motta},\ and\ \citenamefont
  {Iacopi}}]{mishra2016graphene}%
  \BibitemOpen
  \bibfield  {author} {\bibinfo {author} {\bibfnamefont {N.}~\bibnamefont
  {Mishra}}, \bibinfo {author} {\bibfnamefont {J.}~\bibnamefont {Boeckl}},
  \bibinfo {author} {\bibfnamefont {N.}~\bibnamefont {Motta}}, \ and\ \bibinfo
  {author} {\bibfnamefont {F.}~\bibnamefont {Iacopi}},\ }\href@noop {}
  {\bibfield  {journal} {\bibinfo  {journal} {physica status solidi (a)}\
  }\textbf {\bibinfo {volume} {213}},\ \bibinfo {pages} {2277} (\bibinfo {year}
  {2016})}\BibitemShut {NoStop}%
\bibitem [{\citenamefont {Ellis}\ \emph {et~al.}(2011)\citenamefont {Ellis},
  \citenamefont {Mayer}, \citenamefont {Shambat}, \citenamefont {Sarmiento},
  \citenamefont {Harris}, \citenamefont {Haller},\ and\ \citenamefont
  {Vu{\v{c}}kovi{\'c}}}]{ellis2011ultralow}%
  \BibitemOpen
  \bibfield  {author} {\bibinfo {author} {\bibfnamefont {B.}~\bibnamefont
  {Ellis}}, \bibinfo {author} {\bibfnamefont {M.~A.}\ \bibnamefont {Mayer}},
  \bibinfo {author} {\bibfnamefont {G.}~\bibnamefont {Shambat}}, \bibinfo
  {author} {\bibfnamefont {T.}~\bibnamefont {Sarmiento}}, \bibinfo {author}
  {\bibfnamefont {J.}~\bibnamefont {Harris}}, \bibinfo {author} {\bibfnamefont
  {E.~E.}\ \bibnamefont {Haller}}, \ and\ \bibinfo {author} {\bibfnamefont
  {J.}~\bibnamefont {Vu{\v{c}}kovi{\'c}}},\ }\href@noop {} {\bibfield
  {journal} {\bibinfo  {journal} {Nature photonics}\ }\textbf {\bibinfo
  {volume} {5}},\ \bibinfo {pages} {297} (\bibinfo {year} {2011})}\BibitemShut
  {NoStop}%
\bibitem [{\citenamefont {Rao}\ \emph {et~al.}(1999)\citenamefont {Rao},
  \citenamefont {Tucker}, \citenamefont {Holland}, \citenamefont
  {Papanicolaou}, \citenamefont {Chi}, \citenamefont {Kretchmer},\ and\
  \citenamefont {Ghezzo}}]{rao1999donor}%
  \BibitemOpen
  \bibfield  {author} {\bibinfo {author} {\bibfnamefont {M.~V.}\ \bibnamefont
  {Rao}}, \bibinfo {author} {\bibfnamefont {J.}~\bibnamefont {Tucker}},
  \bibinfo {author} {\bibfnamefont {O.}~\bibnamefont {Holland}}, \bibinfo
  {author} {\bibfnamefont {N.}~\bibnamefont {Papanicolaou}}, \bibinfo {author}
  {\bibfnamefont {P.}~\bibnamefont {Chi}}, \bibinfo {author} {\bibfnamefont
  {J.}~\bibnamefont {Kretchmer}}, \ and\ \bibinfo {author} {\bibfnamefont
  {M.}~\bibnamefont {Ghezzo}},\ }\href@noop {} {\bibfield  {journal} {\bibinfo
  {journal} {Journal of electronic materials}\ }\textbf {\bibinfo {volume}
  {28}},\ \bibinfo {pages} {334} (\bibinfo {year} {1999})}\BibitemShut
  {NoStop}%
\bibitem [{\citenamefont {Fotso}\ \emph {et~al.}(2016)\citenamefont {Fotso},
  \citenamefont {Feiguin}, \citenamefont {Awschalom},\ and\ \citenamefont
  {Dobrovitski}}]{fotso2016suppressing}%
  \BibitemOpen
  \bibfield  {author} {\bibinfo {author} {\bibfnamefont {H.}~\bibnamefont
  {Fotso}}, \bibinfo {author} {\bibfnamefont {A.}~\bibnamefont {Feiguin}},
  \bibinfo {author} {\bibfnamefont {D.}~\bibnamefont {Awschalom}}, \ and\
  \bibinfo {author} {\bibfnamefont {V.}~\bibnamefont {Dobrovitski}},\
  }\href@noop {} {\bibfield  {journal} {\bibinfo  {journal} {Physical Review
  Letters}\ }\textbf {\bibinfo {volume} {116}},\ \bibinfo {pages} {033603}
  (\bibinfo {year} {2016})}\BibitemShut {NoStop}%
\bibitem [{\citenamefont {Bluvstein}\ \emph {et~al.}(2019)\citenamefont
  {Bluvstein}, \citenamefont {Zhang}, \citenamefont {McLellan}, \citenamefont
  {Williams},\ and\ \citenamefont {Jayich}}]{bluvstein2019extending}%
  \BibitemOpen
  \bibfield  {author} {\bibinfo {author} {\bibfnamefont {D.}~\bibnamefont
  {Bluvstein}}, \bibinfo {author} {\bibfnamefont {Z.}~\bibnamefont {Zhang}},
  \bibinfo {author} {\bibfnamefont {C.~A.}\ \bibnamefont {McLellan}}, \bibinfo
  {author} {\bibfnamefont {N.~R.}\ \bibnamefont {Williams}}, \ and\ \bibinfo
  {author} {\bibfnamefont {A.~C.~B.}\ \bibnamefont {Jayich}},\ }\href@noop {}
  {\bibfield  {journal} {\bibinfo  {journal} {Physical Review Letters}\
  }\textbf {\bibinfo {volume} {123}},\ \bibinfo {pages} {146804} (\bibinfo
  {year} {2019})}\BibitemShut {NoStop}%
\bibitem [{\citenamefont {Tran}\ \emph {et~al.}(2019)\citenamefont {Tran},
  \citenamefont {Bradac}, \citenamefont {Solntsev}, \citenamefont {Toth},\ and\
  \citenamefont {Aharonovich}}]{tran2019suppression}%
  \BibitemOpen
  \bibfield  {author} {\bibinfo {author} {\bibfnamefont {T.~T.}\ \bibnamefont
  {Tran}}, \bibinfo {author} {\bibfnamefont {C.}~\bibnamefont {Bradac}},
  \bibinfo {author} {\bibfnamefont {A.~S.}\ \bibnamefont {Solntsev}}, \bibinfo
  {author} {\bibfnamefont {M.}~\bibnamefont {Toth}}, \ and\ \bibinfo {author}
  {\bibfnamefont {I.}~\bibnamefont {Aharonovich}},\ }\href@noop {} {\bibfield
  {journal} {\bibinfo  {journal} {Applied Physics Letters}\ }\textbf {\bibinfo
  {volume} {115}},\ \bibinfo {pages} {071102} (\bibinfo {year}
  {2019})}\BibitemShut {NoStop}%
\bibitem [{\citenamefont {Yang}\ \emph {et~al.}(2007)\citenamefont {Yang},
  \citenamefont {Chen}, \citenamefont {Husko},\ and\ \citenamefont
  {Wong}}]{yang2007digital}%
  \BibitemOpen
  \bibfield  {author} {\bibinfo {author} {\bibfnamefont {X.}~\bibnamefont
  {Yang}}, \bibinfo {author} {\bibfnamefont {C.~J.}\ \bibnamefont {Chen}},
  \bibinfo {author} {\bibfnamefont {C.~A.}\ \bibnamefont {Husko}}, \ and\
  \bibinfo {author} {\bibfnamefont {C.~W.}\ \bibnamefont {Wong}},\ }\href@noop
  {} {\bibfield  {journal} {\bibinfo  {journal} {Applied Physics Letters}\
  }\textbf {\bibinfo {volume} {91}},\ \bibinfo {pages} {161114} (\bibinfo
  {year} {2007})}\BibitemShut {NoStop}%
\bibitem [{\citenamefont {Kiravittaya}\ \emph {et~al.}(2011)\citenamefont
  {Kiravittaya}, \citenamefont {Lee}, \citenamefont {Balet}, \citenamefont
  {Li}, \citenamefont {Francardi}, \citenamefont {Gerardino}, \citenamefont
  {Fiore}, \citenamefont {Rastelli},\ and\ \citenamefont
  {Schmidt}}]{kiravittaya2011tuning}%
  \BibitemOpen
  \bibfield  {author} {\bibinfo {author} {\bibfnamefont {S.}~\bibnamefont
  {Kiravittaya}}, \bibinfo {author} {\bibfnamefont {H.}~\bibnamefont {Lee}},
  \bibinfo {author} {\bibfnamefont {L.}~\bibnamefont {Balet}}, \bibinfo
  {author} {\bibfnamefont {L.}~\bibnamefont {Li}}, \bibinfo {author}
  {\bibfnamefont {M.}~\bibnamefont {Francardi}}, \bibinfo {author}
  {\bibfnamefont {A.}~\bibnamefont {Gerardino}}, \bibinfo {author}
  {\bibfnamefont {A.}~\bibnamefont {Fiore}}, \bibinfo {author} {\bibfnamefont
  {A.}~\bibnamefont {Rastelli}}, \ and\ \bibinfo {author} {\bibfnamefont
  {O.}~\bibnamefont {Schmidt}},\ }\href@noop {} {\bibfield  {journal} {\bibinfo
   {journal} {Journal of Applied Physics}\ }\textbf {\bibinfo {volume} {109}},\
  \bibinfo {pages} {053115} (\bibinfo {year} {2011})}\BibitemShut {NoStop}%
\bibitem [{\citenamefont {Faraon}\ \emph {et~al.}(2008)\citenamefont {Faraon},
  \citenamefont {Englund}, \citenamefont {Bulla}, \citenamefont
  {Luther-Davies}, \citenamefont {Eggleton}, \citenamefont {Stoltz},
  \citenamefont {Petroff},\ and\ \citenamefont
  {Vu{\v{c}}kovi{\'c}}}]{faraon2008local}%
  \BibitemOpen
  \bibfield  {author} {\bibinfo {author} {\bibfnamefont {A.}~\bibnamefont
  {Faraon}}, \bibinfo {author} {\bibfnamefont {D.}~\bibnamefont {Englund}},
  \bibinfo {author} {\bibfnamefont {D.}~\bibnamefont {Bulla}}, \bibinfo
  {author} {\bibfnamefont {B.}~\bibnamefont {Luther-Davies}}, \bibinfo {author}
  {\bibfnamefont {B.~J.}\ \bibnamefont {Eggleton}}, \bibinfo {author}
  {\bibfnamefont {N.}~\bibnamefont {Stoltz}}, \bibinfo {author} {\bibfnamefont
  {P.}~\bibnamefont {Petroff}}, \ and\ \bibinfo {author} {\bibfnamefont
  {J.}~\bibnamefont {Vu{\v{c}}kovi{\'c}}},\ }\href@noop {} {\bibfield
  {journal} {\bibinfo  {journal} {Applied Physics Letters}\ }\textbf {\bibinfo
  {volume} {92}},\ \bibinfo {pages} {043123} (\bibinfo {year}
  {2008})}\BibitemShut {NoStop}%
\bibitem [{\citenamefont {Meesala}\ \emph {et~al.}(2018)\citenamefont
  {Meesala}, \citenamefont {Sohn}, \citenamefont {Pingault}, \citenamefont
  {Shao}, \citenamefont {Atikian}, \citenamefont {Holzgrafe}, \citenamefont
  {G{\"u}ndo{\u{g}}an}, \citenamefont {Stavrakas}, \citenamefont {Sipahigil},
  \citenamefont {Chia} \emph {et~al.}}]{meesala2018strain}%
  \BibitemOpen
  \bibfield  {author} {\bibinfo {author} {\bibfnamefont {S.}~\bibnamefont
  {Meesala}}, \bibinfo {author} {\bibfnamefont {Y.-I.}\ \bibnamefont {Sohn}},
  \bibinfo {author} {\bibfnamefont {B.}~\bibnamefont {Pingault}}, \bibinfo
  {author} {\bibfnamefont {L.}~\bibnamefont {Shao}}, \bibinfo {author}
  {\bibfnamefont {H.~A.}\ \bibnamefont {Atikian}}, \bibinfo {author}
  {\bibfnamefont {J.}~\bibnamefont {Holzgrafe}}, \bibinfo {author}
  {\bibfnamefont {M.}~\bibnamefont {G{\"u}ndo{\u{g}}an}}, \bibinfo {author}
  {\bibfnamefont {C.}~\bibnamefont {Stavrakas}}, \bibinfo {author}
  {\bibfnamefont {A.}~\bibnamefont {Sipahigil}}, \bibinfo {author}
  {\bibfnamefont {C.}~\bibnamefont {Chia}},  \emph {et~al.},\ }\href@noop {}
  {\bibfield  {journal} {\bibinfo  {journal} {Physical Review B}\ }\textbf
  {\bibinfo {volume} {97}},\ \bibinfo {pages} {205444} (\bibinfo {year}
  {2018})}\BibitemShut {NoStop}%
\bibitem [{\citenamefont {Sun}\ \emph {et~al.}(2018)\citenamefont {Sun},
  \citenamefont {Zhang}, \citenamefont {Fischer}, \citenamefont {Burek},
  \citenamefont {Dory}, \citenamefont {Lagoudakis}, \citenamefont {Tzeng},
  \citenamefont {Radulaski}, \citenamefont {Kelaita}, \citenamefont
  {Safavi-Naeini} \emph {et~al.}}]{sun2018cavity}%
  \BibitemOpen
  \bibfield  {author} {\bibinfo {author} {\bibfnamefont {S.}~\bibnamefont
  {Sun}}, \bibinfo {author} {\bibfnamefont {J.~L.}\ \bibnamefont {Zhang}},
  \bibinfo {author} {\bibfnamefont {K.~A.}\ \bibnamefont {Fischer}}, \bibinfo
  {author} {\bibfnamefont {M.~J.}\ \bibnamefont {Burek}}, \bibinfo {author}
  {\bibfnamefont {C.}~\bibnamefont {Dory}}, \bibinfo {author} {\bibfnamefont
  {K.~G.}\ \bibnamefont {Lagoudakis}}, \bibinfo {author} {\bibfnamefont
  {Y.-K.}\ \bibnamefont {Tzeng}}, \bibinfo {author} {\bibfnamefont
  {M.}~\bibnamefont {Radulaski}}, \bibinfo {author} {\bibfnamefont
  {Y.}~\bibnamefont {Kelaita}}, \bibinfo {author} {\bibfnamefont
  {A.}~\bibnamefont {Safavi-Naeini}},  \emph {et~al.},\ }\href@noop {}
  {\bibfield  {journal} {\bibinfo  {journal} {Physical review letters}\
  }\textbf {\bibinfo {volume} {121}},\ \bibinfo {pages} {083601} (\bibinfo
  {year} {2018})}\BibitemShut {NoStop}%
\bibitem [{\citenamefont {Nguyen}\ \emph {et~al.}(2019)\citenamefont {Nguyen},
  \citenamefont {Sukachev}, \citenamefont {Bhaskar}, \citenamefont {Machielse},
  \citenamefont {Levonian}, \citenamefont {Knall}, \citenamefont {Stroganov},
  \citenamefont {Riedinger}, \citenamefont {Park}, \citenamefont {Lon{\v{c}}ar}
  \emph {et~al.}}]{nguyen2019quantum}%
  \BibitemOpen
  \bibfield  {author} {\bibinfo {author} {\bibfnamefont {C.}~\bibnamefont
  {Nguyen}}, \bibinfo {author} {\bibfnamefont {D.}~\bibnamefont {Sukachev}},
  \bibinfo {author} {\bibfnamefont {M.}~\bibnamefont {Bhaskar}}, \bibinfo
  {author} {\bibfnamefont {B.}~\bibnamefont {Machielse}}, \bibinfo {author}
  {\bibfnamefont {D.}~\bibnamefont {Levonian}}, \bibinfo {author}
  {\bibfnamefont {E.}~\bibnamefont {Knall}}, \bibinfo {author} {\bibfnamefont
  {P.}~\bibnamefont {Stroganov}}, \bibinfo {author} {\bibfnamefont
  {R.}~\bibnamefont {Riedinger}}, \bibinfo {author} {\bibfnamefont
  {H.}~\bibnamefont {Park}}, \bibinfo {author} {\bibfnamefont {M.}~\bibnamefont
  {Lon{\v{c}}ar}},  \emph {et~al.},\ }\href@noop {} {\bibfield  {journal}
  {\bibinfo  {journal} {Physical review letters}\ }\textbf {\bibinfo {volume}
  {123}},\ \bibinfo {pages} {183602} (\bibinfo {year} {2019})}\BibitemShut
  {NoStop}%
\bibitem [{\citenamefont {Riedel}\ \emph {et~al.}(2017)\citenamefont {Riedel},
  \citenamefont {S{\"o}llner}, \citenamefont {Shields}, \citenamefont
  {Starosielec}, \citenamefont {Appel}, \citenamefont {Neu}, \citenamefont
  {Maletinsky},\ and\ \citenamefont {Warburton}}]{riedel2017deterministic}%
  \BibitemOpen
  \bibfield  {author} {\bibinfo {author} {\bibfnamefont {D.}~\bibnamefont
  {Riedel}}, \bibinfo {author} {\bibfnamefont {I.}~\bibnamefont {S{\"o}llner}},
  \bibinfo {author} {\bibfnamefont {B.~J.}\ \bibnamefont {Shields}}, \bibinfo
  {author} {\bibfnamefont {S.}~\bibnamefont {Starosielec}}, \bibinfo {author}
  {\bibfnamefont {P.}~\bibnamefont {Appel}}, \bibinfo {author} {\bibfnamefont
  {E.}~\bibnamefont {Neu}}, \bibinfo {author} {\bibfnamefont {P.}~\bibnamefont
  {Maletinsky}}, \ and\ \bibinfo {author} {\bibfnamefont {R.~J.}\ \bibnamefont
  {Warburton}},\ }\href@noop {} {\bibfield  {journal} {\bibinfo  {journal}
  {Physical Review X}\ }\textbf {\bibinfo {volume} {7}},\ \bibinfo {pages}
  {031040} (\bibinfo {year} {2017})}\BibitemShut {NoStop}%
\bibitem [{\citenamefont {Hunger}\ \emph {et~al.}(2010)\citenamefont {Hunger},
  \citenamefont {Steinmetz}, \citenamefont {Colombe}, \citenamefont {Deutsch},
  \citenamefont {H{\"a}nsch},\ and\ \citenamefont {Reichel}}]{hunger2010fiber}%
  \BibitemOpen
  \bibfield  {author} {\bibinfo {author} {\bibfnamefont {D.}~\bibnamefont
  {Hunger}}, \bibinfo {author} {\bibfnamefont {T.}~\bibnamefont {Steinmetz}},
  \bibinfo {author} {\bibfnamefont {Y.}~\bibnamefont {Colombe}}, \bibinfo
  {author} {\bibfnamefont {C.}~\bibnamefont {Deutsch}}, \bibinfo {author}
  {\bibfnamefont {T.~W.}\ \bibnamefont {H{\"a}nsch}}, \ and\ \bibinfo {author}
  {\bibfnamefont {J.}~\bibnamefont {Reichel}},\ }\href@noop {} {\bibfield
  {journal} {\bibinfo  {journal} {New Journal of Physics}\ }\textbf {\bibinfo
  {volume} {12}},\ \bibinfo {pages} {065038} (\bibinfo {year}
  {2010})}\BibitemShut {NoStop}%
\bibitem [{\citenamefont {Barredo}\ \emph {et~al.}(2016)\citenamefont
  {Barredo}, \citenamefont {De~L{\'e}s{\'e}leuc}, \citenamefont {Lienhard},
  \citenamefont {Lahaye},\ and\ \citenamefont {Browaeys}}]{barredo2016atom}%
  \BibitemOpen
  \bibfield  {author} {\bibinfo {author} {\bibfnamefont {D.}~\bibnamefont
  {Barredo}}, \bibinfo {author} {\bibfnamefont {S.}~\bibnamefont
  {De~L{\'e}s{\'e}leuc}}, \bibinfo {author} {\bibfnamefont {V.}~\bibnamefont
  {Lienhard}}, \bibinfo {author} {\bibfnamefont {T.}~\bibnamefont {Lahaye}}, \
  and\ \bibinfo {author} {\bibfnamefont {A.}~\bibnamefont {Browaeys}},\
  }\href@noop {} {\bibfield  {journal} {\bibinfo  {journal} {Science}\ }\textbf
  {\bibinfo {volume} {354}},\ \bibinfo {pages} {1021} (\bibinfo {year}
  {2016})}\BibitemShut {NoStop}%
\bibitem [{\citenamefont {Najer}\ \emph {et~al.}(2019)\citenamefont {Najer},
  \citenamefont {S{\"o}llner}, \citenamefont {Sekatski}, \citenamefont
  {Dolique}, \citenamefont {L{\"o}bl}, \citenamefont {Riedel}, \citenamefont
  {Schott}, \citenamefont {Starosielec}, \citenamefont {Valentin},
  \citenamefont {Wieck} \emph {et~al.}}]{najer2019gated}%
  \BibitemOpen
  \bibfield  {author} {\bibinfo {author} {\bibfnamefont {D.}~\bibnamefont
  {Najer}}, \bibinfo {author} {\bibfnamefont {I.}~\bibnamefont {S{\"o}llner}},
  \bibinfo {author} {\bibfnamefont {P.}~\bibnamefont {Sekatski}}, \bibinfo
  {author} {\bibfnamefont {V.}~\bibnamefont {Dolique}}, \bibinfo {author}
  {\bibfnamefont {M.~C.}\ \bibnamefont {L{\"o}bl}}, \bibinfo {author}
  {\bibfnamefont {D.}~\bibnamefont {Riedel}}, \bibinfo {author} {\bibfnamefont
  {R.}~\bibnamefont {Schott}}, \bibinfo {author} {\bibfnamefont
  {S.}~\bibnamefont {Starosielec}}, \bibinfo {author} {\bibfnamefont {S.~R.}\
  \bibnamefont {Valentin}}, \bibinfo {author} {\bibfnamefont {A.~D.}\
  \bibnamefont {Wieck}},  \emph {et~al.},\ }\href@noop {} {\bibfield  {journal}
  {\bibinfo  {journal} {Nature}\ ,\ \bibinfo {pages} {1}} (\bibinfo {year}
  {2019})}\BibitemShut {NoStop}%
\bibitem [{\citenamefont {Hunger}\ \emph {et~al.}(2012)\citenamefont {Hunger},
  \citenamefont {Deutsch}, \citenamefont {Barbour}, \citenamefont {Warburton},\
  and\ \citenamefont {Reichel}}]{hunger2012laser}%
  \BibitemOpen
  \bibfield  {author} {\bibinfo {author} {\bibfnamefont {D.}~\bibnamefont
  {Hunger}}, \bibinfo {author} {\bibfnamefont {C.}~\bibnamefont {Deutsch}},
  \bibinfo {author} {\bibfnamefont {R.~J.}\ \bibnamefont {Barbour}}, \bibinfo
  {author} {\bibfnamefont {R.~J.}\ \bibnamefont {Warburton}}, \ and\ \bibinfo
  {author} {\bibfnamefont {J.}~\bibnamefont {Reichel}},\ }\href@noop {}
  {\bibfield  {journal} {\bibinfo  {journal} {Aip Advances}\ }\textbf {\bibinfo
  {volume} {2}},\ \bibinfo {pages} {012119} (\bibinfo {year}
  {2012})}\BibitemShut {NoStop}%
\bibitem [{\citenamefont {H{\"a}u{\ss}ler}\ \emph {et~al.}(2019)\citenamefont
  {H{\"a}u{\ss}ler}, \citenamefont {Benedikter}, \citenamefont {Bray},
  \citenamefont {Regan}, \citenamefont {Dietrich}, \citenamefont {Twamley},
  \citenamefont {Aharonovich}, \citenamefont {Hunger},\ and\ \citenamefont
  {Kubanek}}]{haussler2019diamond}%
  \BibitemOpen
  \bibfield  {author} {\bibinfo {author} {\bibfnamefont {S.}~\bibnamefont
  {H{\"a}u{\ss}ler}}, \bibinfo {author} {\bibfnamefont {J.}~\bibnamefont
  {Benedikter}}, \bibinfo {author} {\bibfnamefont {K.}~\bibnamefont {Bray}},
  \bibinfo {author} {\bibfnamefont {B.}~\bibnamefont {Regan}}, \bibinfo
  {author} {\bibfnamefont {A.}~\bibnamefont {Dietrich}}, \bibinfo {author}
  {\bibfnamefont {J.}~\bibnamefont {Twamley}}, \bibinfo {author} {\bibfnamefont
  {I.}~\bibnamefont {Aharonovich}}, \bibinfo {author} {\bibfnamefont
  {D.}~\bibnamefont {Hunger}}, \ and\ \bibinfo {author} {\bibfnamefont
  {A.}~\bibnamefont {Kubanek}},\ }\href@noop {} {\bibfield  {journal} {\bibinfo
   {journal} {Physical Review B}\ }\textbf {\bibinfo {volume} {99}},\ \bibinfo
  {pages} {165310} (\bibinfo {year} {2019})}\BibitemShut {NoStop}%
\bibitem [{\citenamefont {Merkel}\ \emph {et~al.}(2020)\citenamefont {Merkel},
  \citenamefont {Ulanowski},\ and\ \citenamefont
  {Reiserer}}]{merkel2020coherent}%
  \BibitemOpen
  \bibfield  {author} {\bibinfo {author} {\bibfnamefont {B.}~\bibnamefont
  {Merkel}}, \bibinfo {author} {\bibfnamefont {A.}~\bibnamefont {Ulanowski}}, \
  and\ \bibinfo {author} {\bibfnamefont {A.}~\bibnamefont {Reiserer}},\
  }\href@noop {} {\bibfield  {journal} {\bibinfo  {journal} {arXiv preprint
  arXiv:2006.14229}\ } (\bibinfo {year} {2020})}\BibitemShut {NoStop}%
\bibitem [{\citenamefont {Tomm}\ \emph {et~al.}(2020)\citenamefont {Tomm},
  \citenamefont {Javadi}, \citenamefont {Antoniadis}, \citenamefont {Najer},
  \citenamefont {L{\"o}bl}, \citenamefont {Korsch}, \citenamefont {Schott},
  \citenamefont {Valentin}, \citenamefont {Wieck}, \citenamefont {Ludwig} \emph
  {et~al.}}]{tomm2020bright}%
  \BibitemOpen
  \bibfield  {author} {\bibinfo {author} {\bibfnamefont {N.}~\bibnamefont
  {Tomm}}, \bibinfo {author} {\bibfnamefont {A.}~\bibnamefont {Javadi}},
  \bibinfo {author} {\bibfnamefont {N.~O.}\ \bibnamefont {Antoniadis}},
  \bibinfo {author} {\bibfnamefont {D.}~\bibnamefont {Najer}}, \bibinfo
  {author} {\bibfnamefont {M.~C.}\ \bibnamefont {L{\"o}bl}}, \bibinfo {author}
  {\bibfnamefont {A.~R.}\ \bibnamefont {Korsch}}, \bibinfo {author}
  {\bibfnamefont {R.}~\bibnamefont {Schott}}, \bibinfo {author} {\bibfnamefont
  {S.~R.}\ \bibnamefont {Valentin}}, \bibinfo {author} {\bibfnamefont {A.~D.}\
  \bibnamefont {Wieck}}, \bibinfo {author} {\bibfnamefont {A.}~\bibnamefont
  {Ludwig}},  \emph {et~al.},\ }\href@noop {} {\bibfield  {journal} {\bibinfo
  {journal} {arXiv preprint arXiv:2007.12654}\ } (\bibinfo {year}
  {2020})}\BibitemShut {NoStop}%
\bibitem [{\citenamefont {Singaravelu}\ \emph {et~al.}(2019)\citenamefont
  {Singaravelu}, \citenamefont {Devarapu}, \citenamefont {Schulz},
  \citenamefont {Wilmart}, \citenamefont {Malhouitre}, \citenamefont {Olivier}
  \emph {et~al.}}]{singaravelu2019low}%
  \BibitemOpen
  \bibfield  {author} {\bibinfo {author} {\bibfnamefont {P.}~\bibnamefont
  {Singaravelu}}, \bibinfo {author} {\bibfnamefont {G.}~\bibnamefont
  {Devarapu}}, \bibinfo {author} {\bibfnamefont {S.~A.}\ \bibnamefont
  {Schulz}}, \bibinfo {author} {\bibfnamefont {Q.}~\bibnamefont {Wilmart}},
  \bibinfo {author} {\bibfnamefont {S.}~\bibnamefont {Malhouitre}}, \bibinfo
  {author} {\bibfnamefont {S.}~\bibnamefont {Olivier}},  \emph {et~al.},\
  }\href@noop {} {\bibfield  {journal} {\bibinfo  {journal} {Journal of Physics
  D: Applied Physics}\ }\textbf {\bibinfo {volume} {52}},\ \bibinfo {pages}
  {214001} (\bibinfo {year} {2019})}\BibitemShut {NoStop}%
\bibitem [{\citenamefont {Pernice}\ \emph {et~al.}(2012)\citenamefont
  {Pernice}, \citenamefont {Schuck}, \citenamefont {Minaeva}, \citenamefont
  {Li}, \citenamefont {Goltsman}, \citenamefont {Sergienko},\ and\
  \citenamefont {Tang}}]{pernice2012high}%
  \BibitemOpen
  \bibfield  {author} {\bibinfo {author} {\bibfnamefont {W.~H.}\ \bibnamefont
  {Pernice}}, \bibinfo {author} {\bibfnamefont {C.}~\bibnamefont {Schuck}},
  \bibinfo {author} {\bibfnamefont {O.}~\bibnamefont {Minaeva}}, \bibinfo
  {author} {\bibfnamefont {M.}~\bibnamefont {Li}}, \bibinfo {author}
  {\bibfnamefont {G.}~\bibnamefont {Goltsman}}, \bibinfo {author}
  {\bibfnamefont {A.}~\bibnamefont {Sergienko}}, \ and\ \bibinfo {author}
  {\bibfnamefont {H.}~\bibnamefont {Tang}},\ }\href@noop {} {\bibfield
  {journal} {\bibinfo  {journal} {Nature communications}\ }\textbf {\bibinfo
  {volume} {3}},\ \bibinfo {pages} {1} (\bibinfo {year} {2012})}\BibitemShut
  {NoStop}%
\bibitem [{\citenamefont {Najafi}\ \emph {et~al.}(2015)\citenamefont {Najafi},
  \citenamefont {Mower}, \citenamefont {Harris}, \citenamefont {Bellei},
  \citenamefont {Dane}, \citenamefont {Lee}, \citenamefont {Hu}, \citenamefont
  {Kharel}, \citenamefont {Marsili}, \citenamefont {Assefa} \emph
  {et~al.}}]{najafi2015chip}%
  \BibitemOpen
  \bibfield  {author} {\bibinfo {author} {\bibfnamefont {F.}~\bibnamefont
  {Najafi}}, \bibinfo {author} {\bibfnamefont {J.}~\bibnamefont {Mower}},
  \bibinfo {author} {\bibfnamefont {N.~C.}\ \bibnamefont {Harris}}, \bibinfo
  {author} {\bibfnamefont {F.}~\bibnamefont {Bellei}}, \bibinfo {author}
  {\bibfnamefont {A.}~\bibnamefont {Dane}}, \bibinfo {author} {\bibfnamefont
  {C.}~\bibnamefont {Lee}}, \bibinfo {author} {\bibfnamefont {X.}~\bibnamefont
  {Hu}}, \bibinfo {author} {\bibfnamefont {P.}~\bibnamefont {Kharel}}, \bibinfo
  {author} {\bibfnamefont {F.}~\bibnamefont {Marsili}}, \bibinfo {author}
  {\bibfnamefont {S.}~\bibnamefont {Assefa}},  \emph {et~al.},\ }\href@noop {}
  {\bibfield  {journal} {\bibinfo  {journal} {Nature communications}\ }\textbf
  {\bibinfo {volume} {6}},\ \bibinfo {pages} {1} (\bibinfo {year}
  {2015})}\BibitemShut {NoStop}%
\bibitem [{\citenamefont {Eltes}\ \emph {et~al.}(2020)\citenamefont {Eltes},
  \citenamefont {Villarreal-Garcia}, \citenamefont {Caimi}, \citenamefont
  {Siegwart}, \citenamefont {Gentile}, \citenamefont {Hart}, \citenamefont
  {Stark}, \citenamefont {Marshall}, \citenamefont {Thompson}, \citenamefont
  {Barreto} \emph {et~al.}}]{eltes2020integrated}%
  \BibitemOpen
  \bibfield  {author} {\bibinfo {author} {\bibfnamefont {F.}~\bibnamefont
  {Eltes}}, \bibinfo {author} {\bibfnamefont {G.~E.}\ \bibnamefont
  {Villarreal-Garcia}}, \bibinfo {author} {\bibfnamefont {D.}~\bibnamefont
  {Caimi}}, \bibinfo {author} {\bibfnamefont {H.}~\bibnamefont {Siegwart}},
  \bibinfo {author} {\bibfnamefont {A.~A.}\ \bibnamefont {Gentile}}, \bibinfo
  {author} {\bibfnamefont {A.}~\bibnamefont {Hart}}, \bibinfo {author}
  {\bibfnamefont {P.}~\bibnamefont {Stark}}, \bibinfo {author} {\bibfnamefont
  {G.~D.}\ \bibnamefont {Marshall}}, \bibinfo {author} {\bibfnamefont {M.~G.}\
  \bibnamefont {Thompson}}, \bibinfo {author} {\bibfnamefont {J.}~\bibnamefont
  {Barreto}},  \emph {et~al.},\ }\href {\doibase
  https://doi.org/10.1038/s41563-020-0725-5} {\bibfield  {journal} {\bibinfo
  {journal} {Nature Materials}\ ,\ \bibinfo {pages} {1}} (\bibinfo {year}
  {2020})}\BibitemShut {NoStop}%
\bibitem [{\citenamefont {Piggott}\ \emph {et~al.}(2015)\citenamefont
  {Piggott}, \citenamefont {Lu}, \citenamefont {Lagoudakis}, \citenamefont
  {Petykiewicz}, \citenamefont {Babinec},\ and\ \citenamefont
  {Vu{\v{c}}kovi{\'c}}}]{piggott2015inverse}%
  \BibitemOpen
  \bibfield  {author} {\bibinfo {author} {\bibfnamefont {A.~Y.}\ \bibnamefont
  {Piggott}}, \bibinfo {author} {\bibfnamefont {J.}~\bibnamefont {Lu}},
  \bibinfo {author} {\bibfnamefont {K.~G.}\ \bibnamefont {Lagoudakis}},
  \bibinfo {author} {\bibfnamefont {J.}~\bibnamefont {Petykiewicz}}, \bibinfo
  {author} {\bibfnamefont {T.~M.}\ \bibnamefont {Babinec}}, \ and\ \bibinfo
  {author} {\bibfnamefont {J.}~\bibnamefont {Vu{\v{c}}kovi{\'c}}},\ }\href@noop
  {} {\bibfield  {journal} {\bibinfo  {journal} {Nature Photonics}\ }\textbf
  {\bibinfo {volume} {9}},\ \bibinfo {pages} {374} (\bibinfo {year}
  {2015})}\BibitemShut {NoStop}%
\bibitem [{\citenamefont {Molesky}\ \emph {et~al.}(2018)\citenamefont
  {Molesky}, \citenamefont {Lin}, \citenamefont {Piggott}, \citenamefont {Jin},
  \citenamefont {Vuckovi{\'c}},\ and\ \citenamefont
  {Rodriguez}}]{molesky2018inverse}%
  \BibitemOpen
  \bibfield  {author} {\bibinfo {author} {\bibfnamefont {S.}~\bibnamefont
  {Molesky}}, \bibinfo {author} {\bibfnamefont {Z.}~\bibnamefont {Lin}},
  \bibinfo {author} {\bibfnamefont {A.~Y.}\ \bibnamefont {Piggott}}, \bibinfo
  {author} {\bibfnamefont {W.}~\bibnamefont {Jin}}, \bibinfo {author}
  {\bibfnamefont {J.}~\bibnamefont {Vuckovi{\'c}}}, \ and\ \bibinfo {author}
  {\bibfnamefont {A.~W.}\ \bibnamefont {Rodriguez}},\ }\href@noop {} {\bibfield
   {journal} {\bibinfo  {journal} {Nature Photonics}\ }\textbf {\bibinfo
  {volume} {12}},\ \bibinfo {pages} {659} (\bibinfo {year} {2018})}\BibitemShut
  {NoStop}%
\bibitem [{\citenamefont {Su}\ \emph {et~al.}(2020)\citenamefont {Su},
  \citenamefont {Vercruysse}, \citenamefont {Skarda}, \citenamefont {Sapra},
  \citenamefont {Petykiewicz},\ and\ \citenamefont
  {Vu{\v{c}}kovi{\'c}}}]{su2020nanophotonic}%
  \BibitemOpen
  \bibfield  {author} {\bibinfo {author} {\bibfnamefont {L.}~\bibnamefont
  {Su}}, \bibinfo {author} {\bibfnamefont {D.}~\bibnamefont {Vercruysse}},
  \bibinfo {author} {\bibfnamefont {J.}~\bibnamefont {Skarda}}, \bibinfo
  {author} {\bibfnamefont {N.~V.}\ \bibnamefont {Sapra}}, \bibinfo {author}
  {\bibfnamefont {J.~A.}\ \bibnamefont {Petykiewicz}}, \ and\ \bibinfo {author}
  {\bibfnamefont {J.}~\bibnamefont {Vu{\v{c}}kovi{\'c}}},\ }\href@noop {}
  {\bibfield  {journal} {\bibinfo  {journal} {Applied Physics Reviews}\
  }\textbf {\bibinfo {volume} {7}},\ \bibinfo {pages} {011407} (\bibinfo {year}
  {2020})}\BibitemShut {NoStop}%
\bibitem [{\citenamefont {Piggott}\ \emph {et~al.}(2017)\citenamefont
  {Piggott}, \citenamefont {Petykiewicz}, \citenamefont {Su},\ and\
  \citenamefont {Vu{\v{c}}kovi{\'c}}}]{piggott2017fabrication}%
  \BibitemOpen
  \bibfield  {author} {\bibinfo {author} {\bibfnamefont {A.~Y.}\ \bibnamefont
  {Piggott}}, \bibinfo {author} {\bibfnamefont {J.}~\bibnamefont
  {Petykiewicz}}, \bibinfo {author} {\bibfnamefont {L.}~\bibnamefont {Su}}, \
  and\ \bibinfo {author} {\bibfnamefont {J.}~\bibnamefont
  {Vu{\v{c}}kovi{\'c}}},\ }\href@noop {} {\bibfield  {journal} {\bibinfo
  {journal} {Scientific reports}\ }\textbf {\bibinfo {volume} {7}},\ \bibinfo
  {pages} {1} (\bibinfo {year} {2017})}\BibitemShut {NoStop}%
\bibitem [{\citenamefont {Vercruysse}\ \emph {et~al.}(2019)\citenamefont
  {Vercruysse}, \citenamefont {Sapra}, \citenamefont {Su},\ and\ \citenamefont
  {Vuckovic}}]{vercruysse2019dispersion}%
  \BibitemOpen
  \bibfield  {author} {\bibinfo {author} {\bibfnamefont {D.}~\bibnamefont
  {Vercruysse}}, \bibinfo {author} {\bibfnamefont {N.~V.}\ \bibnamefont
  {Sapra}}, \bibinfo {author} {\bibfnamefont {L.}~\bibnamefont {Su}}, \ and\
  \bibinfo {author} {\bibfnamefont {J.}~\bibnamefont {Vuckovic}},\ }\href@noop
  {} {\bibfield  {journal} {\bibinfo  {journal} {IEEE Journal of Selected
  Topics in Quantum Electronics}\ }\textbf {\bibinfo {volume} {26}},\ \bibinfo
  {pages} {1} (\bibinfo {year} {2019})}\BibitemShut {NoStop}%
\bibitem [{\citenamefont {Sapra}\ \emph {et~al.}(2019)\citenamefont {Sapra},
  \citenamefont {Vercruysse}, \citenamefont {Su}, \citenamefont {Yang},
  \citenamefont {Skarda}, \citenamefont {Piggott},\ and\ \citenamefont
  {Vu{\v{c}}kovi{\'c}}}]{sapra2019inverse}%
  \BibitemOpen
  \bibfield  {author} {\bibinfo {author} {\bibfnamefont {N.~V.}\ \bibnamefont
  {Sapra}}, \bibinfo {author} {\bibfnamefont {D.}~\bibnamefont {Vercruysse}},
  \bibinfo {author} {\bibfnamefont {L.}~\bibnamefont {Su}}, \bibinfo {author}
  {\bibfnamefont {K.~Y.}\ \bibnamefont {Yang}}, \bibinfo {author}
  {\bibfnamefont {J.}~\bibnamefont {Skarda}}, \bibinfo {author} {\bibfnamefont
  {A.~Y.}\ \bibnamefont {Piggott}}, \ and\ \bibinfo {author} {\bibfnamefont
  {J.}~\bibnamefont {Vu{\v{c}}kovi{\'c}}},\ }\href@noop {} {\bibfield
  {journal} {\bibinfo  {journal} {IEEE Journal of Selected Topics in Quantum
  Electronics}\ }\textbf {\bibinfo {volume} {25}},\ \bibinfo {pages} {1}
  (\bibinfo {year} {2019})}\BibitemShut {NoStop}%
\bibitem [{\citenamefont {Dory}\ \emph {et~al.}(2019)\citenamefont {Dory},
  \citenamefont {Vercruysse}, \citenamefont {Yang}, \citenamefont {Sapra},
  \citenamefont {Rugar}, \citenamefont {Sun}, \citenamefont {Lukin},
  \citenamefont {Piggott}, \citenamefont {Zhang}, \citenamefont {Radulaski}
  \emph {et~al.}}]{dory2019inverse}%
  \BibitemOpen
  \bibfield  {author} {\bibinfo {author} {\bibfnamefont {C.}~\bibnamefont
  {Dory}}, \bibinfo {author} {\bibfnamefont {D.}~\bibnamefont {Vercruysse}},
  \bibinfo {author} {\bibfnamefont {K.~Y.}\ \bibnamefont {Yang}}, \bibinfo
  {author} {\bibfnamefont {N.~V.}\ \bibnamefont {Sapra}}, \bibinfo {author}
  {\bibfnamefont {A.~E.}\ \bibnamefont {Rugar}}, \bibinfo {author}
  {\bibfnamefont {S.}~\bibnamefont {Sun}}, \bibinfo {author} {\bibfnamefont
  {D.~M.}\ \bibnamefont {Lukin}}, \bibinfo {author} {\bibfnamefont {A.~Y.}\
  \bibnamefont {Piggott}}, \bibinfo {author} {\bibfnamefont {J.~L.}\
  \bibnamefont {Zhang}}, \bibinfo {author} {\bibfnamefont {M.}~\bibnamefont
  {Radulaski}},  \emph {et~al.},\ }\href@noop {} {\bibfield  {journal}
  {\bibinfo  {journal} {Nature communications}\ }\textbf {\bibinfo {volume}
  {10}},\ \bibinfo {pages} {1} (\bibinfo {year} {2019})}\BibitemShut {NoStop}%
\bibitem [{\citenamefont {Gonz{\'a}lez-Tudela}\ \emph
  {et~al.}(2015)\citenamefont {Gonz{\'a}lez-Tudela}, \citenamefont {Hung},
  \citenamefont {Chang}, \citenamefont {Cirac},\ and\ \citenamefont
  {Kimble}}]{gonzalez2015subwavelength}%
  \BibitemOpen
  \bibfield  {author} {\bibinfo {author} {\bibfnamefont {A.}~\bibnamefont
  {Gonz{\'a}lez-Tudela}}, \bibinfo {author} {\bibfnamefont {C.-L.}\
  \bibnamefont {Hung}}, \bibinfo {author} {\bibfnamefont {D.~E.}\ \bibnamefont
  {Chang}}, \bibinfo {author} {\bibfnamefont {J.~I.}\ \bibnamefont {Cirac}}, \
  and\ \bibinfo {author} {\bibfnamefont {H.}~\bibnamefont {Kimble}},\
  }\href@noop {} {\bibfield  {journal} {\bibinfo  {journal} {Nature Photonics}\
  }\textbf {\bibinfo {volume} {9}},\ \bibinfo {pages} {320} (\bibinfo {year}
  {2015})}\BibitemShut {NoStop}%
\bibitem [{\citenamefont {Blais}\ \emph {et~al.}(2020)\citenamefont {Blais},
  \citenamefont {Girvin},\ and\ \citenamefont {Oliver}}]{blais2020quantum}%
  \BibitemOpen
  \bibfield  {author} {\bibinfo {author} {\bibfnamefont {A.}~\bibnamefont
  {Blais}}, \bibinfo {author} {\bibfnamefont {S.~M.}\ \bibnamefont {Girvin}}, \
  and\ \bibinfo {author} {\bibfnamefont {W.~D.}\ \bibnamefont {Oliver}},\
  }\href@noop {} {\bibfield  {journal} {\bibinfo  {journal} {Nature Physics}\
  ,\ \bibinfo {pages} {1}} (\bibinfo {year} {2020})}\BibitemShut {NoStop}%
\bibitem [{\citenamefont {Onodera}\ \emph {et~al.}(2020)\citenamefont
  {Onodera}, \citenamefont {Ng},\ and\ \citenamefont
  {McMahon}}]{onodera2020quantum}%
  \BibitemOpen
  \bibfield  {author} {\bibinfo {author} {\bibfnamefont {T.}~\bibnamefont
  {Onodera}}, \bibinfo {author} {\bibfnamefont {E.}~\bibnamefont {Ng}}, \ and\
  \bibinfo {author} {\bibfnamefont {P.~L.}\ \bibnamefont {McMahon}},\
  }\href@noop {} {\bibfield  {journal} {\bibinfo  {journal} {npj Quantum
  Information}\ }\textbf {\bibinfo {volume} {6}},\ \bibinfo {pages} {1}
  (\bibinfo {year} {2020})}\BibitemShut {NoStop}%
\end{thebibliography}%
\end{document}